\documentclass[opre,nonblindrev]{informs3modified}

\DoubleSpacedXI 

\usepackage{endnotes}
\usepackage{bm}
\usepackage{algorithm}
\usepackage{algorithmic}
\usepackage{bbold}
\usepackage{multirow}

\let\footnote=\endnote
\let\enotesize=\normalsize
\def\notesname{Endnotes}%
\def\makeenmark{\hbox to1.275em{\theenmark.\enskip\hss}}
\def\enoteformat{\rightskip0pt\leftskip0pt\parindent=1.275em
  \leavevmode\llap{\makeenmark}}

\usepackage{natbib}
 \bibpunct[, ]{(}{)}{,}{a}{}{,}%
 \def\bibfont{\small}%
\TheoremsNumberedThrough     
\ECRepeatTheorems

\EquationsNumberedThrough    

\begin{document}


\RUNAUTHOR{Lam, Zhang and Zhang}

\RUNTITLE{Enhanced Bias-Variance Balancing: A Minimax Perspective}

\TITLE{Enhanced Balancing of Bias-Variance Tradeoff in Stochastic Estimation: A Minimax Perspective}

\ARTICLEAUTHORS{%
\AUTHOR{Henry Lam}
\AFF{Department of Industrial Engineering and Operations Research, Columbia University, New York, NY 10027, \EMAIL{khl2114@columbia.edu}} 
\AUTHOR{Xinyu Zhang}
\AFF{Department of Industrial Engineering and Operations Research, Columbia University, New York, NY 10027, \EMAIL{zhang.xinyu@columbia.edu}}
\AUTHOR{Xuhui Zhang}
\AFF{School for the Gifted Young, University of Science and Technology of China, \EMAIL{zxh1998@mail.ustc.edu.cn}}
} 

\ABSTRACT{%
Biased stochastic estimators, such as finite-differences for noisy gradient estimation, often contain parameters that need to be properly chosen to balance impacts from the bias and the variance. While the optimal order of these parameters in terms of the simulation budget can be readily established, the precise best values depend on model characteristics that are typically unknown in advance. We introduce a framework to construct new classes of estimators, based on judicious combinations of simulation runs on sequences of tuning parameter values, such that the estimators consistently outperform a given tuning parameter choice in the conventional approach, regardless of the unknown model characteristics. We argue the outperformance via what we call the asymptotic minimax risk ratio, obtained by minimizing the worst-case asymptotic ratio between the mean square errors of our estimators and the conventional one, where the worst case is over any possible values of the model unknowns. In particular, when the minimax ratio is less than 1, the calibrated estimator is guaranteed to perform better asymptotically. We identify this minimax ratio for general classes of weighted estimators, and the regimes where this ratio is less than 1. Moreover, we show that the best weighting scheme is characterized by a sum of two components with distinct decay rates. We explain how this arises from bias-variance balancing that combats the adversarial selection of the model constants, which can be analyzed via a tractable reformulation of a non-convex optimization problem.
}%



\maketitle

%


\section{Introduction}
\label{sec:intro}

This paper studies biased stochastic estimators which, in the simplest form, are expressed as follows. To estimate a target quantity of interest $\theta\in\mathbb R$, we use Monte Carlo simulation where each simulation run outputs
\begin{equation}
\hat\theta(\delta)=\theta+b(\delta)+v(\delta)\label{framework}
\end{equation}
Here $v(\delta)$ represents the noise of the simulation and satisfies $E[v(\delta)]=0$, and $b(\delta)$ is the bias given by $E[\hat\theta(\delta)]-\theta$. We obtain the final estimate by averaging $n$ independent runs produced by \eqref{framework}:
\begin{equation}
\frac{1}{n}\sum_{j=1}^n\hat\theta_j(\delta)\label{simple average}
\end{equation}
where $\hat\theta_j(\cdot)$ denotes an independent run.

The simulation runs in \eqref{framework} are specified by a parameter $\delta$ that typically impacts the bias and the variance in an antagonistic fashion. A common example is finite-difference schemes for black-box or zeroth-order noisy gradient estimation, in which $\delta$ is the perturbation size for the function input of interest. As $\delta$ increases, bias increases while variance decreases (and vice versa). To minimize the mean square error (MSE), the best choice of $\delta$, in terms of the simulation budget, balances the magnitudes of the two error sources. In central finite-difference for instance, this optimal $\delta$ turns out to be of order $n^{-1/6}$, whereas in forward or backward finite-difference it is of order $n^{-1/4}$ (e.g., \cite{glasserman2013monte} Chapter 7; \cite{asmussen2007stochastic} Chapter 7; \cite{fu2006gradient}; \cite{l1991overview}).


While the above tradeoff and the optimal order of $\delta$ in $n$ is well understood in the literature, the precise best choice of $\delta$ depends on other, typically unknown, model characteristics (i.e., the ``constants" inside $b(\delta)$ and $v(\delta)$). For example, choosing $\delta=dn^{-1/6}$ in a central finite-difference, and considering only the first-order error term, the best choice of $d$ depends on third-order derivative information and the variance of the noise that are typically unavailable in advance.




Our goal in this paper is to develop a framework that enhances the standard estimator \eqref{simple average} regarding the choice of $\delta$ subject to the ambiguity of the model characteristics. A key idea we will undertake is to consider estimators beyond the form of naive sample average, in a way that reduces the impact of this uncertainty. Under this framework, we derive new estimators that consistently improve \eqref{simple average} at a given choice of $\delta$, regardless of these unknowns. This improvement is in terms of the asymptotic MSE as the simulation budget increases. More specifically, we consider the asymptotic ratio between the MSEs of any proposed estimator and \eqref{simple average}:
\begin{equation}
R=\limsup_{n\to\infty}\frac{\text{MSE of a proposed estimator}}{\text{MSE of the conventional estimator \eqref{simple average}}}\label{AR}
\end{equation}
The proposed estimator can be parametrized by possibly many tuning parameters. The asymptotic ratio $R$ thus contains these parameters, the unknown model characteristics, and the $\delta$ in \eqref{simple average}. Regarding \eqref{simple average} and its $\delta$ as a ``baseline", we calibrate the tuning parameters in the proposed estimator to minimize the worst-case asymptotic MSE ratio, where the worst case is over all possible model characteristics and choices of $\delta$. On a high level, this can be expressed as
\begin{equation}
R^*=\min_{\substack{\text{calibration}\\\text{strategy}}}\max_{\substack{\text{model}\\\text{characteristics}},\delta}R\label{optimal ratio}
\end{equation}
This minimized worst-case ratio $R^*$ provides a performance guarantee on our calibrated proposed estimator relative to \eqref{simple average} -- The MSE of our estimator is asymptotically at most $R^*$ of \eqref{simple average} at the chosen $\delta$, independent of any possible model specifications. In particular, if $R^*<1$, our estimator is guaranteed to strictly improve over \eqref{simple average}. For convenience, we call $R^*$ the \emph{asymptotic minimax risk ratio (AMRR)}.


As our main contributions, we systematically identify the AMRR $R^*$, achieve $R^*<1$, and construct a scheme that consistently outperforms the conventional choice \eqref{simple average}, over the class of weighted estimator in the form
\begin{equation}
\sum_{j=1}^nw_j\hat\theta_j(\delta_j)\label{weighted}
\end{equation}
where $\delta_j,j=1,\ldots,n$ is a suitable sequence of tuning parameters, and $w_j,j=1,\ldots,n$ is any weighting sequence. When $w_i$'s are the uniform weights, \eqref{weighted} is precisely the ``recursive estimator" introduced in \cite{glynn1992asymptotic}. This latter estimator selects $\delta_j$ as if, roughly speaking, the current simulation run $j$ is the last one in the budget. In other words, given that $\delta=\Theta(n^{-\alpha})$ achieves the optimal MSE order for \eqref{simple average}, it selects $\delta_j=\Theta(j^{-\alpha})$, and it can be shown to exhibit the same optimal MSE order (the term ``recursive" refers to the fact that it can be obtained by iteratively reweighting existing estimates and new runs that depend only on the current run index). This construction can be generalized to stochastic approximation (SA) type recursions (e.g., \cite{dlz18}), and their averaging version (as in \cite{polyak1992acceleration}). Our main results show that, in general, the optimal weighting scheme to obtain $R^*$ is in the form
\begin{equation}
w_j=\frac{\lambda_1}{j^{\beta_1}}+\frac{\lambda_2}{j^{\beta_2}}\label{optimal weight}
\end{equation}
where $\beta_1,\beta_2>0$ are two distinct decay rates. The two coefficients $\lambda_1,\lambda_2$ depends on the budget $n$, in a way that none of the two terms in \eqref{optimal weight} is asymptotically negligible when used in the weighted estimator. This weighting scheme, and an associated transformation from $\delta$ to $\{\delta_j\}_j$, give rise to an explicitly identifiable $R^*$ that decreases with an ``inflation" factor imposed on the $\delta$-transformation. This reveals that, for instance, in the central finite-difference scheme, $R^*$ is $0.67$ when the multiplicative constants in $\delta$ and $\{\delta_j\}_j$ are the same. Since $R^*<1$, the weighted estimator using \eqref{optimal weight} always outperforms \eqref{simple average} in terms of asymptotic MSE, independent of the unknown constants in $b(\delta)$ and $v(\delta)$. In contrast, the corresponding $R^*$ is $1.08$ when the weights are obtained via standard SA recursion, either with or without averaging, indicating that such a restriction on the weighting sequence could lead to subpar performance in the MSE.

Our main analyses build on the insight that, to maintain a low worst-case risk ratio, one typically must calibrate the proposed estimator such that it maintains the relative magnitudes of bias and variance in a similar manner as the conventional scheme \eqref{simple average}. We will show that any distortion away from such a balancing allows an ``adversary" to enlarge the risk ratio, thus leading to suboptimal outcomes. This balancing requirement generally leads to a non-convex constrained optimization problem which, upon a reformulation, reveals a tractable structure and solution to the minimax problem in \eqref{optimal ratio}.


The remainder of the paper is as follows. Section \ref{sec:lit} first reviews some related works. Section \ref{sec:setting} describes the problem settings and reviews some established results on biased estimation. Section \ref{sec:framework} presents our minimax framework and investigation on a special class of estimators. Section \ref{sec:main} presents our main results and explains their implications on general weighted estimators and AMRR. Section \ref{sec:generalizations} discusses how our results carry to multivariate settings. Section \ref{sec:numerics} reports our numerical experiments. Section \ref{sec:conclusion} concludes the paper. All proofs are in the Appendix.

\section{Related Literature}\label{sec:lit}
Our study is related to several lines of work. The minimax formulation that we use to analyze and construct estimators resembles robust optimization (e.g., \cite{ben2009robust,bertsimas2011theory,ben2002robust}) and robust control (e.g., \cite{zhou1998essentials}) that advocates decision-making against the worst-case scenario. Such ideas also have roots in game theory (\cite{cesa2006prediction}). Related notions have also been used in online optimization, in which decision is made at each step under a noisily observed dynamical process (e.g., \cite{flaxman2005online,shalev2012online,hazan2016introduction}). The performance in this literature is often measured by the regret that indicates the suboptimality of a decision relative to the best decision assuming complete information (see, e.g.,  \cite{besbes2009dynamic,besbes2011minimax} for applications in revenue management). Instead of using an ``oracle" best as the benchmark in our minimax formulation, we use the sample average as our benchmark, and focus on improving this conventional estimator by analyzing the risk ratio. In this regard, we note that a ratio formulation and a non-oracle-best benchmark has been used in \cite{agrawal2012price}, but in a different context in quantifying the impact of correlation in mean estimation, and their benchmark is an independent distribution with the worst-case being evaluated over a class of dependent models. Ratios between MSEs also appear in \cite{pasupathy2010choosing} in studying the tradeoff between error tolerance and sample size in so-called retrospective approximation, which is a technique for solving stochastic root-finding or optimization problems via imposing a sequence of sample average approximation problems.

A main application of our work is finite-difference stochastic gradient estimation (e.g., \cite{glasserman2013monte} Chapter 7; \cite{asmussen2007stochastic} Chapter 7; \cite{fu2006gradient}; \cite{l1991overview}), typically used when there is only a noisy simulation oracle to evaluate the function value or model output. Variants of the finite-difference method include the central, forward and backward finite-differences, with different perturbation directions and orders of bias (\cite{zazanis1993convergence,fox1989replication}). In contrast to finite-differences are unbiased derivative estimators, which include the infinitesimal perturbation analysis or pathwise differentiation (\cite{ho1983infinitesimal,heidelberger1988convergence}), the likelihood ratio or the score function method (\cite{glynn1990likelihood,rubinstein1986score,reiman1989sensitivity}), measure-valued or weak differentiation (\cite{heidergott2008measure,heidergott2010gradient}), and other variants such as the push-out method (\cite{rubinstein1992sensitivity,l1990unified}), conditional and smoothed perturbation analysis (\cite{gong1987smoothed,hong2009estimating,fu1992extensions,glasserman1990smoothed,fu2009conditional}) and the generalized likelihood ratio method (\cite{peng2018new}). In multivariate settings, \cite{spall1992multivariate,spall1997one} study simultaneous perturbation to estimate gradients used in SA, by randomly generating a perturbation direction vector and properly weighting with the perturbation sizes to control estimation bias. \cite{nesterov2017random} proposes Gaussian smoothing with a different adjustment and investigates finite-sample behaviors in related optimization. \cite{flaxman2005online} suggests uniform sampling. Our framework can be applied to these procedures, as will be discussed in Section \ref{sec:generalizations}.

The main skeleton of our proposed estimators uses a sequentialized choice of the tuning parameter, which appears in \cite{glynn1992asymptotic} in their discussion of subcanonical estimators. A special case of our scheme, discussed in Section \ref{sec:framework}, resembles the idea of SA in stochastic optimization and root-finding that iteratively updates noisy estimates (\cite{kushner2003stochastic,borkar2009stochastic,pasupathy2011stochastic,nemirovski2009robust,polyak1992acceleration,ruppert1988efficient}). Our analyses there utilize the classical asymptotic techniques in \cite{fabian1968asymptotic} and \cite{chung1954stochastic}, and also \cite{polyak1992acceleration} in the averaging case.





We close this section by briefly comparing our work to multi-level Monte Carlo (\cite{giles2008multilevel}). This approach aims to reduce variance in simulation in the presence of a parameter selection like $\delta$, by stratifying the simulation budget into different $\delta$ values. Of particular relevance is the randomized level selection (\cite{rhee2015unbiased,blanchet2015unbiased,rychlik1990unbiased,mcleish2010general}) that can turn biased estimators in the form of \eqref{framework} into unbiased estimators with canonical square-root convergence. This approach has been applied in the simulation of stochastic differential equations and nonlinear functions of expectations, and requires a proper coupling between simulation runs at consecutive levels to control the simulation effort. In investigating \eqref{framework}, we do not assume any problem structure that allows such coupling, and our performance is benchmarked against the conventional (biased) sample-average scheme.

\section{Background and Problem Setting}\label{sec:setting}

We elaborate our problem and notations in the introduction. We are interested in estimating $\theta\in\mathbb{R}$. Given a tuning parameter $\delta\in\mathbb R_+$, we run Monte Carlo simulation where each run outputs
\begin{equation}
\hat{\theta}(\delta)=\theta+b(\delta)+v(\delta)\label{output}
\end{equation}
with $b(\delta)=B\delta^{q_{1}}+o(\delta^{q_{1}})$ as $\delta\to0$, $v(\delta)=\frac{\varepsilon(\delta)}{\delta^{q_{2}}}$, and $q_{1},q_{2}>0$. We assume that:
\begin{assumption}
We have
\begin{enumerate}
\item $B\in\mathbb{R}$ is a non-zero constant.

\item $\varepsilon(\delta)\in\mathbb{R}$ is a random variable such that $E\varepsilon(\delta)=0$ and $\sigma^2(\delta)=Var(\varepsilon(\delta))\to\sigma^2>0$ as $\delta\to0$.

\end{enumerate}\label{main assumption}
\end{assumption}
The above assumptions dictate that the order of the bias $b(\delta)$ is $\delta^{q_1}$, while the order of the variance is $\delta^{-2q_2}$. The former is ensured by the first assumption and the latter by the second one.

As an example, in estimating the derivative of a function $f(x)$ with unbiased noisy function evaluation, the central finite-difference (CFD) scheme elicits the output
$$\frac{\hat f(x+\delta)-\hat f(x-\delta)}{2\delta}$$
where $\hat f(\cdot)$ is an unbiased evaluation of $f(\cdot)$, and $\delta>0$ is the perturbation size. Given that $f$ is thrice continuously differentiable with non-zero $f'''(x)$, the bias term has order $q_1=2$. Typically, the order of the variance is $q_2=1$. Suppose we do not apply common random numbers (CRN) in generating $\hat f(x+\delta)$ and $\hat f(x-\delta)$, and that $Var(\hat f(x\pm\delta))\to Var(\hat f(x))$ as $\delta\to0$, then $\sigma^2=Var(\hat f(x))/2$. Suppose we are able to apply CRN so that $Cov(\hat f(x+\delta),\hat f(x-\delta))\to\rho$ as $\delta\to0$ (i.e., we cannot fully eliminate the first-order variance as $\delta$ shrinks), then we have $\sigma^2=(Var(\hat f(x))-\rho)/2$.

Similarly, the forward finite-difference (FFD) scheme elicits the output
$$\frac{\hat f(x+\delta)-\hat f(x)}{\delta}$$
Given that $f$ is twice continuously differentiable with non-zero $f''(x)$, the bias term has order $q_1=1$. Analogous conditions on the noise as above guarantees that $q_2=1$. The same discussion holds for the backward finite-difference (BFD) scheme.

Given the capability to output independent runs of \eqref{output}, say $\hat\theta_j(\delta)$, the conventional approach to obtain an estimate of $\theta$ is to take their sample average. Denote this as $\bar\theta_n=(1/n)\sum_{j=1}^n\hat\theta_j(\delta)$. The MSE of $\bar\theta_n$, denoted $\text{MSE}_0=E(\bar\theta_n-\theta)^2$, can be expressed as
\begin{equation}
\text{MSE}_0=\text{bias}^2+\text{variance}=B^2\delta^{2q_{1}}+\frac{\sigma^2}{n\delta^{2q_2}}+\text{higher-order terms}\label{first-order tradeoff}
\end{equation}
Considering the first order term, the bias increases with $\delta$ and the variance decreases with $\delta$. Minimizing the MSE requires balancing these two errors to the same order, namely by choosing $\delta=\Theta(n^{-\alpha})$ where $\alpha=1/(2(q_1+q_2))$, which solves the equation $-2\alpha q_1=-1+2\alpha q_2$. This leads to an optimal MSE order $n^{-q_1/(q_1+q_2)}$. For example, in CFD and under the conditions we discussed above, we have $\delta=\Theta(n^{-1/6})$, leading to an optimal MSE order $n^{-2/3}$; in FFD or BFD we have $\delta=\Theta(n^{-1/4})$, leading to an optimal MSE order $n^{-1/2}$.

In order to fully optimize the first-order MSE, including the coefficient, one needs to choose
$$\delta= \left(\frac{\sigma^2q_2}{B^2q_1}\right)^{1/(2(q_1+q_2))}\frac{1}{n^{1/(2(q_1+q_2))}}$$
(e.g., by applying the first-order optimality condition on the leading terms in \eqref{first-order tradeoff}). This gives an optimal first-order MSE
\begin{equation}
B^{2q_2/(q_1+q_2)}\sigma^{2q_1/(q_1+q_2)}\left(\left(\frac{q_2}{q_1}\right)^{q_1/(q_1+q_2)}+\left(\frac{q_1}{q_2}\right)^{q_2/(q_1+q_2)}\right)\frac{1}{n^{q_1/(q_1+q_2)}}\label{oracle}
\end{equation}
The above choice of $\delta$ depends on the ``constants" in the bias and variance terms, namely $B$ and $\sigma^2$. While $q_1$ and $q_2$ are often obtainable, constants like $B$ and $\sigma^2$ are unknown a priori and can affect the performance of the simulation estimator, despite choosing an optimal order on $n$ in $\delta$ using the knowledge of $q_1$ and $q_2$. Suppose we choose $\delta=d/n^\alpha$ for some $d>0$, where $\alpha=1/(2(q_1+q_2))$ is optimally chosen. Then the first-order MSE is
\begin{equation}
\left(B^2d^{2q_1}+\frac{\sigma^2}{d^{2q_2}}\right)\frac{1}{n^{q_1/(q_1+q_2)}}\label{MSE suboptimal}
\end{equation}
which can be arbitrarily suboptimal relative to the best coefficient in \eqref{oracle}. Our goal in this paper is to improve on this suboptimality, by considering estimators beyond the conventional sample average that consistently outperforms the constant showing up in \eqref{MSE suboptimal}.

The following theorem, which follow straightforwardly from \cite{fox1989replication}, summarizes the above discussion on the optimal order of the MSE:

\begin{theorem}\label{offmse single}
Under Assumption \ref{main assumption}, suppose that $\lim_{n\to\infty}\delta n^{\alpha}=d>0$, the sample-average-based estimator $\bar\theta_n$ exhibits the asymptotic MSE
$$E(\bar{\theta}_{n}-\theta)^{2}=d^{2q_{1}}B^{2}n^{-2\alpha q_1}+\frac{\sigma^2}{d^{2q_{2}}}n^{2\alpha q_2-1}+o(n^{-2\alpha q_1}+n^{2\alpha q_2-1})\textrm{ as }n\to\infty$$
Choosing $\alpha=1/(2(q_1+q_2))$ achieves the optimal MSE order, and the asymptotic MSE is
$$E(\bar{\theta}_n-\theta)^2=\big(d^{2q_1}B^2+\frac{\sigma^2}{d^{2q_2}}\big)n^{-q_1/(q_1+q_2)}+o(n^{-q_1/(q_1+q_2)})\text{ as }n\to\infty$$
\end{theorem}

Lastly, we mention that, in practice, there are other considerations in obtaining good estimators, such as issues regarding the finiteness of the sample that can affect the accuracy of the asymptotic results. These considerations are beyond the scope of this work, which focuses mainly on a theoretical framework on improving the asymptotic constant.

\section{A Minimax Comparison Framework}\label{sec:framework}
We introduce a framework to assess, and calibrate, estimators beyond the sample-average-based estimator $\bar\theta_n$. This framework compares the asymptotic MSEs using $\bar\theta_n$ as a baseline based on a minimax argument. Section \ref{sec:comparison} presents this framework, and Section \ref{sec:initial} provides an initial study on a special type of estimators.

\subsection{Asymptotic Risk Ratio}\label{sec:comparison}
Consider an estimator $\hat\theta_n$ for $\theta$ using $n$ simulation runs in the form \eqref{output}, where the tuning parameter $\delta$ in each run can be arbitrarily chosen. Our goal is to calibrate $\hat\theta_n$ that performs well, or outperforms, $\bar\theta_n$ in the first-order coefficient of the MSE, presuming that both $\hat\theta_n$ and $\bar\theta_n$ have the optimal order of errors. Let $\text{MSE}_1$ denote the MSE of $\hat\theta_n$ for convenience.

The estimator $\hat\theta_n$ can depend on other tuning parameters in addition to the $\delta$ in each run. We denote the collection of all the parameters that $\hat\theta_n$ involves as $\nu$, so that $\hat\theta_n=\hat\theta_n(\nu)$. Correspondingly, $\text{MSE}_1$ also depends on $\nu$.

We suppose knowledge on the order of the bias and noise, namely $q_1$ and $q_2$ in \eqref{output}. However, we do not know the constants $B$ and $\sigma^2$. To make the discussion more precise, for fixed $q_1,q_2>0$, we denote the class of simulation outputs
\begin{eqnarray}
\Theta&=&\big\{\hat\theta(\cdot):\hat\theta(\delta)=\theta+b(\delta)+v(\delta)\text{\ such that\ }\notag\\
&&{}b(\delta)=B\delta^{q_1}+o(\delta^{q_1})\text{\ and\ }v(\delta)=\frac{\varepsilon(\delta)}{\delta^{q_2}}\text{\ where\ }Var(\epsilon(\delta))\to\sigma^2,\text{\ as\ }\delta\to0,{}\notag\\
&&{}\text{\ for arbitrary non-zero $B$ and positive $\sigma^2$}\big\}\label{simulation class}
\end{eqnarray}
In other words, $\Theta$ is the set of outputs with bias of order $\delta^{q_1}$ and noise of order $\delta^{-q_2}$, with arbitrary constants $B$, $\sigma^2$, and the higher-order error terms.

The MSE of $\bar\theta_n$, $\text{MSE}_0$, depends on $\hat\theta(\cdot)$ evaluated at chosen $\delta$. To highlight this dependence, we write $\text{MSE}_0=\text{MSE}_0(\hat\theta(\cdot),\delta)$. Similarly, $\text{MSE}_1$ depends on $\hat\theta(\cdot)$ and $\nu$, so that $\text{MSE}_1=\text{MSE}_1(\hat\theta(\cdot),\nu)$. We consider the \emph{asymptotic risk ratio}
\begin{equation}
R(\hat\theta(\cdot),\nu,\delta)=\limsup_{n\to\infty}\frac{\text{MSE}_1(\hat\theta(\cdot),\nu)}{\text{MSE}_0(\hat\theta(\cdot),\delta)}\label{asymptotic risk ratio expression}
\end{equation}
that measures the performance of $\hat\theta_n$ relative to $\bar\theta_n$ as a baseline. Since we only know $\hat\theta(\cdot)$ is in $\Theta$ but not its exact forms (i.e., the constants), we consider the worst-case scenario of $R$, and search for the best parameters in $\hat\theta_n$ that minimize this worst-case risk. Namely, we aim to solve
\begin{equation}
\min_{\nu}\max_{\hat\theta(\cdot)\in\Theta}R(\hat\theta(\cdot),\nu,\delta)\label{minimax risk prelim}
\end{equation}
Note that \eqref{minimax risk prelim}, and the best choice of $\nu$, depend on the $\delta$ used in $\bar\theta_n$. We now take a further viewpoint that an arbitrary user may select any $\delta$, and we look for a strategy to calibrate $\hat\theta_n$ that is guaranteed to perform well no matter how $\delta$ is chosen. To write this more explicitly, we let $\nu=\nu(\delta)$ be dependent on $\delta$, and we search for the best collection of parameters that is potentially a function $\nu(\cdot)$ on $\delta$:
\begin{equation}
R^*=\min_{\nu(\cdot)\in\Lambda}\max_{\hat\theta(\cdot)\in\Theta,\delta\in\mathbb R}R(\hat\theta(\cdot),\nu(\cdot),\delta)\label{minimax risk}
\end{equation}
where $\Lambda$ denotes the set of admissible functions $\nu(\cdot)$. This set $\Lambda$ depends on the class of estimators $\hat\theta_n$ we use, which will be described in detail. Moreover, as we will see, \eqref{minimax risk prelim} and \eqref{minimax risk} are closely related; in fact, under the settings we consider, solving either of them simultaneously solves another. In the following, we will focus on \eqref{minimax risk} and discuss the immediate implications on \eqref{minimax risk prelim} where appropriate. We shall call $R^*$ the asymptotic minimax risk ratio (AMRR).



\subsection{An Initial Example: Recursive Estimators}\label{sec:initial}
For convenience, let us from now on set $\delta=d(n+n_0)^{-\alpha}$ as the tuning parameter in the sample-average-based estimator $\bar\theta_n$, where $\alpha=1/(2(q_1+q_2))$ so that it achieves the optimal MSE order. The number $n_0$ can be any fixed integer to prevent $\delta$ from being too big at the early stage, and does not affect our asymptotic analyses.

To construct our proposed estimator $\hat\theta_n$, we will first use the idea of the recursive estimator studied in Section 5 of \cite{glynn1992asymptotic}. At run $j$, we simulate $\hat\theta_j(\delta_j)$, where $\delta_j=\tilde d(j+n_0)^{-\alpha}$ for some constant $\tilde d$, and $\alpha$ is the same as in $\bar\theta_n$, i.e., the parameter is chosen as if the current simulation run is the last one in the budget if a conventional sample-average-based estimator is used. The estimator in \cite{glynn1992asymptotic} uses the average of $\hat\theta_j(\delta_j)$, namely $(1/n)\sum_{j=1}^n\hat\theta_j(\delta_j)$. As shown in  \cite{glynn1992asymptotic}, this estimator exhibits the optimal MSE order like $\bar\theta_n$. Moreover, as they have also noted, this estimator admits a recursive representation $\hat\theta_n=(1-1/n)\hat\theta_{n-1}+(1/n)\hat\theta_n(\delta_n)$, where each update depends only on the parameter indexed by the current run number, rather than the budget. Thus, the optimal MSE order is achieved in an ``online" fashion as $n$ increases, independent of the final budget.

The initial class of estimators that we will consider is a generalization of \cite{glynn1992asymptotic}. Specifically, we consider estimators defined via the recursion
\begin{equation}
\hat\theta_n^{rec}=\left(1-\gamma_n\right)\hat\theta_{n-1}^{rec}+\gamma_n\hat\theta_n(\delta_n)\label{recursive def}
\end{equation}
where $\delta_n=\tilde d(n+n_0)^{-\alpha}$ is defined as before and $\alpha>0$, and $\gamma_n$ is in the form $c(n+n_0)^{-\beta}$ for some $c>0$ and $\beta>0$. $\hat\theta_0^{rec}$ can be arbitrary. Moreover, we also consider averaging  $\hat\theta_n^{rec}$ in the form
\begin{equation}
\hat\theta_n^{avg}=\frac{1}{n}\sum_{j=1}^n\hat\theta_j^{rec}\label{averaging def}
\end{equation}
which resembles the standard Polyak-Ruppert averaging in SA (\cite{polyak1992acceleration}).

Our first result is that, in terms of the AMRR, the class of estimators $\hat\theta_n^{rec}$ and $\hat\theta_n^{avg}$ are quite restrictive and cannot bring in much improvement over $\bar\theta_n$. To elicit this result, we begin with some consistency properties of $\hat\theta_n^{rec}$:


\begin{proposition}\label{ondiv single}
Under Assumption \ref{main assumption}, we have:
\begin{enumerate}
\item If $\beta\leq1$ and $\alpha<\beta/(2q_{2})$, the estimator $\hat\theta_n^{rec}$ is $L_2$-consistent for $\theta$, i.e.,
$$\lim_{n\to\infty}E(\hat\theta_n^{rec}-\theta)^2=0$$
\item If $\beta\leq1$ and $\alpha\geq\beta/(2q_{2})$, or if $\beta>1$, the error of $\hat\theta_n^{rec}$ in estimating $\theta$ is bounded away from zero in $L_{2}$-norm as $n\to\infty$, i.e.,
$$\liminf_{n\to\infty}E(\hat\theta_{n}^{rec}-\theta)^{2}>0$$\end{enumerate}
\end{proposition}




Proposition \ref{ondiv single} shows that $\hat\theta_n^{rec}$ estimates $\theta$ sensibly only when $\beta\leq1$ and $\alpha<\beta/(2q_2)$. We thus focus on this case subsequently. The following describes the convergence rate:

\begin{theorem}\label{onmse single}
Under Assumption \ref{main assumption}, the MSE of $\hat\theta_n^{rec}$ in estimating $\theta$ behaves as follows:
\begin{enumerate}
\item For $\beta<1$ and $\alpha<\beta/(2q_2)$,
{\small
$$E(\hat\theta_n^{rec}-\theta)^{2}=d^{2q_{1}}B^{2}n^{-2q_{1}\alpha}+\frac{c\sigma^2}{2d^{2q_{2}}}n^{2q_{2}\alpha-\beta} +o(n^{-2q_{1}\alpha}+n^{2q_{2}\alpha-\beta})\text{ as }n\to\infty$$
}
\item For $\beta=1$, $\alpha=1/(2(q_1+q_2))$ and $c>q_1/(2(q_1+q_2))$,
{\small
$$E(\hat\theta_n^{rec}-\theta)^2=\left(\big(\frac{cd^{q_{1}}}{c-q_{1}/(2(q_1+q_2))}\big)^{2}B^{2}+\frac{c^{2}\sigma^2}{(2c-q_1/(q_1+q_2))d^{2q_{2}}} \right)n^{-q_1/(q_1+q_2)}+o(n^{-q_1/(q_1+q_2)})\text{ as }n\to\infty$$
}
\item For $\beta=1$, $\alpha=1/(2(q_1+q_2))$ and $c\leq q_1/(2(q_1+q_2))$, or for $\beta=1$ and $\alpha\neq1/(2(q_1+q_2))$,
$$\limsup_{n\to\infty}n^{q_1/(q_1+q_2)}E(\hat\theta_n^{rec}-\theta)^2=\infty$$
\end{enumerate}
\end{theorem}

The proofs of the above results, which are detailed in Appendix \ref{sec:proof recursive}, utilize the classical asymptotic techniques for recursive sequences in \cite{fabian1968asymptotic} and a slight modification of Chung's lemma (i.e., Lemma \ref{chung new} in Appendix \ref{sec:proof recursive}).



We now look at the AMRR for $\hat\theta_n^{rec}$. First, Theorem \ref{onmse single} shows that the choice $\beta=1,\alpha=1/(2(q_1+q_2))$ is the unique choice that gives rise to the optimal MSE order $n^{-q_1/(q_1+q_2)}$. Moreover, given this choice of $\alpha$, we need $c>q_1/(2(q_1+q_2))$, in addition to $\beta=1$. We will focus on these configurations for $\hat\theta_n^{rec}$ that achieve the same MSE order as the conventional estimator $\bar\theta_n$ with the same $\alpha$.


Suppose we set $\tilde d=d$, but allow the free selection of $c$ within the range that gives rise to the optimal MSE order. We thus can write $\hat\theta_n^{rec}=\hat\theta_n^{rec}(d,c)$, defined via \eqref{recursive def} with $\gamma_n=c(n+n_0)^{-1}$ where $c>q_1/(2(q_1+q_2))$. 
The integer $n_0$ does not affect any asymptotic and can be taken as any given value. The following characterizes the AMRR and the configuration that attains it:

\begin{theorem}\label{comd single}
Under Assumption \ref{main assumption}, let $\text{MSE}_1^{rec}(\hat\theta(\cdot),d,c)$ be the MSE of $\hat\theta_n^{rec}(d,c)$, and 
$$R^{rec}(\hat\theta(\cdot),d,c)=\limsup_{n\to\infty}\frac{\text{MSE}_1^{rec}(\hat\theta(\cdot),d,c)}{\text{MSE}_0(\hat\theta(\cdot),d)}$$
We have
$$\min_{c>\frac{q_1}{2(q_1+q_2)}}\,\max_{\hat\theta(\cdot)\in\Theta,d>0} R^{rec}(\hat\theta(\cdot),d,c)=\frac{q_{1}^{2}}{16(q_{1}+q_{2})^{2}}+\frac{q_{1}}{2(q_{1}+q_{2})}+1$$
which is attained by choosing $c=\frac{5q_{1}+4q_{2}}{2(q_{1}+q_{2})}$. 

\end{theorem}


Next, we provide more flexibility in the choice of $\tilde d$ in $\hat\theta_n^{rec}(\tilde d,c)$. 
In particular, rather than setting $\tilde d=d$, we allow $\tilde d$ to depend on $d$ in any arbitrary fashion, i.e., $\tilde d=g(d)$ where $g(\cdot):\mathbb R_+\to\mathbb R_+$ is any function. Let $\mathcal F$ be the space of any functions from $\mathbb R_+$ to $\mathbb R_+$. We have the following results on the AMRR of this enhanced scheme:

\begin{theorem}\label{uncomd single}
Under Assumption \ref{main assumption}, let $\text{MSE}_1^{rec}(\hat\theta(\cdot),\tilde d,c)$ be the MSE of $\hat\theta_n^{rec}(\tilde d,c)$, and 
$$R^{rec}(\hat\theta(\cdot),d,\tilde d,c)=\limsup_{n\to\infty}\frac{\text{MSE}_1^{rec}(\hat\theta(\cdot),\tilde d,c)}{\text{MSE}_0(\hat\theta(\cdot),d)}$$
We have
$$\min_{g(\cdot)\in\mathcal F,c>\frac{q_1}{2(q_1+q_2)}}\,\max_{\hat\theta(\cdot)\in\Theta, d>0} R^{rec}(\hat\theta(\cdot),d,g(d),c)=2^{\frac{2q_{2}}{q_{1}+q_{2}}}(\frac{q_{1}+2q_{2}}{q_{1}+q_{2}})^{-\frac{q_{1}+2q_{2}}{q_{1}+q_{2}}}$$
which is attained by choosing $g(d)=(\frac{q_{1}+2q_{2}}{4(q_{1}+q_{2})})^{\frac{1}{2(q_{1}+q_{2})}}d$ and $c=1$. 
\end{theorem}


We note that Theorem \ref{uncomd single} indicates $c=1$ is optimal in this enhanced scheme, while the optimal $\tilde d$ is chosen as a constant factor $((q_{1}+2q_{2})/(4(q_{1}+q_{2})))^{1/(2(q_{1}+q_{2}))}$ of $d$.


Next we look at $\hat\theta_n^{avg}$. It turns out that the AMRR depicted for $\hat\theta_n^{rec}$ in Theorem \ref{uncomd single} applies also to $\hat\theta_n^{avg}$. To this end, we first state the MSE of $\hat\theta_n^{avg}$:
\begin{theorem}\label{averagemse}
Under Assumption \ref{main assumption}, the MSE of $\hat\theta_n^{avg}$ in estimating $\theta$ behaves as follows:
\begin{enumerate}
\item For $\beta<1$ and $\alpha\leq1/(2(q_1+q_2))$,
$$E(\hat\theta_n^{avg}-\theta)^2=\big(\frac{d^{q_{1}}}{1-q_{1}\alpha}\big)^{2}B^{2}n^{-2q_{1}\alpha}+\frac{\sigma^2}{(1+2q_{2}\alpha)d^{2q_{2}}} n^{2q_{2}\alpha-1}+o(n^{-2q_{1}\alpha}+n^{2q_{2}\alpha-1})\text{ as }n\to\infty$$
\item For $\beta<1$ and $\alpha>1/(2(q_1+q_2))$,
$$E(\hat\theta_n^{avg}-\theta)^2=\frac{\sigma^2}{(1+2q_{2}\alpha)d^{2q_{2}}} n^{2q_{2}\alpha-1}+o(n^{2q_{2}\alpha-1})\text{ as }n\to\infty$$
\end{enumerate}
\end{theorem}

Comparing Theorem \ref{averagemse} with Theorem \ref{onmse single}, we see that the first-order MSE of $\hat\theta_n^{avg}$ in the considered regime exactly equals that of $\hat\theta_n^{rec}$ when $c=1$ and $\beta=1$. Like before, $\alpha=1/(2(q_1+q_2))$ is the unique choice that optimizes the MSE order for $\hat\theta_n^{avg}$. Thus, we will focus on this choice of $\alpha$ in $\hat\theta_n^{avg}$. Note that then $\hat\theta_n^{avg}=\hat\theta_n^{avg}(\tilde d,c,\beta)$ depends on $\tilde d,c,\beta$. This leads us to the following AMRR:
\begin{theorem}\label{avg single}
Under Assumption \ref{main assumption}, let $\text{MSE}_1^{avg}(\hat\theta(\cdot),\tilde d,c,\beta)$ be the MSE of $\hat\theta_n^{avg}=\hat\theta_n^{avg}(\tilde d,c,\beta)$. Let
 $$R^{avg}(\hat\theta(\cdot),d,\tilde d,c,\beta)=\limsup_{n\to\infty}\frac{\text{MSE}_1^{avg}(\hat\theta(\cdot),\tilde d,c,\beta)}{\text{MSE}_0(\hat\theta(\cdot),d)}$$
We have
$$\min_{g(\cdot)\in\mathcal F,c>0,0<\beta<1}\,\max_{\hat\theta(\cdot)\in\Theta, d>0} R^{avg}(\hat\theta(\cdot),d,g(d),c,\beta)=2^{\frac{2q_{2}}{q_{1}+q_{2}}}(\frac{q_{1}+2q_{2}}{q_{1}+q_{2}})^{-\frac{q_{1}+2q_{2}}{q_{1}+q_{2}}}$$
which is attained by choosing $g(d)=(\frac{q_{1}+2q_{2}}{4(q_{1}+q_{2})})^{\frac{1}{2(q_{1}+q_{2})}}d$, and any $c>0$ and $0<\beta<1$.
\end{theorem}

The minimax ratios stated in Theorems \ref{comd single}, \ref{uncomd single} and \ref{avg single} remain the same, in a uniform fashion, when the parameter $d$ in $\bar\theta_n$ is fixed instead of being chosen by an adversarial user. In other words, the minimax risk ratio of $\hat\theta_n^{rec}$ or $\hat\theta_n^{avg}$
compared to $\bar\theta_n$ would not improve with a finer calibration on the tuning parameters $\tilde d,c,\beta$ catered to each specific $d$. This is described in the following result:

\begin{theorem}\label{fixed parameter}
We have the following:
\begin{enumerate}
\item Under the conditions and notations in Theorem \ref{comd single}, we have, for any fixed $d$,
$$\min_{c>\frac{q_1}{2(q_1+q_2)}}\,\max_{\hat\theta(\cdot)\in\Theta} R^{rec}(\hat\theta(\cdot),d,c)=\frac{q_{1}^{2}}{16(q_{1}+q_{2})^{2}}+\frac{q_{1}}{2(q_{1}+q_{2})}+1$$
which is attained by choosing $c=\frac{5q_{1}+4q_{2}}{2(q_{1}+q_{2})}$.
\item Under the conditions and notations in Theorem \ref{uncomd single}, we have, for any fixed $d$,
$$\min_{\tilde d>0,c>\frac{q_1}{2(q_1+q_2)}}\,\max_{\hat\theta(\cdot)\in\Theta} R^{rec}(\hat\theta(\cdot),d,\tilde d,c)=2^{\frac{2q_{2}}{q_{1}+q_{2}}}(\frac{q_{1}+2q_{2}}{q_{1}+q_{2}})^{-\frac{q_{1}+2q_{2}}{q_{1}+q_{2}}}$$
which is attained by choosing $\tilde d=(\frac{q_{1}+2q_{2}}{4(q_{1}+q_{2})})^{\frac{1}{2(q_{1}+q_{2})}}d$ and $c=1$.
\item Under the conditions and notations in Theorem \ref{avg single}, we have, for any fixed $d$,
$$\min_{\tilde d>0,c>0,0<\beta<1}\,\max_{\hat\theta(\cdot)\in\Theta} R^{avg}(\hat\theta(\cdot),d,\tilde d,c,\beta)=2^{\frac{2q_{2}}{q_{1}+q_{2}}}(\frac{q_{1}+2q_{2}}{q_{1}+q_{2}})^{-\frac{q_{1}+2q_{2}}{q_{1}+q_{2}}}$$
which is attained by choosing $\tilde d=(\frac{q_{1}+2q_{2}}{4(q_{1}+q_{2})})^{\frac{1}{2(q_{1}+q_{2})}}d$, and any $c>0$ and $0<\beta<1$. 
\end{enumerate}
\end{theorem}

Theorem \ref{fixed parameter} is consistent with Theorems \ref{comd single}, \ref{uncomd single} and \ref{avg single} in that the optimal strategies in calibrating the $\tilde d$ in $\hat\theta_n^{rec}$ and $\hat\theta_n^{avg}$
remain as a constant scaling on $d$, regardless of what the specific value of $d$ is.




To get a numerical sense of the above results, Tables \ref{table:q1is2} and \ref{table:q1is1} show the AMRR and optimal configurations of $\hat\theta_n^{rec}$ and $\hat\theta_n^{avg}$. Table \ref{table:q1is2} illustrates the scenario $q_1=2$ and $q_2=1$ (the CFD case). Restricting $\tilde d=d$ in $\hat\theta_n^{rec}$ 
(i.e., Theorem \ref{comd single}), the AMRR is $1.38$, attained by setting $c=2.33$ in $\hat\theta_n^{rec}$. 
In contrary, if we allow $\tilde d$ to arbitrarily depend on $d$ (i.e., Theorem \ref{uncomd single}), the AMRR is reduced to $1.08$, attained by setting $g(d)=0.83d$, and $c=1$ in  $\hat\theta_n^{rec}$. Similarly, the AMRR for $\hat\theta_n^{avg}$ (i.e., Theorem \ref{avg single}) is also $1.08$, attained again by setting $g(d)=0.83d$ but now with any $c>0$ and $0<\beta<1$.

Analogously, Table \ref{table:q1is1} illustrates the scenario $q_1=1$ and $q_2=1$ (the FFD and BFD cases). If we restrict $\tilde d=d$ in $\hat\theta_n^{rec}$ 
(i.e., Theorem \ref{comd single}), the AMRR becomes $1.27$, attained by setting $c=2.25$ in $\hat\theta_n^{rec}$. 
In contrary, if we allow $\tilde d$ to arbitrarily depend on $d$ (i.e., Theorems \ref{uncomd single} and \ref{avg single}), the AMRR is $1.09$, attained by setting $g(d)=0.78d$, and $c=1$ in  $\hat\theta_n^{rec}$ or $c>0,0<\beta<1$ in $\hat\theta_n^{avg}$.

Note that, in all cases considered above, the AMRR is greater than 1, implying that without knowledge on the model characteristics, the estimators $\hat\theta_n^{rec}$ and $\hat\theta_n^{avg}$ can have a higher MSE than the baseline $\bar\theta_n$ asymptotically.

\begin{table}[htb]
\begin{small}
\begin{center}
\begin{tabular}{|c|c|c|c|}
  \hline
   & $\hat\theta_n^{rec}$ ($d$ unadjusted) & $\hat\theta_n^{rec}$ ($d$ optimized) & $\hat\theta_n^{avg}$ \\\hline
AMRR& 1.38 &  1.08 & 1.08\\\hline
Optimal Configuration & $c=2.33$, $\beta=1$ & $\tilde d=0.83d,c=1,\beta=1$&$\tilde d=0.83d,c>0,0<\beta<1$\\\hline
\end{tabular}
\end{center}\end{small}
\caption{\label{table.label} AMRR and optimal configurations for the case $q_1=2,q_2=1$} 
\label{table:q1is2}
\end{table}

\begin{table}[htb]
\begin{small}
\begin{center}
\begin{tabular}{|c|c|c|c|}
  \hline
   & $\hat\theta_n^{rec}$ ($d$ unadjusted) & $\hat\theta_n^{rec}$ ($d$ optimized) & $\hat\theta_n^{avg}$ \\\hline
AMRR& 1.27 &  1.09 & 1.09\\\hline
Optimal Configuration & $c=2.25,\beta=1$ & $\tilde d=0.78d,c=1,\beta=1$&$\tilde d=0.78d,c>0,0<\beta<1$\\\hline
\end{tabular}
\end{center}\end{small}
\caption{\label{table.label} AMRR and optimal configurations for the case $q_1=1,q_2=1$} 
\label{table:q1is1}
\end{table}



\subsection{Maintaining Bias-Variance Balance}\label{sec:maintain balance}
We provide an intuitive explanation on the minimax results in Section \ref{sec:initial}. More specifically, we demonstrate that a key argument to obtain the minimax calibration strategy of a proposed class of estimators is to balance bias and variance in a similar manner as the baseline estimator, in terms of the factors multiplying the unknown first-order constants $B$ and $\sigma^2$. This insight is general and will be helpful in optimally calibrating wider classes of estimators, such as the general weighted estimators presented in the next section.

To explain, let us recall the notation in \eqref{asymptotic risk ratio expression} that in general, the asymptotic risk ratio between a proposed estimator with parameter $\nu$ and a baseline estimator (where we hide its parameter for now) can be expressed as
$$R(\hat\theta(\cdot),\nu)=\limsup_{n\to\infty}\frac{\text{MSE}_1(\hat\theta(\cdot),\nu)}{\text{MSE}_0(\hat\theta(\cdot))}$$
Suppose that both estimators have the same MSE order, which is obtained optimally by balancing the orders of the bias and variance. Then the limit in the above expression becomes
\begin{equation}
R(\hat\theta(\cdot),\nu)=\frac{\text{bias}_1(\nu)^2+\text{var}_1(\nu)}{\text{bias}_0^2+\text{var}_0}\label{explanation interim}
\end{equation}
where $\text{bias}_1(\nu)$ and $\text{var}_1(\nu)$ refer to the first-order coefficient in the bias and variance terms of the proposed estimator, and similarly $\text{bias}_0$ and $\text{var}_0$ refer to the corresponding quantities of the baseline estimator. Furthermore, with the model constants $B$ and $\sigma^2$, we can further write \eqref{explanation interim} as
$$R(\hat\theta(\cdot),\nu)=\frac{C_1^{bias}(\nu)B^2+C_1^{var}(\nu)\sigma^2}{C_0^{bias}B^2+C_0^{var}\sigma^2}$$
where $C_1^{bias}(\nu)$ and $C_1^{var}(\nu)$ are the coefficients in front of $B^2$ and $\sigma^2$ in the first-order MSE of the proposed estimator, and $C_0^{bias}$ and $C_0^{var}$ are the corresponding quantities of the baseline estimator.

Now, given these coefficients, an adversary who attempts to maximize $R(\hat\theta(\cdot),\nu)$ would select either an arbitrarily big $B^2$ or $\sigma^2$ , depending on which ratio $C_1^{bias}(\nu)/C_0^{bias}$ or $C_1^{var}(\nu)/C_0^{var}$ is larger respectively, which leads to a worst-case  ratio $\max\{C_1^{bias}(\nu)/C_0^{bias},C_1^{var}(\nu)/C_0^{var}\}$. This typically enforces the minimizer to calibrate $\nu$ such that the two ratios are exactly the same, i.e., we choose $\nu$ such that
\begin{equation}
\frac{C_1^{bias}(\nu)}{C_0^{bias}}=\frac{C_1^{var}(\nu)}{C_0^{var}}=S\label{explanation interim1}
\end{equation}
for some constant $S$. With this observation, the solution to solve for AMRR can be formulated as minimizing $S$ subject to the constraint \eqref{explanation interim1}, namely
\begin{equation}
\min_\nu\ \ S\text{\ \ subject to\ \ }\frac{C_1^{bias}(\nu)}{C_0^{bias}}=\frac{C_1^{var}(\nu)}{C_0^{var}}=S\label{explanation interim3}
\end{equation}
which gives the AMRR $R^*$, and an optimal solution for \eqref{explanation interim3} is the minimax calibration for the proposed estimator. This line of analysis applies similarly when the baseline estimator contains its own tuning parameter $\delta$, and that the proposed estimator is calibrated in a way dependent on $\delta$ (either in formulation \eqref{minimax risk prelim} or \eqref{minimax risk}).


Now let us consider $\hat\theta_n^{rec}$ in Theorem \ref{comd single}. From Theorems \ref{offmse single} and \ref{onmse single}, since we assume both the parameters of $\bar\theta_n$ and $\hat\theta_n^{rec}$ are chosen to exhibit optimal MSE order, we can write
\begin{align*}
R^{rec}(\hat\theta(\cdot),d,c)&=\limsup_{n\to\infty}\frac{\text{MSE}_1^{rec}(\hat\theta(\cdot),d,c)}{\text{MSE}_0(\hat\theta(\cdot),d)}\\
&=\limsup_{n\to\infty}\frac{\left((\frac{cd^{q_{1}}}{c-\frac{q_{1}}{2(q_{1}+q_{2})}})^{2}B^{2}+\frac{c^{2}\sigma^2}{2d^{2q_{2}}(c-\frac{q_{1}}{2(q_{1}+q_{2})})}\right)n^{-\frac{q_{1}}{q_{1}+q_{2}}}+o(n^{-\frac{q_{1}}{q_{1}+q_{2}}})}{\big(d^{2q_{1}}B^{2}+\frac{\sigma^2}{d^{2q_{2}}}\big)n^{-\frac{q_{1}}{q_{1}+q_{2}}}+o(n^{-\frac{q_{1}}{q_{1}+q_{2}}})}\\
&=\frac{(\frac{cd^{q_{1}}}{c-\frac{q_{1}}{2(q_{1}+q_{2})}})^{2}B^{2}+\frac{c^{2}}{2d^{2q_{2}}(c-\frac{q_{1}}{2(q_{1}+q_{2})})}\sigma^2}{d^{2q_{1}}B^{2}+\frac{1}{d^{2q_{2}}}\sigma^2}
\end{align*}
We set
$$\frac{(\frac{cd^{q_{1}}}{c-\frac{q_{1}}{2(q_{1}+q_{2})}})^{2}}{d^{2q_1}}=\frac{\frac{c^{2}}{2d^{2q_{2}}(c-\frac{q_{1}}{2(q_{1}+q_{2})})}}{\frac{1}{d^{2q_2}}}$$
and notice that $d$ can be all cancelled out, giving
$$(\frac{c}{c-\frac{q_{1}}{2(q_{1}+q_{2})}})^{2}=\frac{c^{2}}{2(c-\frac{q_{1}}{2(q_{1}+q_{2})})}$$
which upon solving leads to $c=\frac{5q_{1}+4q_{2}}{2(q_{1}+q_{2})}$ and both sides of the equation being $\frac{q_{1}^{2}}{16(q_{1}+q_{2})^{2}}+\frac{q_{1}}{2(q_{1}+q_{2})}+1$, thus giving the corresponding result in Theorem \ref{comd single}. Note that, since $d$ is cancelled out in the above derivation, the same result in Theorem \ref{fixed parameter} holds immediately for the setting of any fixed $d$.


For $\hat\theta_n^{rec}$ in Theorem \ref{uncomd single}, we can write $$R^{rec}(\hat\theta(\cdot),d,\tilde d,c)=\frac{(\frac{c\tilde d^{q_{1}}}{c-\frac{q_{1}}{2(q_{1}+q_{2})}})^{2}B^{2}+\frac{c^{2}}{2\tilde d^{2q_{2}}(c-\frac{q_{1}}{2(q_{1}+q_{2})})}\sigma^2}{d^{2q_{1}}B^{2}+\frac{1}{d^{2q_{2}}}\sigma^2}$$
and we set
\begin{equation}
\frac{(\frac{c\tilde d^{q_{1}}}{c-\frac{q_{1}}{2(q_{1}+q_{2})}})^{2}}{ d^{2q_1}}=\frac{\frac{c^{2}}{2\tilde d^{2q_{2}}(c-\frac{q_{1}}{2(q_{1}+q_{2})})}}{\frac{1}{d^{2q_2}}}\label{explanation interim2}
\end{equation}
However, the $d$ is not cancelled out here. Nonetheless, we can rewrite \eqref{explanation interim2} in terms of the ratio $\tilde d/d$, as
$$(\frac{c}{c-\frac{q_{1}}{2(q_{1}+q_{2})}})^{2}\left(\frac{\tilde d}{d}\right)^{2q_1}=\frac{c^{2}}{2(c-\frac{q_{1}}{2(q_{1}+q_{2})})}\frac{1}{\left(\frac{\tilde d}{d}\right)^{2q_2}}$$
Optimizing jointly over $c$ and $\eta=\tilde d/d$ gives $c=1$ and $\eta=(\frac{q_{1}+2q_{2}}{4(q_{1}+q_{2})})^{\frac{1}{2(q_{1}+q_{2})}}$, and the value on both sides of the equation is $2^{\frac{2q_{2}}{q_{1}+q_{2}}}(\frac{q_{1}+2q_{2}}{q_{1}+q_{2}})^{-\frac{q_{1}+2q_{2}}{q_{1}+q_{2}}}$. This shows the result for $\hat\theta_n^{rec}$ in Theorem \ref{uncomd single}. Moreover, note that regardless of whether $d$ is chosen by the adversary or fixed in advance, we choose $\tilde d$ as $\eta d$, and thus we also show the corresponding results in Theorem \ref{fixed parameter}. 
Appendix \ref{sec:proof recursive} further details the above arguments.



\section{General Weighted Estimators}\label{sec:main}
We now consider a substantially more general class of estimators than $\hat\theta_n^{rec}$ and $\hat\theta_n^{avg}$. Namely, given we generate $\hat\theta_j(\delta_j),j=1,\ldots,n$ where $\delta_j=\tilde d(j+n_0)^{-\alpha}$ with the optimally chosen $\alpha=1/(2(q_1+q_2))$ and $n_0$ is any fixed integer, we consider
\begin{equation}
\hat\theta_n^{gen}=\sum_{j=1}^nw_{j,n}\hat\theta_j(\delta_j)\label{general weighted estimator}
\end{equation}
where $w^{(n)}=(w_{j,n})_{j=1,\ldots,n}$ is any weighting sequence.

In the following, we will first present our main result on the AMRR of \eqref{general weighted estimator} relative to $\bar\theta_n$ with $\delta=d(n+n_0)^{-\alpha}$, and the associated characterization of the optimal weighting scheme as a sum of two distinct decaying components (Section \ref{sec:AMRR main}). Then we will describe the key developments of the result that relies on analyzing a non-convex constrained optimization (Section \ref{sec:opt}).

\subsection{Optimal Weighted Estimators and Two-Decay Characterization}\label{sec:AMRR main}
The estimator $\hat\theta_n^{gen}$ in \eqref{general weighted estimator} contains the tuning parameter $\tilde d$ and the weighting sequence $w^{(n)}$. While $\tilde d$ is chosen independent of $n$ in the asymptotic (as it appears in the asymptotic risk ratio that is independent of $n$), the sequence $\{w^{(n)}\}_{n=1,2,\ldots}$ is a triangular array of $w_{j,n}$ as $n\to\infty$. For convenience, we denote $W=\{w^{(n)}\}_{n=1,2,\ldots}$ as this array. We write $\text{MSE}_1^{gen}(\hat\theta(\cdot),\tilde d,w^{(n)})$ as the MSE of $\hat\theta_n^{gen}=\hat\theta_n^{gen}(\tilde d,w^{(n)})$, and recall $\text{MSE}_0(\hat\theta(\cdot),d)$ as the MSE of the baseline estimator $\bar\theta_n=\bar\theta_n(d)$. We define
$$R^{gen}(\hat\theta(\cdot),d,\tilde d,W)=\limsup_{n\to\infty}\frac{\text{MSE}_1^{gen}(\hat\theta(\cdot),\tilde d,w^{(n)})}{\text{MSE}_0(\hat\theta(\cdot),d)}$$
as the asymptotic risk ratio between $\hat\theta_n^{gen}$ and $\bar\theta_n$.

Moreover, we impose a condition on the magnitude of $\tilde d$ relative to $d$. In particular, we restrict $\tilde d$ to be at most $Kd$ for some constant $K>0$. Suppose we consider calibration of $\tilde d$ as a function $g(\cdot)$ on $d$. This is equivalent to requiring $g(d)\leq Kd$ for any $d$, for a maximal inflation factor $K>0$. This assumption makes sense since our AMRR calculation relies on asymptotic arguments and thus, if $K$ is too large, the inherited large magnitude of the tuning parameter $\delta$ in the proposed estimator can affect the finite-sample behavior significantly and discount the accuracy of the asymptotic calculation. Relatedly, we will see that if $K$ is unrestricted, $\hat\theta_n^{gen}$ can achieve zero AMRR, which does not reveal useful practical information; in fact, as $K\to\infty$, we will have an AMRR that gradually decays to zero.


Denote
$$\mathcal F_K=\{g(\cdot):g(d)\leq Kd\}$$
$\mathcal W$ as the space of any triangular array, and $\Theta$ as in \eqref{simulation class}. We consider the AMRR
$$\min_{g(\cdot)\in\mathcal F_K,W\in\mathcal W}\max_{\hat\theta(\cdot)\in\Theta,d>0}R^{gen}(\hat\theta(\cdot),d,g(d),W)$$
We have the following identification of the AMRR and the characterization of optimal calibration:

\begin{theorem}\label{main thm}
Under Assumption \ref{main assumption}, we have the following:
\begin{enumerate}
\item The AMRR of $\hat\theta_n^{gen}$ satisfies\begin{equation}
\min_{g(\cdot)\in\mathcal F_K,W\in\mathcal W}\max_{\hat\theta(\cdot)\in\Theta,d>0}R^{gen}(\hat\theta(\cdot),d,g(d),W)=\frac{q_1}{q_1+q_2}\frac{1}{K^{2q_2}}\label{interim5}
\end{equation}
\item The weights $W^*=(w_{j,n}^*)_{\substack{j=1,\ldots,n\\n=1,2,\ldots}}$ that achieve \eqref{interim5} is given by
$$w_{j,n}^*=\frac{\lambda_1^*}{(j+n_0)^{(q_1+2q_2)/2(q_1+q_2)}}+\frac{\lambda_2^*}{(j+n_0)^{q_2/(q_1+q_2)}}$$
where $\lambda_1^*,\lambda_2^*$ are solved by
\begin{equation}
\left[\begin{array}{c}\lambda_1\\\lambda_2\end{array}\right]=\left[\begin{array}{cc}\xi_{11}&\xi_{12}\\\xi_{21}&\xi_{22}\end{array}\right]\left[\begin{array}{c}a^*\\1\end{array}\right]\label{revised interim2}
\end{equation}
and $a^*$ is an optimal solution to
\begin{equation}
\min_{a:(K^{2(q_1+q_2)}-\xi_{11})a^2-2\xi_{12}a-\xi_{22}\geq0}|a|^{2q_2/(q_1+q_2)}\left(\xi_{11}a^2+2\xi_{12}a+\xi_{22}\right)^{q_1/(q_1+q_2)}\label{solve a}
\end{equation}
where
$$\left[\begin{array}{cc}\xi_{11}&\xi_{12}\\\xi_{21}&\xi_{22}\end{array}\right]=\left[\begin{array}{cc}\phi\left(1\right)&\phi\left(\frac{q_1+2q_2}{2(q_1+q_2)}\right)\\\phi\left(\frac{q_1+2q_2}{2(q_1+q_2)}\right)&\phi\left(\frac{q_2}{q_1+q_2}\right)\end{array}\right]^{-1}$$
and $\phi(\kappa)=\sum_{j=1}^n1/(j+n_0)^\kappa$. Moreover, $g(\cdot)$ is defined by $g(d)=Kd$.
\end{enumerate}
\end{theorem}

Next, we also note the same result if we fix $d$ in the baseline estimator $\bar\theta_n$, uniformly for any $d$:

\begin{corollary}\label{main cor}
Under the conditions and notations in Theorem \ref{main thm}, we have, for any fixed $d$,
$$\min_{\substack{\tilde d=g(d):g(\cdot)\in\mathcal F_K\\W\in\mathcal W}}\max_{\hat\theta(\cdot)\in\Theta}R^{gen}(\hat\theta(\cdot),d,\tilde d,W)=\frac{q_1}{q_1+q_2}\frac{1}{K^{2q_2}}$$
which is attained by  the weights $W^*=(w_{j,n}^*)_{\substack{j=1,\ldots,n\\n=1,2,\ldots}}$ and setting $\tilde d=Kd$ that achieve the AMRR in part 2 of Theorem \ref{main thm}.
\end{corollary}


We discuss several implications of Theorem \ref{main thm}. First, the optimal weighting sequence $w_{j,n}^*$ comprises two components, each with a different decay rate, i.e., $(q_1+2q_2)/(2(q_1+q_2))$ and $q_2/(q_1+q_2)$ respectively.
The coefficients in these decays, namely $\lambda_1^*$ and $\lambda_2^*$, depend on $n$ that is solved via a linear system of equations, which ensures that none of the two components in $w_{j,n}^*$ is asymptotically negligible.

To illustrate the latter point, we demonstrate the asymptotic behaviors of $\lambda_1^*,\lambda_2^*$, which are revealed by first understanding the behavior of $a^*$ and using \eqref{revised interim2}. Note that $\phi(\kappa)\sim\frac{1}{1-\kappa}n^{1-\kappa}$ for $\kappa<1$ and $\sim\log n$ for $\kappa=1$. Thus, the matrix
\begin{align}
\left[\begin{array}{cc}\xi_{11}&\xi_{12}\\\xi_{21}&\xi_{22}\end{array}\right]&=\left[\begin{array}{cc}\phi\left(1\right)&\phi\left(\frac{q_1+2q_2}{2(q_1+q_2)}\right)\\\phi\left(\frac{q_1+2q_2}{2(q_1+q_2)}\right)&\phi\left(\frac{q_2}{q_1+q_2}\right)\end{array}\right]^{-1}\notag\\
&\sim\left[\begin{array}{cc}\log n&\frac{2(q_1+q_2)}{q_1}n^{q_1/(2(q_1+q_2))}\\\frac{2(q_1+q_2)}{q_1}n^{q_1/(2(q_1+q_2))}&\frac{q_1+q_2}{q_1}n^{q_1/(q_1+q_2)}\end{array}\right]^{-1}\notag\\
&=\frac{1}{\frac{q_1+q_2}{q_1}n^{q_1/(q_1+q_2)}\log n-\frac{4(q_1+q_2)^2}{q_1^2}n^{q_1/(q_1+q_2)}}\left[\begin{array}{cc}\frac{q_1+q_2}{q_1}n^{q_1/(q_1+q_2)}&-\frac{2(q_1+q_2)}{q_1}n^{q_1/(2(q_1+q_2))}\\-\frac{2(q_1+q_2)}{q_1}n^{q_1/(2(q_1+q_2))}&\log n\end{array}\right]\label{revised interim1}
\end{align}
where the asymptotic equivalence ``$\sim$" is on every entry of the matrix.

Now, conjecturing that $a^*$ is of order $1/n^{q_1/(2(q_1+q_2))}$, we write $a=\tilde a/n^{q_1/(2(q_1+q_2))}$. By plugging in \eqref{revised interim1}, we have
\begin{align*}\xi_{11}a^2+2\xi_{12}a+\xi_{22}&=\left[\frac{\tilde a}{n^{q_1/(2(q_1+q_2))}}\ \ 1\right]\left[\begin{array}{cc}\xi_{11}&\xi_{12}\\\xi_{21}&\xi_{22}\end{array}\right]\left[\begin{array}{c}\frac{\tilde a}{n^{q_1/(2(q_1+q_2))}}\\1\end{array}\right]\\
&\sim\frac{1}{n^{q_1/(q_1+q_2)}}\left[\tilde a\ \ 1\right]\left[\begin{array}{cc}0&0\\0&\frac{q_1}{q_1+q_2}\end{array}\right]\left[\begin{array}{c}\tilde a\\1\end{array}\right]\\
&=\frac{q_1}{q_1+q_1}\frac{1}{n^{q_1/(q_1+q_2)}}
\end{align*}
Thus, as $n\to\infty$, an ``asymptotic" version of \eqref{solve a}, when multiplying the objective value by $n^{q_1/(q_1+q_2)}$, becomes
$$\min_{\tilde a: K^{2(q_1+q_2)}\tilde a^2\geq\frac{q_1}{q_1+q_2}}|\tilde a|^{2q_2/(q_1+q_2)}\left(\frac{q_1}{q_1+q_2}\right)^{q_1/(q_1+q_2)}$$
which gives $|\tilde a|=\sqrt{q_1/(q_1+q_2)}(1/K^{q_1+q_2})$. This implies that
\begin{equation}
a^*\sim\sqrt{\frac{q_1}{q_1+q_2}}\frac{1}{K^{q_1+q_2}}\frac{1}{n^{q_1/(2(q_1+q_2))}}\label{revised interim3}
\end{equation}
Thus, putting \eqref{revised interim1} and \eqref{revised interim3} into \eqref{revised interim2}, we obtain that
\begin{equation}
\lambda_1^*\sim \left(\sqrt{\frac{q_1}{q_1+q_2}}\frac{1}{K^{q_1+q_2}}-2\right)\frac{1}{n^{q_1/(2(q_1+q_2))}\log n}\label{revised interim4}
\end{equation}
and
\begin{equation}
\lambda_2^*\sim\frac{q_1}{q_1+q_2}\frac{1}{n^{q_1/(q_1+q_2)}}\label{revised interim5}
\end{equation}

We can now see that both terms in $w_{j,n}^*$, namely $\frac{\lambda_1^*}{(j+n_0)^{(q_1+2q_2)/2(q_1+q_2)}}$ and $\frac{\lambda_2^*}{(j+n_0)^{q_2/(q_1+q_2)}}$, contribute to the first-order bias. Note that the first-order bias is of order $\sum_{j=1}^nw_{j,n}\delta_j^{q_1}$, where $\delta_j=\tilde d(j+n_0)^{-\alpha}$ and $\alpha=1/(2(q_1+q_2))$. Thus, using \eqref{revised interim4}, the bias contribution from the first component in $w_{j,n}^*$ gives rise to an order
\begin{align}
\frac{1}{n^{q_1/(2(q_1+q_2))}\log n}\sum_{j=1}^n\frac{1}{(j+n_0)^{(q_1+2q_2)/(2(q_1+q_2))}}\frac{1}{(j+n_0)^{q_1/(2(q_1+q_2))}}&=\frac{1}{n^{q_1/(2(q_1+q_2))}\log n}\sum_{j=1}^n\frac{1}{j+n_0}\notag\\
&=\Theta\left(\frac{1}{n^{q_1/(2(q_1+q_2))}}\right)\label{revised interim6}
\end{align}
On the other hand, using \eqref{revised interim5}, the bias contribution from the second component in $w_{j,n}^*$ gives rise to an order
\begin{align*}
\frac{1}{n^{q_1/(q_1+q_2)}}\sum_{j=1}^n\frac{1}{(j+n_0)^{q_2/(q_1+q_2)}}\frac{1}{(j+n_0)^{q_1/(2(q_1+q_2))}}&=\frac{1}{n^{q_1/(q_1+q_2)}}\sum_{j=1}^n\frac{1}{(j+n_0)^{(q_1+2q_2)/(2(q_1+q_2))}}\\
&=\Theta\left(\frac{1}{n^{q_1/(2(q_1+q_2))}}\right)
\end{align*}
which is the same order as \eqref{revised interim6}. Thus both terms in $w_{j,n}^*$ contribute significantly to the first-order bias term.

Similarly, the first-order variance is of order $\sum_{j=1}^nw_{j,n}^2\delta_j^{-2q_2}$. Using \eqref{revised interim4}, the contribution from the first component in $w_{j,n}^*$ gives rise to an order
\begin{align}
\frac{1}{n^{q_1/(q_1+q_2)}(\log n)^2}\sum_{j=1}^n\frac{1}{(j+n_0)^{(q_1+2q_2)/(q_1+q_2)}}(j+n_0)^{q_2/(q_1+q_2)}&=\frac{1}{n^{q_1/(q_1+q_2)}(\log n)^2}\sum_{j=1}^n\frac{1}{j+n_0}\notag\\
&=\Theta\left(\frac{1}{n^{q_1/(q_1+q_2)}\log n}\right)\label{revised interim7}
\end{align}
and, using \eqref{revised interim5}, the contribution from the second component gives rise to an order
\begin{align*}
\frac{1}{n^{2q_1/(q_1+q_2)}}\sum_{j=1}^n\frac{1}{(j+n_0)^{2q_2/(q_1+q_2)}}(j+n_0)^{q_2/(q_1+q_2)}&=\frac{1}{n^{2q_1/(q_1+q_2)}}\sum_{j=1}^n\frac{1}{(j+n_0)^{q_2/(q_1+q_2)}}\\
&=\Theta\left(\frac{1}{n^{q_1/(q_1+q_2)}}\right)
\end{align*}
which has an order larger than \eqref{revised interim7} by a logarithmic factor. Thus, considering also the cross term between the two components in $w_{j,n}^*$ in the expansion of the variance, the first-order variance is of order $1/n^{q_1/(q_1+q_2)}$, which is the same as the squared bias.

Next we present some basic numerical values of the AMRR. Table \ref{table:gamma_K} shows the values of the AMRR for various maximal inflation factor $K$ when $q_1=2$ and $q_2=1$ (the CFD case). The AMRR is non-increasing in $K$, as advocated in Theorem \ref{main thm} and making intuitive sense since increasing $K$ places more optimizing power for the proposed estimator and hence drives down the AMRR. The critical threshold of $K$ above which  $\hat\theta_n^{gen}$ is guaranteed to improve over $\bar\theta_n$ is $K=\sqrt{2/3}=0.82$. In particular, when $K=1$ (we only allow choosing $\tilde d$ as large as $d$ at most), we have the AMRR equal to $2/3$, which is strictly less than 1. In other words, no matter what are the values of the model unknowns, the optimized calibration of $\hat\theta_n^{gen}$, in particular the two-decay weights $\{w_{j,n}^*\}_{j=1,\ldots,n}$ and setting $\tilde d=d$, would achieve a better MSE than $\bar\theta_n$ asymptotically.

\begin{figure}
  \centering
  \includegraphics[width=5in]{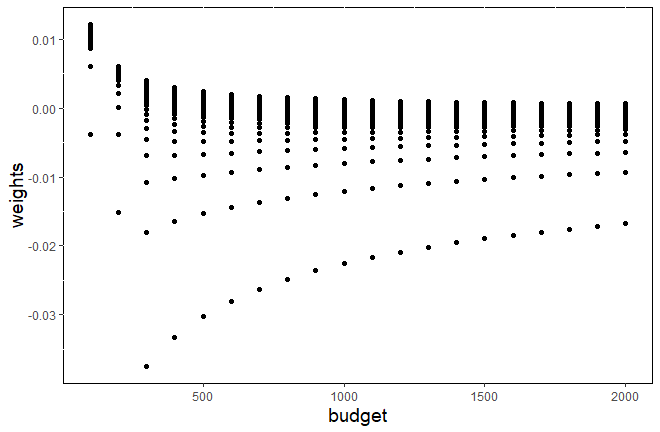}\\
  \caption{Distribution of weights, with $K=1$, and budget $n$ from $100$ to $2000$,  when $q_1=2,q_2=1$}\label{fig3}
\end{figure}

\begin{figure}
  \centering
  \includegraphics[width=5in]{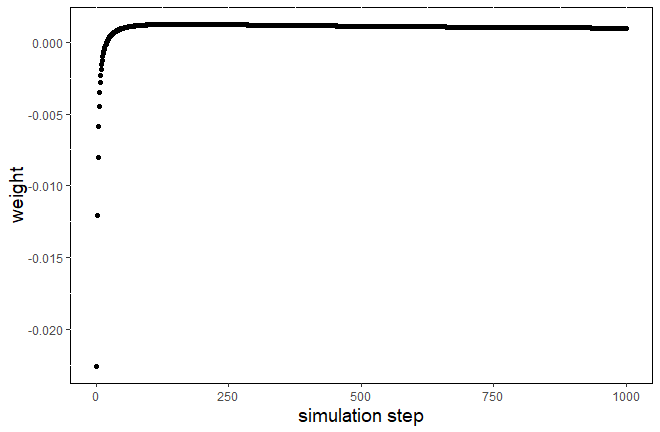}\\
  \caption{Distribution of weights against simulation step, with $K=1$, and budget $n=1000$,  when $q_1=2,q_2=1$}\label{fig4}
\end{figure}

Figures \ref{fig3} and \ref{fig4} show the behaviors of the optimal weights for $K=1$. Figure \ref{fig3} shows that in general the weights range across positive and negative numbers, with higher concentration around 0 as the budget increases. Figure \ref{fig4} shows that, against the simulation run index, the weight starts from the most negative and gradually increases to the positive region. Lastly, Table \ref{table:gamma_Kq1is1} shows the AMRR when $q_1=1,q_2=1$ (the FFD and BFD cases) as a comparison. The AMRR in this case has the same decay rate and is smaller than that for $q_1=2,q_2=1$ across all $K$.

\begin{table}[htb]
\begin{small}
\begin{tabular}{|c|c|c|c|c|c|c|c|c|c|c|c|c|c|c|c|c|}
  \hline
  $K$ & 0.5 & 0.6 & 0.7 & 0.8& 0.9 & 1.0 &1.1& 1.2& 1.3& 1.4& 1.5& 1.6& 1.7& 1.8& 1.9 & 2.0\\\hline
AMRR& 2.67& 1.85& 1.36& 1.04& 0.82& 0.67& 0.55& 0.46& 0.39&
 0.34& 0.30& 0.26& 0.23& 0.21& 0.18& 0.17\\\hline
\end{tabular}
\end{small}
\caption{\label{table.label} AMRR for general weighted estimators, against $K$,  when $q_1=2,q_2=1$} 
\label{table:gamma_K}
\end{table}





\begin{table}[htb]
\begin{small}
\begin{tabular}{|c|c|c|c|c|c|c|c|c|c|c|c|c|c|c|c|c|}
  \hline
  $K$ & 0.5 & 0.6 & 0.7 & 0.8& 0.9 & 1.0 &1.1& 1.2& 1.3& 1.4& 1.5& 1.6& 1.7& 1.8& 1.9 & 2.0\\\hline
AMRR& 2.00& 1.39& 1.02&0.78& 0.62& 0.50& 0.41& 0.34& 0.30& 0.26& 0.22&0.20& 0.17& 0.15& 0.14& 0.13 \\\hline
\end{tabular}
\end{small}
\caption{\label{table.label} AMRR for general weighted estimators, against $K$, when $q_1=1,q_2=1$} 
\label{table:gamma_Kq1is1}
\end{table}




\subsection{Constrained Optimization for Bias-Variance Balancing}\label{sec:opt}

We explain intuitively the key arguments that lead to the optimal two-decay weights $w_{j,n}^*$ and the identification of the AMRR in the form depicted in Theorem \ref{main thm}. We first note that to avoid arbitrarily large value of $R^{gen}$, the sequence $w_{j,n}$ must sum up to 1 (up to a vanishing error), since otherwise the scenario where $\hat\theta(\cdot)$ has no bias and noise but $\theta$ is arbitrarily big will blow up $R^{gen}$.

Thus, for simplicity let us assume that $\sum_{j=1}^nw_{j,n}=1$. Also, for convenience, we shorthand $w_j$ as $w_{j,n}$, and $w$ as $w^{(n)}$ when no confusion arises. Moreover, without loss of generality, here we assume $n_0=0$ for notational convenience. Considering the bias and variance of $\sum_{j=1}^nw_j\hat\theta_j(\delta_j)$, we can write
\begin{eqnarray}
\text{MSE}_1^{gen}(\hat\theta(\cdot),\tilde d,w)&=&\left(\sum_{j=1}^nw_jb(\delta_j)\right)^2+\sum_{j=1}^nw_j^2Var(v(\delta_j))\notag\\
&=&\left(\sum_{j=1}^nw_j\left(B\frac{\tilde d^{q_1}}{j^{\alpha q_1}}+o\left(\frac{1}{j^{\alpha q_1}}\right)\right)\right)^2+\sum_{j=1}^nw_j^2\frac{\sigma^2(1+o(1))j^{2\alpha q_2}}{\tilde d^{2q_2}}\notag\\
&=&\left(B\tilde d^{q_1}\sum_{j=1}^n\frac{w_j}{j^{\alpha q_1}}\right)^2+\frac{\sigma^2}{\tilde d^{2q_2}}\sum_{j=1}^nj^{2\alpha q_2}w_j^2+\text{error}\label{higher-order}
\end{eqnarray}
Recall the discussion in Section \ref{sec:maintain balance}. To control the adversary from increasing $R^{gen}$, we attempt to maintain the relative balance of bias and variance in a similar manner as the baseline. More specifically, presuming that $\hat\theta_n^{gen}$ exhibits the optimal MSE order $n^{-q_1/(q_1+q_2)}$, we keep the ratios of the coefficients in front of $B^2$ and $\sigma^2$ of the first-order MSE terms, between $\hat\theta_n^{gen}$ and $\bar\theta_n$, to be the same. The coefficient of the squared bias term is roughly
$$n^{q_1/(q_1+q_2)}\left(\tilde d^{q_1}\sum_{j=1}^n\frac{w_j}{j^{\alpha q_1}}\right)^2$$
while the coefficient of the variance term is roughly
$$n^{q_1/(q_1+q_2)}\frac{1}{\tilde d^{2q_2}}\sum_{j=1}^nj^{2\alpha q_2}w_j^2$$
Thus, similar to \eqref{explanation interim1}, we would like to ensure
\begin{equation}
n^{q_1/(q_1+q_2)}\left(\left(\frac{\tilde d}{d}\right)^{q_1}\sum_{j=1}^n\frac{w_j}{j^{\alpha q_1}}\right)^2=n^{q_1/(q_1+q_2)}\frac{1}{\left(\frac{\tilde d}{d}\right)^{2q_2}}\sum_{j=1}^nj^{2\alpha q_2}w_j^2\label{balance general}
\end{equation}
Denoting $\eta=\tilde d/d$, and dropping $n^{q_1/(q_1+q_2)}$ on both sides of \eqref{balance general}, we consider the optimization problem
\begin{equation}
\begin{array}{ll}
\min_{w,\eta}&S\\
\text{subject to}&S=\left(\eta^{q_1}\sum_{j=1}^n\frac{w_j}{j^{\alpha q_1}}\right)^2=\frac{1}{\eta^{2q_2}}\sum_{j=1}^nj^{2\alpha q_2}w_j^2\\
&\eta\leq K\\
&\sum_{j=1}^nw_j=1
\end{array}\label{opt new}
\end{equation}
Note that the first constraint is the bias-variance-balancing condition as in \eqref{explanation interim3}. The second and third constraints capture the inflation condition $g(\cdot)\in\mathcal F_K$ and $\sum_{j=1}^nw_j=1$. Denote the optimal value of \eqref{opt new} as $S_n^*$. Then roughly speaking, the AMRR would be $\lim_{n\to\infty}n^{q_1/(q_1+q_2)}S_n^*$. The associated optimal solution $w,\eta$ turns out to dominate any other possibilities, in particular those obtained by allowing any of the bias and variance terms dominate another.


In the rest of this subsection, we will explain how \eqref{opt new} leads to the two-decay representation of $w_{j,n}^*$, and leave other details to Appendix \ref{sec:proof main}. Note that \eqref{opt new} is non-convex. However, we can reformulate it into a convex program together with a simple one-dimensional line search over a region that consists of at most two intervals.

To this end, first notice that from the first constraint in \eqref{opt new}, we have
\begin{equation}\eta=\left(\frac{\sum_{j=1}^nj^{2\alpha q_2}w_j^2}{\left(\sum_{j=1}^n\frac{w_j}{j^{\alpha q_1}}\right)^2}\right)^{1/(2(q_1+q_2))}\label{interim1 new}
\end{equation}
so that the second constraint is equivalent to
$$\sum_{j=1}^nj^{2\alpha q_2}w_j^2\leq K^{2(q_1+q_2)}\left(\sum_{j=1}^n\frac{w_j}{j^{\alpha q_1}}\right)^2$$
Moreover, by plugging in \eqref{interim1 new}, the objective function of \eqref{opt new} becomes
$$\left|\sum_{j=1}^n\frac{w_j}{j^{\alpha q_1}}\right|^{2q_2/(q_1+q_2)}\left(\sum_{j=1}^nj^{2\alpha q_2}w_j^2\right)^{q_1/(q_1+q_2)}$$
Therefore, \eqref{opt new} can be rewritten as
\begin{equation}
\begin{array}{ll}
\min_{w}&\left|\sum_{j=1}^n\frac{w_j}{j^{\alpha q_1}}\right|^{2q_2/(q_1+q_2)}\left(\sum_{j=1}^nj^{2\alpha q_2}w_j^2\right)^{q_1/(q_1+q_2)}\\
\text{subject to}&\sum_{j=1}^nj^{2\alpha q_2}w_j^2\leq K^{2(q_1+q_2)}\left(\sum_{j=1}^n\frac{w_j}{j^{\alpha q_1}}\right)^2\\
&\sum_{j=1}^nw_j=1
\end{array}\label{opt1 new}
\end{equation}

To reduce \eqref{opt1 new} into a more tractable form, we introduce the variable $a=\sum_{j=1}^n\frac{w_j}{j^{\alpha q_1}}$, and write \eqref{opt1 new} as
$$\begin{array}{ll}
\min_{w,a}&|a|^{2q_2/(q_1+q_2)}\left(\sum_{j=1}^nj^{2\alpha q_2}w_j^2\right)^{q_1/(q_1+q_2)}\\
\text{subject to}&\sum_{j=1}^nj^{2\alpha q_2}w_j^2\leq K^{2(q_1+q_2)}a^2\\
&\sum_{j=1}^n\frac{w_j}{j^{\alpha q_1}}=a\\
&\sum_{j=1}^nw_j=1
\end{array}$$
which can be further rewritten as
\begin{equation}\min_a|a|^{2q_2/(q_1+q_2)}Z_n^*(a)^{2q_1/(q_1+q_2)}\label{opt6 new}
\end{equation}
where
\begin{equation}
\begin{array}{lll}
Z_n^*(a)=&\min_{w}&\left(\sum_{j=1}^nj^{2\alpha q_2}w_j^2\right)^{1/2}\\
&\text{subject to}&\sum_{j=1}^nj^{2\alpha q_2}w_j^2\leq K^{2(q_1+q_2)}a^2\\
&&\sum_{j=1}^n\frac{w_j}{j^{\alpha q_1}}=a\\
&&\sum_{j=1}^nw_j=1
\end{array}\label{opt3 new}
\end{equation}

Note that \eqref{opt3 new} is a quadratic program. We write it in a simpler form as
\begin{equation}
\begin{array}{ll}
\min_{w}&\|\Sigma^{1/2}w\|\\
\text{subject to}&\|\Sigma^{1/2}w\|^2\leq K^{2(q_1+q_2)}a^2\\
&\mu'w=a\\
&\mathbb 1'w=1
\end{array}\label{opt4 new}
\end{equation}
where $\Sigma=\text{diag}(j^{2\alpha q_2})_{j=1,\ldots,n}\in\mathbb R^{n\times n}$, $\mu=(1/j^{\alpha q_1})_{j=1,\ldots,n}\in\mathbb R^n$, $\mathbb 1=(1)_j\in\mathbb R^n$, and $\|\cdot\|$ is the $L_2$-norm. We can further separate out the first constraint in \eqref{opt4 new}. To this end, denote
\begin{equation}
\begin{array}{lll}
\tilde Z_n^*(a)=&\min_{w}&\|\Sigma^{1/2}w\|\\
&\text{subject to}&\mu'w=a\\
&&\mathbb 1'w=1
\end{array}\label{opt5 new}
\end{equation}
and note that
\begin{equation}
Z_n^*(a)=\left\{\begin{array}{ll}\tilde Z_n^*(a)&\text{\ if\ }{\tilde Z_n^*(a)}^{2}\leq K^{2(q_1+q_2)}a^2\\\infty&\text{\ otherwise}\end{array}\right.\label{interim2 new}
\end{equation}
Putting in \eqref{interim2 new}, optimization problem \eqref{opt6 new} becomes\begin{equation}
\min_{a:{\tilde Z_n^*(a)}^2\leq K^{2(q_1+q_2)}a^2}|a|^{2q_2/(q_1+q_2)}\tilde Z_n^*(a)^{2q_1/(q_1+q_2)}\label{opt7 new}
\end{equation}
Thus, our strategy to solve \eqref{opt new} is to first solve for an optimal solution $w^*(a)=(w_j^*(a^*))_{j=1,\ldots,n}$ to \eqref{opt5 new} and obtain $\tilde Z_n^*(a)$, and then conduct a line search for $a$ in \eqref{opt7 new}. An optimal calibration configuration is given by the weighting sequence $w^*(a^*)$, where $a^*$ is an optimal solution to \eqref{opt7 new}, and $\eta^*$, where
$$\eta^*=\left(\frac{\sum_{j=1}^nj^{2\alpha q_2}{w_j^*(a^*)}^2}{\left(\sum_{j=1}^n\frac{w_j^*(a^*)}{j^{\alpha q_1}}\right)^2}\right)^{1/(2(q_1+q_2))}
$$
by using \eqref{interim1 new}.


The two-decay characterization of the weighting sequence arises from the solution to \eqref{opt5 new}. To illustrate, consider the Lagrangian
$$\|\Sigma^{1/2}w\|-\lambda_1(\mu'w-a)-\lambda_2(\mathbb 1'w-1)$$
Differentiating with respect to $w$ and equating to 0, we get
$$\frac{\Sigma w}{\|\Sigma^{1/2}w\|}-\lambda_1\mu-\lambda_2\mathbb 1=0$$
which gives
$$w=\Sigma^{-1}(\lambda_1\mu+\lambda_2\mathbb 1)=\lambda_1\Sigma^{-1}\mu+\lambda_2\Sigma^{-1}\mathbb 1$$
for some $\lambda_1,\lambda_2$ (scaled by $\|\Sigma^{1/2}w\|$ compared to the ones displayed before). Note that this is equivalent to
$$w_j=\frac{\lambda_1}{j^{\alpha(q_1+2q_2)}}+\frac{\lambda_2}{j^{2\alpha q_2}}$$
for $j=1,\ldots,n$. This is precisely the form of $w_{j,n}^*$ in Theorem \ref{main thm}. By identifying $\lambda_1$ and $\lambda_2$ using the constraints in \eqref{opt5 new}, and writing out $\eta^*$ and $\tilde Z_n^*(a)$, we arrive at the depicted choices of $w$ and $g(\cdot)$ in the theorem. The remainder of the argument comprises an analysis to show that no other choices of $w$ and $g(\cdot)$ can give a better asymptotic minimax ratio, via comparing with an alternate optimization problem, and demonstrating that the residual error induced by $w_{j,n}^*$ and $\eta^*$ in \eqref{higher-order} is indeed of higher order. Appendix \ref{sec:proof main} shows the details.


\section{Multivariate Generalizations}\label{sec:generalizations}
All results we have presented apply to the multivariate version of \eqref{output}. For convenience, we adopt the notations there. We are interested in estimating $\theta\in\mathbb R^p$. Given a tuning parameter $\delta\in\mathbb R_+$, we can run Monte Carlo simulation where each simulation run outputs
\begin{equation}
\hat{\theta}(\delta)=\theta+b(\delta)+v(\delta)\label{multi output}
\end{equation}
with $b(\delta)=B\delta^{q_{1}}+o(\delta^{q_{1}})$ as $\delta\to0$, $v(\delta)=\frac{\varepsilon(\delta)}{\delta^{q_{2}}}$, and $q_1,q_2>0$. We assume that:
\begin{assumption}
We have
\begin{enumerate}
\item $B\in\mathbb{R}^{p}$ is a non-zero constant vector.

\item $\varepsilon(\delta)\in\mathbb{R}^{p}$ is a family of random vectors such that $E\varepsilon(\delta)=0$ and $\lim_{\delta\to0}Cov(\varepsilon(\delta))=\Sigma$ for some positive semidefinite matrix $\Sigma$ with $tr(\Sigma)>0$.

\end{enumerate}\label{multi assumption}
\end{assumption}

The constructions of the considered estimators are generalized in a natural manner. Namely, the sample-average-based estimator $\bar\theta_n$ is obtained by taking the average of $n$ vectors of $\hat\theta(\delta)$. The recursive estimator \eqref{recursive def} is obtained in a vectorized form, where the step size $\gamma_n\in\mathbb R_+$ is still in the form $c(n+n_0)^{-\beta}$ and $\delta_n=\tilde d(n+n_0)^{-\alpha}$. Similar vectorization holds for the averaging estimator \eqref{averaging def}. Lastly, the general weighted estimator in \eqref{general weighted estimator} can also be defined in a vectorized form, with $\{w_{j,n}\}_{j=1,\ldots,n,\ n=1,2,\ldots}$ still a triangular array of weights.

To gauge the error of an estimator $\hat\theta_n$, we use the MSE given by $E\|\hat\theta_n-\theta\|^2$. Note that we can decompose this into bias and variance in $L_2$, namely $\|E\hat\theta_n-\theta\|^2+tr(Cov(\hat\theta_n))$. With this definition of MSE, the asymptotic risk ratios \eqref{asymptotic risk ratio expression} and \eqref{minimax risk} can be similarly defined. Then all the results in Sections \ref{sec:setting}, \ref{sec:framework} and \ref{sec:main} hold with only cosmetic changes. Appendices \ref{sec:proof standard} and \ref{sec:proof recursive} show the multivariate version of the theorems and proofs in Sections \ref{sec:setting} and \ref{sec:framework}, while it will  be clear from the developments in Appendix \ref{sec:proof main} that the multivariate analog of Theorem \ref{main thm} follows from its proof directly (essentially, by replacing $B^2$ with $\|B\|^2$ and $\sigma^2$ with $\text{tr}(\Sigma)$).

Multivariate estimators in the form \eqref{multi output} arise in, for example, zeroth order gradient estimator using simultaneous perturbation (\cite{spall1992multivariate}). To estimate $\nabla f(x)$, a sample output would involve first simulating a random vector, say $h=(h_i)_{i=1,\ldots,p}\in\mathbb R^p$, then generating two unbiased simulation runs $\hat f(x+\delta h)$ and $\hat f(x-\delta h)$, and finally outputting, for each direction $i$,
\begin{equation}
\frac{\hat f(x+\delta h)-\hat f(x-\delta h)}{2\delta h_i}\label{sp}
\end{equation}
where $\delta>0$ is the perturbation size. This scheme satisfies \eqref{multi output} with $q_1=2,q_2=1$ by choosing $h$ to have mean-zero, independent components with finite inverse second moments, and under enough smoothness conditions on $f$. One can also use several variants of \eqref{sp} to obtain similar conclusions, for example the one-sided version $\hat f(x+\delta h)/(\delta h_i)$ (\cite{spall1997one}), or $\hat f(x+\delta h)h_i/\delta$ by choosing $h$ to satisfy other types of conditions, as in Gaussian smoothing (\cite{nesterov2017random}) or uniform sampling (\cite{flaxman2005online}).

Moreover, one important application of the above multivariate estimators concerns input uncertainty quantification (e.g., \cite{barton2012input,henderson2003input,chick2006bayesian,song2014advanced,lam2016advanced}). In particular, a common estimation target in this problem is the output variance of a simulation experiment that is contributed from the statistical noises of the input models calibrated from external data sources, which is typically expressed in the form $\nabla\psi(x)'\Lambda\nabla\psi(x)$ where $\Lambda$ is the sampling covariance of the estimates of the input parameter vector $x\in\mathbb R^p$, $\nabla\psi(x)$ is the gradient of the simulation performance measure with respect to $x$, and $'$ denotes transpose. Thus, this is in the form of $G(\theta)$ where $\theta=\nabla\psi(x)$ and $G(\theta)=\theta'\Lambda\theta$. Our results applies to estimate $G(\theta)$ with a plug-in of $\theta$ and a standard application of the delta method to control the inherited error.

\section{Numerical Results}\label{sec:numerics}


We conduct a simple experiment to demonstrate the theoretical results in this paper. We consider a generic $M/M/1$ queueing system. The arrival and service rates are both set to be 4, so that the system is critically loaded. We consider a transient performance measure of the expected averaged system time of the first 10 customers, and are interested in the gradient of this quantity with respect to the arrival and service rates. The true derivatives with respect to these rates are $0.0946$ and  $-0.2501$ respectively.


We consider two settings. The first setting uses CFD to estimate the derivative with respect to the arrival rate. The second setting uses simultaneous perturbation (described in Section \ref{sec:generalizations}), with the perturbation vector $h$ being independent symmetric variables on $\pm1$, to estimate the gradient with respect to the arrival and service rates simultaneously. In each setting, we consider three estimators: 1) the conventional sample-average-based estimator $\bar\theta_n$; 2) the recursive estimator $\hat\theta_n^{rec}$; and 3) the general weighted estimator $\hat\theta_n^{gen}$. In $\bar\theta_n$, we set $\delta=d(n+n_0)^{-1/6}$ where $d=1$ or $2$. In $\hat\theta_n^{rec}$, we set $c=1$, $\delta_j=\tilde d(j+n_0)^{-1/6}$ for the $j$-th simulation run, where $\tilde d=3^{-1/6}d=0.83d$. For $\hat\theta_n^{gen}$, we set $\delta_j=\tilde d(j+n_0)^{-1/6}$ where $\tilde d=\eta^*d$, and use weights $w_{j,n}^*$, with $\eta^*$ and $w_{j,n}^*$ both chosen according to Theorem \ref{main thm}. We set the ``burn-in" step size $n_0=500$. For $\hat\theta_n^{gen}$, we further consider different values of $K$ from $1$ to $4$. We consider the run-length varying among $n=10^4$, $2\times10^4$, $3\times10^4$, $5\times10^4$, $8\times10^4$ and $10^5$.

Tables \ref{table:comparison single d1} and \ref{table:comparison single d2} show the performances of the three estimators, using $d=1$ and $d=2$ in the baseline estimator $\bar\theta_n$ respectively. The tables demonstrate both the empirical MSE of each estimator and, for $\hat\theta_n^{rec}$ and $\hat\theta_n^{gen}$, the risk ratio compared with $\bar\theta_n$ (i.e., ratio between the empirical MSEs of the considered estimator and $\bar\theta_n$; shown in the bracket). In each parameter configuration, the empirical MSE is calculated by independently repeating the experiment $1,000$ times. When $d=1$, we see that the recursive estimator $\hat\theta_n^{rec}$ has generally a larger MSE compared to $\bar\theta_n$, among all the budget $n$ we consider, with an inflation ranging from $99\%$ to $111\%$. This is roughly consistent with the implications of Theorems \ref{uncomd single} and \ref{fixed parameter} that the AMRR for $\hat\theta_n^{rec}$ in this case is $108\%$ (also shown in Table \ref{table:q1is2}).

In contrast, the general weighted estimator $\hat\theta_n^{gen}$ has a significantly smaller MSE than $\bar\theta_n$. Moreover, the risk ratio is consistent with the implication of Theorem \ref{main thm} and Corollary \ref{main cor}. More concretely, we see that the risk ratio is estimated to be around $68-81\%$ when $K=1$, and $17-21\%$ when $K=2$, for our considered range of $n$. Table \ref{table:gamma_K} shows that the AMRR is $67\%$ when $K=1$, and $17\%$ when $K=2$, which largely match the experimental ratios. Moreover, these ratios appear to be quite stable for the various values of $n$. However, when $K$ is further increased to 3 or 4, although the improvement of $\hat\theta_n^{gen}$ persists, we see a general deterioration in the improvement, with the risk ratio rising back to $74\%$ for $K=3$ and $56\%$ for $K=4$ when $n=10,000$, the smallest budget we consider. Moreover, the risk ratios for $K=3$ and $K=4$ generally decrease with $n$ towards the AMRR. These hint a manifestation of finite-sample behaviors as $K$ is now relatively large compared to the budget $n$. From Table \ref{table:comparison single d1}, it seems that taking $K=2$ is a safe and notably good choice.

Table \ref{table:comparison single d2} shows a similar pattern as Table  \ref{table:comparison single d1}, when $d$ is now taken to be 2. The risk ratio of $\hat\theta_n^{rec}$ compared to $\bar\theta_n$ ranges from $104\%$ to $129\%$, which again roughly match the AMRR $108\%$ in this case. The risk ratio of $\hat\theta_n^{gen}$ when $K=1$ ranges from $75\%$ to $83\%$, while when $K=2$ ranges from $19\%$ to $21\%$, again roughly consistent with the AMRRs of $67\%$ and $17\%$ respectively. Like the case of $d=1$, the improvement of $\hat\theta_n^{gen}$ over $\bar\theta_n$ persists when $K=3$ and $4$, but the improvement generally deteriorates, and in this case also becomes more variable.

\begin{table}[htb]
\begin{center}
\begin{tabular}{|c|c|c|c|c|c|c|}
  \hline
 \multirow{2}{*}{$n$}  &\multirow{2}{*}{$\bar\theta_n$} & \multirow{2}{*}{$\hat\theta_n^{rec}$}&\multicolumn{4}{c|}{$\hat\theta_n^{gen}$}\\
 \cline{4-7}
 &&&$K=1$& $K=2$&$K=3$&$K=4$\\\hline
10000&1.79E-4 &1.97E-4 (110\%)&1.21E-4 (68\%)&3.05E-5 (17\%)&	1.32E-4 (74\%)&	9.98E-5 (56\%)\\
20000&1.02E-4&	1.09E-4 (107\%)&8.07E-5 (79\%)&1.94E-5  (19\%)&7.40E-5 (73\%)&	4.71E-5 (46\%)\\
30000&7.59E-5&	8.31E-5 (110\%)&	6.12E-5 (81\%)&1.59E-5 (21\%)&		5.03E-5 (66\%)&	2.71E-5 (36\%)\\
50000&5.67E-5&	6.29E-5 (111\%)&	4.03E-5 (71\%)&	1.16E-5 (20\%)&	2.79E-5 (49\%)&	1.62E-5 (29\%)\\
80000&4.11E-5&	4.34E-5 (106\%)&	3.10E-5 (75\%)&8.21E-6 (20\%)&			1.94E-5 (47\%)&	1.00E-5 (24\%)\\
100000&3.82E-5&	3.79E-5 (99\%)&	2.79E-5 (73\%)&	7.10E-6 (19\%)&	1.47E-5 (39\%)&	8.67E-6 (23\%)\\\hline
\end{tabular}
\end{center}
\caption{\label{table.label} Empirical MSE among estimators for the derivative with respect to the arrival rate, $d=1$. Bracketed numbers represent the risk ratios between the considered estimators and the baseline $\bar\theta_n$.} 
\label{table:comparison single d1}
\end{table}


\begin{table}[htb]
\begin{center}
\begin{tabular}{|c|c|c|c|c|c|c|}
  \hline
 \multirow{2}{*}{$n$}  &\multirow{2}{*}{$\bar\theta_n$} & \multirow{2}{*}{$\hat\theta_n^{rec}$}&\multicolumn{4}{c|}{$\hat\theta_n^{gen}$}\\
 \cline{4-7}
 &&&$K=1$& $K=2$&$K=3$&$K=4$\\\hline
10000&4.38E-5&4.64E-5 (106\%)&	3.47E-5 (79\%)&8.28E-6 (19\%)&3.73E-6 (9\%)&2.83E-5 (65\%)\\
20000&2.43E-5	&2.71E-5 (112\%)&	1.85E-5 (76\%)&4.73E-6 (19\%)&2.14E-6 (9\%)&1.78E-5 (73\%)\\
30000&1.91E-5&	2.45E-5 (129\%)&	1.57E-5 (83\%)&3.96E-6 (21\%)&1.74E-5 (91\%)&1.18E-5 (62\%)\\
50000&1.54E-5&	1.60E-5 (104\%)&	1.18E-5 (77\%)&2.87E-6 (19\%)&9.79E-6 (63\%)&6.03E-6 (39\%)\\
80000&1.11E-5&	1.21E-5 (109\%)&	8.52E-6 (77\%)&2.13E-6 (19\%)&5.79E-6 (52\%)&4.18E-6 (38\%)\\
100000&9.19E-6&	1.08E-5 (117\%)&	6.88E-6 (75\%)&1.77E-6 (19\%)&5.44E-6 (59\%)&3.58E-6 (39\%)\\\hline
\end{tabular}
\end{center}
\caption{\label{table.label} Empirical MSE among estimators for the derivative with respect to the arrival rate, $d=2$. Bracketed numbers represent the risk ratios between the considered estimators and the baseline $\bar\theta_n$.} 
\label{table:comparison single d2}
\end{table}


In the second set of experiments, we test the gradient estimator for both the arrival and service rates using simultaneous perturbation. Tables \ref{table:comparison multi} and \ref{table:comparison multi2} show the performances of $\hat\theta_n^{rec}$ and $\hat\theta_n^{gen}$ relative to $\bar\theta_n$ when $d=1$ and $d=2$ respectively. We use $K=1$ or $2$ in $\hat\theta_n^{gen}$, as they are observed to perform reasonably well in the single-variate case. We see that $\hat\theta_n^{rec}$ continues to have generally a larger MSE than $\bar\theta_n$, with risk ratios ranging from $100\%$ to $115\%$ when $d=1$ and $103\%$ to $115\%$ when $d=2$ for the considered budgets, which are consistent with the AMRR of $108\%$. When $d=1$, the risk ratios of $\hat\theta_n^{gen}$ range from $71\%$ to $83\%$  for $K=1$, and improve to $35\%$ to $46\%$ for $K=2$. When $d=2$, the risk ratios of $\hat\theta_n^{gen}$ range from $94\%$ to $102\%$ for $K=1$, and from $57\%$ to $67\%$ for $K=2$. These trends are consistent with the AMRR for $K=1$ and $K=2$ respectively, although the experimental numbers seem to increase generally and are above the theoretical calculation as $d$ increases. Lastly, the risk ratios across different $n$ in the considered range seem to be quite stable, hinting that these budgets are sufficient to observe the studied asymptotic behaviors in this case.


\begin{table}[htb]
\begin{center}
\begin{tabular}{|c|c|c|c|c|}
  \hline
 \multirow{2}{*}{$n$}  &\multirow{2}{*}{$\bar\theta_n$} & \multirow{2}{*}{$\hat\theta_n^{rec}$}&\multicolumn{2}{c|}{$\hat\theta_n^{gen}$}\\
 \cline{4-5}
 &&&$K=1$& $K=2$\\\hline
10000&1.69E-4&	1.94E-4 (115\%)&	1.35E-4 (80\%)&6.62E-5 (39\%)\\
20000&1.13E-4&	1.26E-4 (112\%)&	8.60E-5 (76\%)&4.45E-5 (39\%)\\
30000&9.40E-5&	9.41E-5 (100\%)&	6.70E-5 (71\%)&3.33E-5 (35\%)\\
50000&5.99E-5&	6.08E-5 (102\%)&	4.70E-5 (78\%)&2.56E-5 (43\%)\\
80000&4.62E-5&	5.30E-5 (115\%)&	3.52E-5 (76\%)&1.92E-5 (41\%)\\
100000&3.83E-5&	4.29E-5 (112\%)&	3.17E-5 (83\%)&1.77E-5 (46\%)\\\hline
\end{tabular}
\end{center}
\caption{\label{table.label} Empirical MSE among estimators for the gradient with respect to the arrival and service rates, $d=1$. Bracketed numbers represent the risk ratios between the considered estimators and the baseline $\bar\theta_n$} 
\label{table:comparison multi}
\end{table}

			\begin{table}[htb]
\begin{center}
\begin{tabular}{|c|c|c|c|c|}
  \hline
 \multirow{2}{*}{$n$}  &\multirow{2}{*}{$\bar\theta_n$} & \multirow{2}{*}{$\hat\theta_n^{rec}$}&\multicolumn{2}{c|}{$\hat\theta_n^{gen}$}\\
 \cline{4-5}
 &&&$K=1$& $K=2$\\\hline
10000&5.85E-5&	6.28E-5 (107\%)&	5.50E-5 (94\%)&3.35E-5 (57\%)\\
20000&3.27E-5& 3.76E-5 (115\%)&	3.33E-5 (102\%)&2.19E-5 (67\%)\\
30000&2.67E-5&	2.82E-5 (106\%)&	2.52E-5 (94\%)&1.65E-5 (62\%)\\
50000&1.85E-5&	2.00E-5 (108\%)&	1.81E-5 (98\%)&1.16E-5 (63\%)\\
80000&1.36E-5&	1.40E-5 (103\%)&	1.30E-5 (96\%)&8.06E-6 (59\%)\\
100000&1.18E-5&	1.26E-5 (107\%)&	1.10E-5 (94\%)&6.73E-6 (57\%)\\\hline
\end{tabular}
\end{center}
\caption{\label{table.label} Empirical MSE among estimators for the gradient with respect to the arrival and service rates, $d=2$. Bracketed numbers represent the risk ratios between the considered estimators and the baseline $\bar\theta_n$} 
\label{table:comparison multi2}
\end{table}


\section{Conclusion}\label{sec:conclusion}
We have studied a framework to construct new estimators that, in situations where simulation runs are biased for a target estimation quantity, consistently outperform baseline estimators as the sample averages of the simulation runs with a chosen tuning parameter. One challenge in choosing the latter lies in the often lack of knowledge on the model characteristics that affect the bias-variance tradeoff. To mitigate the adversarial impact of this ambiguity, we propose a minimax analysis on the asymptotic risk ratio that compares the mean square errors between proposed estimators and the baseline. In particular, we identify the asymptotic minimax risk ratio (AMRR) and the optimal configurations for recursive estimators and their standard averaging versions. We show that, in typical cases, the AMRR for these estimators are not small enough to justify any outperformance against the standard baseline. We then consider a more general class of weighted estimators, and identify the AMRR that can be significantly reduced to a level that implies that the resulting optimal estimator asymptotically outperforms the baseline, regardless of any realizations of the unknown model characteristics. Moreover, we provide an explicit characterizations of the optimal weights in a two-decay-rate form, and argue how this arises from a balancing of bias-variance that matches the baseline in order to control an adversarial enlargement of the risk ratio.

Our work opens the door to multiple lines of expansion, in terms of both the formulating framework and the techniques. For example, our approach can be used to find better estimators for problems where simulation runtime is significantly affected by the tuning parameters, in addition to bias and variance. This arises in the discretization schemes in, e.g., the simulation of stochastic differential equations. The statistical inference and construction of confidence intervals/regions of our weighted estimators, which involves analyzing central limit behaviors and the proper design of data-driven schemes like sectioning, are also of interest. Lastly, we plan to study our enhanced estimators, in the context of finite-differences for zeroth-order gradient estimation, for iterative algorithms in black-box stochastic optimization.

\ACKNOWLEDGMENT{We gratefully acknowledge support from the National Science Foundation under grants CMMI-1542020, CMMI-1523453 and CAREER CMMI-1653339/1834710.}


\bibliographystyle{informs2014} 
\bibliography{onlineestimate_ref} 

\begin{thebibliography}{57}
\providecommand{\natexlab}[1]{#1}
\providecommand{\url}[1]{\texttt{#1}}
\providecommand{\urlprefix}{URL }

\bibitem[{Agrawal et~al.(2012)Agrawal, Ding, Saberi, \protect\BIBand{}
  Ye}]{agrawal2012price}
Agrawal S, Ding Y, Saberi A, Ye Y (2012) Price of correlations in stochastic
  optimization. \emph{Operations Research} 60(1):150--162.

\bibitem[{Asmussen \protect\BIBand{} Glynn(2007)}]{asmussen2007stochastic}
Asmussen S, Glynn PW (2007) \emph{Stochastic simulation: algorithms and
  analysis}, volume~57 (Springer Science \& Business Media).

\bibitem[{Barton(2012)}]{barton2012input}
Barton RR (2012) Input uncertainty in outout analysis. \emph{Proceedings of the
  Winter Simulation Conference}, 6 (Winter Simulation Conference).

\bibitem[{Ben-Tal et~al.(2009)Ben-Tal, El~Ghaoui, \protect\BIBand{}
  Nemirovski}]{ben2009robust}
Ben-Tal A, El~Ghaoui L, Nemirovski A (2009) \emph{Robust optimization},
  volume~28 (Princeton University Press).

\bibitem[{Ben-Tal \protect\BIBand{} Nemirovski(2002)}]{ben2002robust}
Ben-Tal A, Nemirovski A (2002) Robust optimization--methodology and
  applications. \emph{Mathematical Programming} 92(3):453--480.

\bibitem[{Bertsimas et~al.(2011)Bertsimas, Brown, \protect\BIBand{}
  Caramanis}]{bertsimas2011theory}
Bertsimas D, Brown DB, Caramanis C (2011) Theory and applications of robust
  optimization. \emph{SIAM review} 53(3):464--501.

\bibitem[{Besbes \protect\BIBand{} Zeevi(2009)}]{besbes2009dynamic}
Besbes O, Zeevi A (2009) Dynamic pricing without knowing the demand function:
  Risk bounds and near-optimal algorithms. \emph{Operations Research}
  57(6):1407--1420.

\bibitem[{Besbes \protect\BIBand{} Zeevi(2011)}]{besbes2011minimax}
Besbes O, Zeevi A (2011) On the minimax complexity of pricing in a changing
  environment. \emph{Operations research} 59(1):66--79.

\bibitem[{Blanchet \protect\BIBand{} Glynn(2015)}]{blanchet2015unbiased}
Blanchet JH, Glynn PW (2015) Unbiased monte carlo for optimization and
  functions of expectations via multi-level randomization. \emph{Winter
  Simulation Conference (WSC), 2015}, 3656--3667 (IEEE).

\bibitem[{Borkar(2009)}]{borkar2009stochastic}
Borkar VS (2009) \emph{Stochastic approximation: a dynamical systems
  viewpoint}, volume~48 (Springer).

\bibitem[{Cesa-Bianchi \protect\BIBand{} Lugosi(2006)}]{cesa2006prediction}
Cesa-Bianchi N, Lugosi G (2006) \emph{Prediction, learning, and games}
  (Cambridge university press).

\bibitem[{Chick(2006)}]{chick2006bayesian}
Chick SE (2006) Bayesian ideas and discrete event simulation: why, what and
  how. \emph{Proceedings of the 38th conference on Winter simulation}, 96--105
  (Winter Simulation Conference).

\bibitem[{Chung(1954)}]{chung1954stochastic}
Chung KL (1954) On a stochastic approximation method. \emph{The Annals of
  Mathematical Statistics} 463--483.

\bibitem[{Duplay et~al.(2018)Duplay, Lam, \protect\BIBand{} Zhang}]{dlz18}
Duplay D, Lam H, Zhang X (2018) Achieving optimal bias-variance tradeoff in
  online derivative estimation. \emph{Winter Simulation Conference (WSC), 2018}
  (IEEE).

\bibitem[{Fabian(1967)}]{fabian1967stochastic}
Fabian V (1967) Stochastic approximation of minima with improved asymptotic
  speed. \emph{The Annals of Mathematical Statistics} 191--200.

\bibitem[{Fabian(1968)}]{fabian1968asymptotic}
Fabian V (1968) On asymptotic normality in stochastic approximation. \emph{The
  Annals of Mathematical Statistics} 39(4):1327--1332.

\bibitem[{Flaxman et~al.(2005)Flaxman, Kalai, \protect\BIBand{}
  McMahan}]{flaxman2005online}
Flaxman AD, Kalai AT, McMahan HB (2005) Online convex optimization in the
  bandit setting: gradient descent without a gradient. \emph{Proceedings of the
  sixteenth annual ACM-SIAM symposium on Discrete algorithms}, 385--394
  (Society for Industrial and Applied Mathematics).

\bibitem[{Fox \protect\BIBand{} Glynn(1989)}]{fox1989replication}
Fox BL, Glynn PW (1989) Replication schemes for limiting expectations.
  \emph{Probability in the Engineering and Informational Sciences}
  3(3):299--318.

\bibitem[{Fu(2006)}]{fu2006gradient}
Fu MC (2006) Gradient estimation. \emph{Handbooks in operations research and
  management science} 13:575--616.

\bibitem[{Fu et~al.(2009)Fu, Hong, \protect\BIBand{} Hu}]{fu2009conditional}
Fu MC, Hong LJ, Hu JQ (2009) Conditional monte carlo estimation of quantile
  sensitivities. \emph{Management Science} 55(12):2019--2027.

\bibitem[{Fu \protect\BIBand{} Hu(1992)}]{fu1992extensions}
Fu MC, Hu JQ (1992) Extensions and generalizations of smoothed perturbation
  analysis in a generalized semi-markov process framework. \emph{IEEE
  Transactions on Automatic Control} 37(10):1483--1500.

\bibitem[{Giles(2008)}]{giles2008multilevel}
Giles MB (2008) Multilevel monte carlo path simulation. \emph{Operations
  Research} 56(3):607--617.

\bibitem[{Glasserman(2013)}]{glasserman2013monte}
Glasserman P (2013) \emph{Monte Carlo methods in financial engineering},
  volume~53 (Springer Science \& Business Media).

\bibitem[{Glasserman \protect\BIBand{} Gong(1990)}]{glasserman1990smoothed}
Glasserman P, Gong WB (1990) Smoothed perturbation analysis for a class of
  discrete-event systems. \emph{IEEE Transactions on Automatic Control}
  35(11):1218--1230.

\bibitem[{Glynn(1990)}]{glynn1990likelihood}
Glynn PW (1990) Likelihood ratio gradient estimation for stochastic systems.
  \emph{Communications of the ACM} 33(10):75--84.

\bibitem[{Glynn \protect\BIBand{} Whitt(1992)}]{glynn1992asymptotic}
Glynn PW, Whitt W (1992) The asymptotic efficiency of simulation estimators.
  \emph{Operations research} 40(3):505--520.

\bibitem[{Gong \protect\BIBand{} Ho(1987)}]{gong1987smoothed}
Gong WB, Ho YC (1987) Smoothed (conditional) perturbation analysis of discrete
  event dynamical systems. \emph{IEEE Transactions on Automatic Control}
  32(10):858--866.

\bibitem[{Hazan et~al.(2016)}]{hazan2016introduction}
Hazan E, et~al. (2016) Introduction to online convex optimization.
  \emph{Foundations and Trends{\textregistered} in Optimization}
  2(3-4):157--325.

\bibitem[{Heidelberger et~al.(1988)Heidelberger, Cao, Zazanis,
  \protect\BIBand{} Suri}]{heidelberger1988convergence}
Heidelberger P, Cao XR, Zazanis MA, Suri R (1988) Convergence properties of
  infinitesimal perturbation analysis estimates. \emph{Management Science}
  34(11):1281--1302.

\bibitem[{Heidergott et~al.(2010)Heidergott, Pflug, Farenhorst-Yuan
  et~al.}]{heidergott2010gradient}
Heidergott B, Pflug G, Farenhorst-Yuan T, et~al. (2010) Gradient estimation for
  discrete-event systems by measure-valued differentiation. \emph{ACM
  Transactions on Modeling and Computer Simulation (TOMACS)} 20(1):5.

\bibitem[{Heidergott \protect\BIBand{}
  V{\'a}zquez-Abad(2008)}]{heidergott2008measure}
Heidergott B, V{\'a}zquez-Abad FJ (2008) Measure-valued differentiation for
  markov chains. \emph{Journal of Optimization Theory and Applications}
  136(2):187--209.

\bibitem[{Henderson(2003)}]{henderson2003input}
Henderson SG (2003) Input model uncertainty: Why do we care and what should we
  do about it? \emph{Proceedings of the 35th conference on Winter simulation:
  driving innovation}, 90--100 (Winter Simulation Conference).

\bibitem[{Ho et~al.(1983)Ho, Cao, \protect\BIBand{}
  Cassandras}]{ho1983infinitesimal}
Ho YC, Cao X, Cassandras C (1983) Infinitesimal and finite perturbation
  analysis for queueing networks. \emph{Automatica} 19(4):439--445.

\bibitem[{Hong(2009)}]{hong2009estimating}
Hong LJ (2009) Estimating quantile sensitivities. \emph{Operations research}
  57(1):118--130.

\bibitem[{Kushner \protect\BIBand{} Yin(2003)}]{kushner2003stochastic}
Kushner H, Yin GG (2003) \emph{Stochastic approximation and recursive
  algorithms and applications}, volume~35 (Springer Science \& Business Media).

\bibitem[{Lam(2016)}]{lam2016advanced}
Lam H (2016) Advanced tutorial: Input uncertainty and robust analysis in
  stochastic simulation. \emph{Winter Simulation Conference (WSC), 2016},
  178--192 (IEEE).

\bibitem[{L'Ecuyer(1990)}]{l1990unified}
L'Ecuyer P (1990) A unified view of the ipa, sf, and lr gradient estimation
  techniques. \emph{Management Science} 36(11):1364--1383.

\bibitem[{L'Ecuyer(1991)}]{l1991overview}
L'Ecuyer P (1991) An overview of derivative estimation. \emph{Simulation
  Conference, 1991. Proceedings., Winter}, 207--217 (IEEE).

\bibitem[{McLeish(2010)}]{mcleish2010general}
McLeish D (2010) A general method for debiasing a monte carlo estimator.
  \emph{arXiv preprint arXiv:1005.2228} .

\bibitem[{Nemirovski et~al.(2009)Nemirovski, Juditsky, Lan, \protect\BIBand{}
  Shapiro}]{nemirovski2009robust}
Nemirovski A, Juditsky A, Lan G, Shapiro A (2009) Robust stochastic
  approximation approach to stochastic programming. \emph{SIAM Journal on
  optimization} 19(4):1574--1609.

\bibitem[{Nesterov \protect\BIBand{} Spokoiny(2017)}]{nesterov2017random}
Nesterov Y, Spokoiny V (2017) Random gradient-free minimization of convex
  functions. \emph{Foundations of Computational Mathematics} 17(2):527--566.

\bibitem[{Pasupathy(2010)}]{pasupathy2010choosing}
Pasupathy R (2010) On choosing parameters in retrospective-approximation
  algorithms for stochastic root finding and simulation optimization.
  \emph{Operations Research} 58(4-part-1):889--901.

\bibitem[{Pasupathy \protect\BIBand{} Kim(2011)}]{pasupathy2011stochastic}
Pasupathy R, Kim S (2011) The stochastic root-finding problem: Overview,
  solutions, and open questions. \emph{ACM Transactions on Modeling and
  Computer Simulation (TOMACS)} 21(3):19.

\bibitem[{Peng et~al.(2018)Peng, Fu, Hu, \protect\BIBand{}
  Heidergott}]{peng2018new}
Peng Y, Fu MC, Hu JQ, Heidergott B (2018) A new unbiased stochastic derivative
  estimator for discontinuous sample performances with structural parameters.
  \emph{Operations Research} 66(2):487--499.

\bibitem[{Polyak \protect\BIBand{} Juditsky(1992)}]{polyak1992acceleration}
Polyak BT, Juditsky AB (1992) Acceleration of stochastic approximation by
  averaging. \emph{SIAM Journal on Control and Optimization} 30(4):838--855.

\bibitem[{Reiman \protect\BIBand{} Weiss(1989)}]{reiman1989sensitivity}
Reiman MI, Weiss A (1989) Sensitivity analysis for simulations via likelihood
  ratios. \emph{Operations Research} 37(5):830--844.

\bibitem[{Rhee \protect\BIBand{} Glynn(2015)}]{rhee2015unbiased}
Rhee Ch, Glynn PW (2015) Unbiased estimation with square root convergence for
  sde models. \emph{Operations Research} 63(5):1026--1043.

\bibitem[{Rubinstein(1986)}]{rubinstein1986score}
Rubinstein RY (1986) The score function approach for sensitivity analysis of
  computer simulation models. \emph{Mathematics and Computers in Simulation}
  28(5):351--379.

\bibitem[{Rubinstein(1992)}]{rubinstein1992sensitivity}
Rubinstein RY (1992) Sensitivity analysis of discrete event systems by the
  “push out” method. \emph{Annals of Operations Research} 39(1):229--250.

\bibitem[{Ruppert(1988)}]{ruppert1988efficient}
Ruppert D (1988) Efficient estimations from a slowly convergent robbins-monro
  process. Technical report, Cornell University Operations Research and
  Industrial Engineering.

\bibitem[{Rychlik(1990)}]{rychlik1990unbiased}
Rychlik T (1990) Unbiased nonparametric estimation of the derivative of the
  mean. \emph{Statistics \& probability letters} 10(4):329--333.

\bibitem[{Shalev-Shwartz et~al.(2012)}]{shalev2012online}
Shalev-Shwartz S, et~al. (2012) Online learning and online convex optimization.
  \emph{Foundations and Trends{\textregistered} in Machine Learning}
  4(2):107--194.

\bibitem[{Song et~al.(2014)Song, Nelson, \protect\BIBand{}
  Pegden}]{song2014advanced}
Song E, Nelson BL, Pegden CD (2014) Advanced tutorial: Input uncertainty
  quantification. \emph{Simulation Conference (WSC), 2014 Winter}, 162--176
  (IEEE).

\bibitem[{Spall(1992)}]{spall1992multivariate}
Spall JC (1992) Multivariate stochastic approximation using a simultaneous
  perturbation gradient approximation. \emph{IEEE transactions on automatic
  control} 37(3):332--341.

\bibitem[{Spall(1997)}]{spall1997one}
Spall JC (1997) A one-measurement form of simultaneous perturbation stochastic
  approximation. \emph{Automatica} 33(1):109--112.

\bibitem[{Zazanis \protect\BIBand{} Suri(1993)}]{zazanis1993convergence}
Zazanis MA, Suri R (1993) Convergence rates of finite-difference sensitivity
  estimates for stochastic systems. \emph{Operations research} 41(4):694--703.

\bibitem[{Zhou \protect\BIBand{} Doyle(1998)}]{zhou1998essentials}
Zhou K, Doyle JC (1998) \emph{Essentials of robust control}, volume 104
  (Prentice hall Upper Saddle River, NJ).

\end{thebibliography}

\begin{APPENDICES}

\section{Proofs for Section \ref{sec:setting}}\label{sec:proof standard}

We will prove a multivariate version of Theorem \ref{offmse single}.

\begin{theorem}\label{offmse}
Under Assumption \ref{multi assumption}, suppose that $\lim_{n\to\infty}\delta n^{\alpha}=d>0$, the sample-average-based estimator $\bar\theta_n$ exhibits the asymptotic MSE
$$E\|\bar{\theta}_{n}-\theta\|^{2}=d^{2q_{1}}\|B\|^{2}n^{-2\alpha q_1}+\frac{tr(\Sigma)}{d^{2q_{2}}}n^{2\alpha q_2-1}+o(n^{-2\alpha q_1}+n^{2\alpha q_2-1})\textrm{ as }n\to\infty$$
Choosing $\alpha=1/(2(q_1+q_2))$ achieves the optimal MSE order, and the asymptotic MSE is
$$E\|\bar{\theta}_n-\theta\|^2=\big(d^{2q_1}\|B\|^2+\frac{tr(\Sigma)}{d^{2q_2}}\big)n^{-q_1/(q_1+q_2)}+o(n^{-q_1/(q_1+q_2)})\text{ as }n\to\infty$$
\end{theorem}

\proof{Proof of theorem \ref{offmse}.}
By the bias-variance decomposition, we have
\begin{align*}
E\|\bar\theta_n-\theta\|^2&=\|E\bar\theta_n-\theta\|^2+tr(Cov(\bar\theta_n))\\
&=\|b(\delta)\|^2+\frac{1}{n}tr(Cov(v(\delta)))\\
&=\|B\|^2\delta^{2q_1}+o(\delta^{2q_1})+\frac{1}{n}\frac{tr(\Sigma)+o(1)}{\delta^{2q_2}}
\end{align*}
Setting $\delta=\frac{d+o(1)}{n^{\alpha}}$, we obtain
\begin{align*}
E\|\bar\theta_n-\theta\|^2&=\|B\|^2\frac{(d+o(1))^{2q_1}}{n^{2\alpha q_1}}+o(n^{-2\alpha q_1})+\frac{tr(\Sigma)+o(1)}{(d+o(1))^{2q_2}}n^{2\alpha q_2-1}\\
&=\left(\|B\|^2d^{2q_1}+o(1)\right)n^{-2\alpha q_1}+\left(\frac{tr(\Sigma)}{d^{2q_2}}+o(1)\right)n^{2\alpha q_2-1}
\end{align*}
To achieve the optimal MSE order, we solve $-2\alpha q_1=2\alpha q_2-1$. Thus $\alpha=1/(2(q_1+q_2))$ and the optimal order is $n^{-q_1/(q_1+q_2)}$.\hfill\Halmos
\endproof

\proof{Proof of Theorem \ref{offmse single}.}
The proof follows immediately by considering dimension 1 in Theorem \ref{offmse}.\hfill\Halmos
\endproof

\section{Proofs for Section \ref{sec:initial}}\label{sec:proof recursive}



We provide and prove multivariate versions of the results, from which the ones in Section \ref{sec:initial} follow immediately.




Frequently used in the subsequent proofs is the following result adapted from Lemma 4.2, a version of Chung's Lemma, in \cite{fabian1967stochastic}:
\begin{lemma}[Chung's Lemma]
For $v_{n},c_{n},b_{n}$ real numbers, and $0<\alpha\leq1$, suppose $\lim_{n\to\infty}c_{n}=c>0$, and consider the iteration
\begin{equation}\label{chung}
v_{n+1}=(1-\frac{c_{n}}{n^{\alpha}})v_{n}+\frac{b_{n}}{n^{\alpha}}
\end{equation}
If $b_{n}\to0$, then $v_{n}\to0$; if $b_{n}\to b>0$, then $v_{n}\to b/c$; and if $b_{n}\to\infty$, then $v_{n}\to \infty$.\label{chung new}
\end{lemma}

\proof{Proof of Lemma \ref{chung new}.}
Our version of Chung's lemma is different in appearance from Lemma 4.2 in \cite{fabian1968asymptotic}, and thus we repeat the proof here. First, if $b_{n}\to b$ where $b\geq 0$ is a real number, then for given $0<\epsilon<c$, we can choose $n_{1}$ sufficiently large such that, for all $n\geq n_{1}$, we have $\frac{c_{n}}{n^{\alpha}}<1$, $b_{n}<b+\epsilon$ and $c-\epsilon<c_{n}<c+\epsilon$. Now let $n\geq n_{1}$. If $v_{n}\geq\frac{b+2\epsilon}{c-\epsilon}$, then from the iteration (\ref{chung})
$$v_{n+1}\leq v_{n}-\frac{b+2\epsilon}{c-\epsilon}(c-\epsilon)\frac{1}{n^{\alpha}}+(b+\epsilon)\frac{1}{n^{\alpha}}\leq v_{n}-\frac{\epsilon}{n^{\alpha}}$$
On the other hand, if $v_{n}\leq\frac{b+2\epsilon}{c-\epsilon}$, then since the right hand side of the iteration (\ref{chung}) is an increasing function of $v_{n}$, we have
$$v_{n+1}\leq\frac{b+2\epsilon}{c-\epsilon}-\frac{b+2\epsilon}{c-\epsilon}(c-\epsilon)\frac{1}{n^{\alpha}}+(b+\epsilon)\frac{1}{n^{\alpha}}\leq\frac{b+2\epsilon}{c-\epsilon}$$
Combined with the fact that $\sum\limits_{n=1}^{\infty}\frac{1}{n^{\alpha}}$ diverges, we have $\limsup_{n\to\infty}v_{n}\leq\frac{b+2\epsilon}{c-\epsilon}$. Since $\epsilon$ is arbitrary, we get
\begin{equation}\label{cb1}
\limsup_{n\to\infty}v_{n}\leq\frac{b}{c}
\end{equation}
If $b=0$, $v_{n+1}\geq v_{n}+\frac{\epsilon}{n^{\alpha}}$ for $v_n\leq -\frac{2\epsilon}{c-\epsilon}$ and $v_{n+1}\geq -\frac{2\epsilon}{c-\epsilon}$ for $v_n\geq -\frac{2\epsilon}{c-\epsilon}$. Therefore we have $\liminf\limits_{n\rightarrow \infty} v_n \geq 0$ and $\limsup\limits_{n\rightarrow \infty} v_n \leq 0$. We conclude that $\lim_{n\to\infty}v_{n}=0$. By the same analysis, if $ b_{n}\to b>0$, where $b$ possibly take the value of $\infty$, we would have
\begin{equation}\label{cb2}
\liminf_{n\to\infty}v_{n}\geq\frac{b}{c}
\end{equation}
Thus if $b=\infty$, we conclude that $\lim_{n\to\infty}v_{n}\to\infty$, and if $0<b<\infty$, combining (\ref{cb1}) and (\ref{cb2}), we get $\lim_{n\to\infty} v_{n}=\frac{b}{c}$.\hfill\Halmos
\endproof

We now consider multivariate versions of our results and their proofs:

\begin{proposition}\label{ondiv}
Under Assumption \ref{multi assumption}, we have:
\begin{enumerate}
\item If $\beta\leq1$ and $\alpha<\beta/(2q_{2})$, the estimator $\hat\theta_n^{rec}$ is $L_2$-consistent for $\theta$, i.e.,
$$\lim_{n\to\infty}E\|\hat\theta_n^{rec}-\theta\|^2=0$$
\item If $\beta\leq1$ and $\alpha\geq\beta/(2q_{2})$, or if $\beta>1$, the error of $\hat\theta_n^{rec}$ in estimating $\theta$ is bounded away from zero in $L_{2}$ norm as $n\to\infty$, i.e.,
$$\liminf_{n\to\infty}E\|\hat\theta_{n}^{rec}-\theta\|^{2}>0$$\end{enumerate}
\end{proposition}

\proof{Proof of Proposition \ref{ondiv}.}
We first prove the proposition for $\beta\leq1$. From the recursion
\begin{equation}\label{recursionequ}
\hat\theta_n^{rec}=(1-\gamma_n)\hat\theta_{n-1}^{rec}+\gamma_n\hat\theta(\delta_n)
\end{equation}
we have
$$E\hat\theta_n^{rec}-\theta=(1-\gamma_n)\left(E\hat\theta_{n-1}^{rec}-\theta\right)+\gamma_n\left(E\hat\theta(\delta_n)-\theta\right)$$
Since $E\hat\theta(\delta_n)-\theta=b(\delta_n)\to0\text{ as }n\to\infty$, we have $E\hat\theta_n^{rec}-\theta\to0$ by Chung's lemma.
Note that $E\|\hat\theta_{n}^{rec}-\theta\|^{2}=\|E\hat\theta_{n}^{rec}-\theta\|^{2}+tr(Cov(\hat\theta_{n}^{rec}))$. Thus the convergence will depend on the variance term.
Taking covariance of \eqref{recursionequ}, by independence we have
\begin{equation}\label{coviter}
Cov(\hat\theta_{n}^{rec})=(1-\gamma_n)^{2}Cov(\hat\theta_{n-1}^{rec})+\gamma_n^{2}Cov(\hat{\theta}(\delta_{n}))
\end{equation}
Since
$$Cov(\hat{\theta}(\delta_{n}))=\frac{1}{\delta_{n}^{2q_{2}}}Cov(\epsilon(\delta_{n}))$$
we have
$$\lim_{n\to\infty}\frac{Cov(\hat{\theta}(\delta_{n}))}{n^{2q_{2}\alpha}}=\frac{\Sigma}{d^{2q_{2}}}$$
We now rewrite the iteration (\ref{coviter}) as
$$tr(Cov(\hat\theta_{n}^{rec}))=(1-(2+o(1))\gamma_n)tr(Cov(\hat\theta_{n-1}^{rec}))+\gamma_ns_n$$
where $s_{n}=c\frac{tr(\Sigma)}{d^{2q_{2}}}n^{2q_{2}\alpha-\beta}+o(n^{2q_{2}\alpha-\beta})$. We note that $\lim_{n\to\infty}s_{n}=\infty$ if $\alpha>\beta/(2q_{2})$, $\lim_{n\to\infty}s_{n}=c\frac{tr(\Sigma)}{d^{2q_{2}}}>0$ if $\alpha=\beta/(2q_{2})$, and $\lim_{n\to\infty}s_n=0$ if $\alpha<\beta/(2q_2)$. Thus by Chung's lemma
$$\lim_{n\to\infty}tr(Cov(\hat\theta_{n}^{rec}))\to\infty\textrm{ if }\alpha>\beta/(2q_{2})$$
$$\lim_{n\to\infty}tr(Cov(\hat\theta_{n}^{rec}))=c\frac{tr(\Sigma)}{2d^{2q_{2}}}\textrm{ if }\alpha=\beta/(2q_{2})$$
and
$$\lim_{n\to\infty}tr(Cov(\hat\theta_{n}^{rec}))=0\textrm{ if }\alpha<\beta/(2q_{2})$$
This completes the proof for $\beta\leq1$.

Next consider $\beta>1$, we now argue that choosing $\gamma_{n}=c/n^{\beta}$ does not lead to convergence. We note that $\hat\theta_{n}^{rec}$ is a linear combination of $\hat\theta_{0}^{rec},\hat\theta_i(\delta_{i}),i=1,\cdots,n$, i.e.
$$\hat\theta_{n}^{rec}=a_{0}\hat\theta_{0}^{rec}+\sum_{i=1}^{n}a_{i}\hat\theta_i(\delta_{i})$$
where $a_{0}=\prod_{j=1}^{n}(1-\gamma_{j})$ and $a_{i}=\gamma_{i}\prod_{j=i+1}^{n}(1-\gamma_{j})$. Since $\sum_{n=1}^{\infty}\gamma_{n}=\sum_{n=1}^{\infty}\frac{c}{n^{\beta}}<\infty$, by the relation between infinite product and infinite sum, we get
$$\lim_{n\to\infty}a_{i}\textrm{ exists and is positive for any }i$$
Since by independence
$$tr(Cov(\hat\theta_{n}^{rec}))=a_{0}^2tr(Cov(\hat\theta_{0}^{rec}))+\sum_{i=1}^{n}a_{i}^2tr(Cov(\hat\theta(\delta_{i})))$$
we have that
$$\liminf_{n\to\infty}tr(Cov(\hat\theta_n^{rec}))>0$$
\hfill\Halmos\endproof




\begin{theorem}\label{onmse}
Under Assumption \ref{multi assumption}, the MSE of $\hat\theta_n^{rec}$ in estimating $\theta$ behaves as follows:
\begin{enumerate}
\item For $\beta<1$ and $\alpha<\beta/(2q_2)$,
{\small
$$E\|\hat\theta_n^{rec}-\theta\|^{2}=d^{2q_{1}}\|B\|^{2}n^{-2q_{1}\alpha}+\frac{c}{2d^{2q_{2}}}tr(\Sigma)n^{2q_{2}\alpha-\beta} +o(n^{-2q_{1}\alpha}+n^{2q_{2}\alpha-\beta})\text{ as }n\to\infty$$
}
\item For $\beta=1$, $\alpha=1/(2(q_1+q_2))$ and $c>q_1/(2(q_1+q_2))$,
{\small
$$E\|\hat\theta_n^{rec}-\theta\|^2=\left(\big(\frac{cd^{q_{1}}}{c-q_{1}/(2(q_1+q_2))}\big)^{2}\|B\|^{2}+\frac{c^{2}}{(2c-q_1/(q_1+q_2))d^{2q_{2}}} tr(\Sigma)\right)n^{-q_1/(q_1+q_2)}+o(n^{-q_1/(q_1+q_2)})\text{ as }n\to\infty$$
}
\item For $\beta=1$, $\alpha=1/(2(q_1+q_2))$ and $c\leq q_1/(2(q_1+q_2))$, or for $\beta=1$ and $\alpha\neq1/(2(q_1+q_2))$,
$$\limsup_{n\to\infty}n^{q_1/(q_1+q_2)}E\|\hat\theta_n^{rec}-\theta\|^2=\infty$$
\end{enumerate}
\end{theorem}

\proof{Proof of Theorem \ref{onmse}.}
Taking expectation of \eqref{recursionequ} and rearranging terms, we have
\begin{equation}\label{recmean}
E(\hat\theta_n^{rec}-\theta)=(1-\gamma_{n})E(\hat\theta_{n-1}^{rec}-\theta)+\gamma_{n}(E\hat\theta(\delta_n)-\theta)=(1-\gamma_{n})E(\hat\theta_{n-1}^{rec}-\theta)+\gamma_{n}(B\delta_{n}^{q_{1}}+o(\delta_{n}^{q_{1}}))
\end{equation}
If $\gamma_{n}=c/n$ and $\alpha\leq1/(2(q_1+q_2))$, we multiply~(\ref{recmean}) by $n^{q_{1}\alpha}$ to get
\begin{align*}
n^{q_{1}\alpha}E(\hat\theta_n^{rec}-\theta)&=(\frac{n}{n-1})^{q_{1}\alpha}(1-\frac{c}{n})(n-1)^{q_{1}\alpha}E(\hat\theta_{n-1}^{rec}-\theta)+\frac{c}{n}(Bd^{q_{1}}+o(1))\\
&=(1-\frac{c-q_{1}\alpha+o(1)}{n})(n-1)^{q_{1}\alpha}E(\hat\theta_{n-1}^{rec}-\theta)+\frac{c}{n}(Bd^{q_{1}}+o(1))
\end{align*}
For $c>q_1\alpha$, by Chung's lemma, $\lim_{n\to\infty}n^{q_{1}\alpha}E(\hat\theta_n^{rec}-\theta)=\frac{cd^{q_{1}}}{c-q_{1}\alpha}B$. Thus
$$E(\hat\theta_n^{rec}-\theta)=\frac{cd^{q_{1}}}{c-q_{1}\alpha}Bn^{-q_{1}\alpha}+o(n^{-q_{1}\alpha})$$
If $\gamma_{n}=c/n$ and $\alpha>1/(2(q_1+q_2))$, we multiply~(\ref{recmean}) by $n^{1/2-q_2\alpha}$ to get
$$n^{1/2-q_2\alpha}E(\hat\theta_n^{rec}-\theta)=(1-\frac{c-1/2+q_2\alpha+o(1)}{n})(n-1)^{1/2-q_2\alpha}E(\hat\theta_{n-1}^{rec}-\theta)+o(\frac{1}{n})$$
For $c>1/2-q_2\alpha$, by Chung's lemma, $\lim_{n\to\infty}n^{1/2-q_2\alpha}E(\hat\theta_n^{rec}-\theta)=0$. Thus
$$E(\hat\theta_n^{rec}-\theta)=o(n^{q_2\alpha-1/2})$$
Similarly, if $\gamma_{n}=c/n^{\beta},\beta<1$, we multiply~(\ref{recmean}) by $n^{q_{1}}\alpha$ to get
$$n^{q_{1}\alpha}E(\hat\theta_n^{rec}-\theta)=(1-\frac{c+o(1)}{n^{\beta}})(n-1)^{q_{1}\alpha}E(\hat\theta_{n-1}^{rec}-\theta)+\frac{c}{n^{\beta}}(Bd^{q_{1}}+o(1))$$
For $c>0$, by Chung's lemma, $\lim_{n\to\infty}n^{q_{1}\alpha}E(\hat\theta_n^{rec}-\theta)=Bd^{q_{1}}$. Thus
\begin{equation}\label{smallbetamean}
E(\hat\theta_n^{rec}-\theta)=Bd^{q_{1}}n^{-q_{1}\alpha}+o(n^{-q_{1}\alpha})
\end{equation}
Next we take covariance of \eqref{recursionequ} and by independence,
\begin{align}\label{coviteration}
Cov(\hat\theta_n^{rec})&=(1-\gamma_{n})^{2}Cov(\hat\theta_{n-1}^{rec})+\gamma_{n}^{2}Cov(\hat\theta(\delta_{n}))\notag\\
&=(1-\gamma_{n})^{2}Cov(\hat\theta_{n-1}^{rec})+\gamma_{n}^{2}\frac{Cov(\epsilon(\delta_{n}))}{\delta_{n}^{2q_{2}}}\notag\\
&=(1-\gamma_{n})^{2}Cov(\hat\theta_{n-1}^{rec})+\gamma_{n}^{2}n^{2q_{2}\alpha}\frac{\Sigma+o(1)}{d^{2q_{2}}}
\end{align}
If $\gamma_{n}=c/n$ and $\alpha\geq1/(2(q_1+q_2))$, we multiply \eqref{coviteration} by $n^{1-2q_{2}\alpha}$ and take trace to get
\begin{align}
n^{1-2q_{2}\alpha}tr(Cov(\hat\theta_n^{rec}))&=(\frac{n}{n-1})^{1-2q_{2}\alpha}(1-\frac{c}{n})^{2}(n-1)^{1-2q_{2}\alpha}tr(Cov(\hat\theta_{n-1}^{rec}))+\frac{c^{2}}{n}\frac{tr(\Sigma)+o(1)}{d^{2q_{2}}}\notag\\
&=(1-\frac{2c+2q_{2}\alpha-1+o(1)}{n})(n-1)^{1-2q_{2}\alpha}tr(Cov(\hat\theta_{n-1}^{rec}))+\frac{c^{2}}{n}\frac{tr(\Sigma)+o(1)}{d^{2q_{2}}}\label{bigalphacov}
\end{align}
For $c>1/2-q_2\alpha$, by Chung's lemma, $\lim_{n\to\infty}n^{1-2q_{2}\alpha}tr(Cov(\hat\theta_n^{rec}))=\frac{c^{2}tr(\Sigma)}{(2c+2q_{2}\alpha-1)d^{2q_{2}}}$. Thus
$$tr(Cov(\hat\theta_n^{rec}))=\frac{c^{2}tr(\Sigma)}{(2c+2q_{2}\alpha-1)d^{2q_{2}}}n^{2q_{2}\alpha-1}+o(n^{2q_{2}\alpha-1})$$
Similarly, if $\gamma_{n}=c/n^{\beta},\beta<1$, we multiply \eqref{coviteration} by $n^{\beta-2q_{2}\alpha}$ and take trace to get
\begin{align*}
n^{\beta-2q_{2}\alpha}tr(Cov(\hat\theta_n^{rec}))&=(\frac{n}{n-1})^{\beta-2q_{2}\alpha}(1-\frac{c}{n^{\beta}})^{2}(n-1)^{\beta-2q_{2}\alpha}tr(Cov(\hat\theta_{n-1}^{rec}))+\frac{c^{2}}{n^{\beta}}\frac{tr(\Sigma)+o(1)}{d^{2q_{2}}}\\
&=(1-\frac{2c+o(1)}{n^{\beta}})(n-1)^{\beta-2q_{2}\alpha}tr(Cov(\hat\theta_{n-1}^{rec}))+\frac{c^{2}}{n^{\beta}}\frac{tr(\Sigma)+o(1)}{d^{2q_{2}}}
\end{align*}
For $c>0$, by Chung's lemma, $\lim_{n\to\infty}n^{\beta-2q_{2}\alpha}tr(Cov(\hat\theta_n^{rec}))=\frac{ctr(\Sigma)}{2d^{2q_{2}}}$. Thus
$$tr(Cov(\hat\theta_n^{rec}))=\frac{ctr(\Sigma)}{2d^{2q_{2}}}n^{2q_{2}\alpha-\beta}+o(n^{2q_{2}\alpha-\beta})$$
In conclusion, if $\gamma_{n}=c/n$, $\alpha=1/(2(q_1+q_2))$ and $c>q_1/(2(q_1+q_2))$, then
\begin{align*}
E\|\hat\theta_n^{rec}-\theta\|^{2}&=\|E\hat\theta_n^{rec}-\theta\|^{2}+tr(Cov(\hat\theta_n^{rec}))\notag\\
&=\big(\frac{cd^{q_{1}}}{c-q_{1}\alpha}\big)^{2}\|B\|^{2}n^{-2q_{1}\alpha}+o(n^{-2q_{1}\alpha})+\frac{c^{2}tr(\Sigma)}{(2c+2q_{2}\alpha-1)d^{2q_{2}}}n^{2q_{2}\alpha-1}+o(n^{2q_{2}\alpha-1})\\
&=\left(\big(\frac{cd^{q_{1}}}{c-q_{1}/(2(q_1+q_2))}\big)^{2}\|B\|^{2}+\frac{c^{2}}{(2c-q_1/(q_1+q_2))d^{2q_{2}}} tr(\Sigma)\right)n^{-q_1/(q_1+q_2)}+o(n^{-q_1/(q_1+q_2)})
\end{align*}
If $\gamma_{n}=c/n$, $\alpha>1/(2(q_1+q_2))$ and $c>1/2-q_2\alpha$, then
\begin{equation}\label{msebigalpha}
E\|\hat\theta_n^{rec}-\theta\|^{2}=\frac{c^{2}tr(\Sigma)}{(2c+2q_{2}\alpha-1)d^{2q_{2}}}n^{2q_{2}\alpha-1}+o(n^{2q_{2}\alpha-1})
\end{equation}
Similarly, if $\gamma_{n}=c/n^{\beta},\beta<1$ and $c>0$, then
$$E\|\hat\theta_n^{rec}-\theta\|^{2}=d^{2q_{1}}\|B\|^{2}n^{-2q_{1}\alpha}+o(n^{-2q_{1}\alpha})+\frac{c}{2d^{2q_{2}}}tr(\Sigma)n^{2q_{2}\alpha-\beta}+o(n^{2q_{2}\alpha-\beta})$$
This completes the proof for part 1 and part 2 of the theorem.

Next we prove part 3 of the theorem. If $\alpha>1/(2(q_1+q_2))$ and $c>1/2-q_2\alpha$, we note from \eqref{msebigalpha} that
$$\lim_{n\to\infty}n^{q_1/(q_1+q_2)}E\|\hat\theta_n^{rec}-\theta\|^2=\infty $$
If $\alpha\geq1/(2(q_1+q_2))$ and $c\leq1/2-q_2\alpha$, and supposing that the sequence $n^{1-2q_2\alpha}tr(Cov(\hat\theta_n^{rec}))$ is bounded, then from \eqref{bigalphacov} we have that
$$n^{1-2q_{2}\alpha}tr(Cov(\hat\theta_n^{rec}))\geq(n-1)^{1-2q_{2}\alpha}tr(Cov(\hat\theta_{n-1}^{rec}))+\frac{C_1+o(1)}{n}$$
for some $C_1>0$, for all large enough $n$. Since $\sum_{n=1}^\infty1/n=\infty$, we get
$$n^{1-2q_{2}\alpha}tr(Cov(\hat\theta_n^{rec}))\to\infty\text{ as }n\to\infty$$
which is a contradiction. Thus
$$\limsup_{n\to\infty}n^{q_1/(q_1+q_2)}E\|\hat\theta_n^{rec}-\theta\|^2\geq\limsup_{n\to\infty}n^{1-2q_2\alpha}tr(Cov(\hat\theta_n^{rec}))=\infty$$
If $\alpha<1/(2(q_1+q_2))$, and supposing that the sequence $n^{q_1/(2(q_1+q_2))}E(\hat\theta_n^{rec}-\theta)$ is bounded, we multiply~(\ref{recmean}) by $n^{q_{1}/(2(q_1+q_2))}$ to get
\begin{eqnarray*}
&&n^{q_1/(2(q_1+q_2))}E(\hat\theta_n^{rec}-\theta)\\
&=&(1-\frac{c-q_{1}/(2(q_1+q_2))+o(1)}{n})(n-1)^{q_{1}/(2(q_1+q_2))}E(\hat\theta_{n-1}^{rec}-\theta)+\frac{c}{n^{1-q_1(1/(2(q_1+q_2))-\alpha)}}(Bd^{q_{1}}+o(1))\\
&=&(n-1)^{q_{1}/(2(q_1+q_2))}E(\hat\theta_{n-1}^{rec}-\theta)+\frac{cBd^{q_1}+o(1)}{n^{1-q_1(1/(2(q_1+q_2))-\alpha)}}
\end{eqnarray*}
Since $\sum_{n=1}^\infty1/n^{1-q_1(1/(2(q_1+q_2))-\alpha)}=\infty$, we get
$$n^{q_1/(2(q_1+q_2))}E(\hat\theta_n^{rec}-\theta)\to\infty\text{ as }n\to\infty$$
which is a contradiction. Thus
$$\limsup_{n\to\infty}n^{q_1/(q_1+q_2)}E\|\hat\theta_n^{rec}-\theta\|^2\geq\limsup_{n\to\infty}n^{q_1/(q_1+q_2)}\|E\hat\theta_n^{rec}-\theta\|^2=\infty$$
This completes the proof for part 3 of the theorem.\hfill\Halmos
\endproof

\proof{Proof of Theorem \ref{onmse single}.}
This follows immediately from Theorem \ref{onmse} by setting the dimension to 1.\hfill\Halmos
\endproof

\proof{Proof of Theorem~\ref{comd single}.}
We have
$$R^{rec}(\hat\theta(\cdot),d,c)=\frac{(\frac{cd^{q_{1}}}{c-\frac{q_{1}}{2(q_{1}+q_{2})}})^{2}B^{2}+\frac{c^{2}}{2d^{2q_{2}}(c-\frac{q_{1}}{2(q_{1}+q_{2})})}\sigma^2}{d^{2q_{1}}B^{2}+\frac{1}{d^{2q_{2}}}\sigma^2}$$
For any $d,B$ and $\sigma^2$, we have
$$R^{rec}(\hat\theta(\cdot),d,c)\leq\max\left\{\frac{(\frac{cd^{q_{1}}}{c-\frac{q_{1}}{2(q_{1}+q_{2})}})^{2}}{d^{2q_1}},\frac{\frac{c^{2}}{2d^{2q_{2}}(c-\frac{q_{1}}{2(q_{1}+q_{2})})}}{\frac{1}{d^{2q_2}}}\right\} =\max\left\{(\frac{c}{c-\frac{q_{1}}{2(q_{1}+q_{2})}})^{2},\frac{c^{2}}{2(c-\frac{q_{1}}{2(q_{1}+q_{2})})}\right\}$$
Note that the right hand side above is approachable by choosing $B$ or $\sigma^2$ to be arbitrarily big. Therefore
\begin{equation}\label{comdmaxexp}
\max_{\hat\theta(\cdot)\in\Theta,d>0}R^{rec}(\hat\theta(\cdot),d,c) =\max\left\{(\frac{c}{c-\frac{q_{1}}{2(q_{1}+q_{2})}})^{2},\frac{c^{2}}{2(c-\frac{q_{1}}{2(q_{1}+q_{2})})}\right\}
\end{equation}
Now suppose that
$$(\frac{c}{c-\frac{q_{1}}{2(q_{1}+q_{2})}})^{2}>\frac{c^{2}}{2(c-\frac{q_{1}}{2(q_{1}+q_{2})})}$$
which is equivalent to $c<\frac{5q_{1}+4q_{2}}{2(q_{1}+q_{2})}$. Since the function $(\frac{c}{c-\frac{q_{1}}{2(q_{1}+q_{2})}})^{2}$ is monotonically decreasing in the region $\frac{q_1}{2(q_1+q_2)}<c<\frac{5q_{1}+4q_{2}}{2(q_{1}+q_{2})}$, we have
$$\max\left\{(\frac{c}{c-\frac{q_{1}}{2(q_{1}+q_{2})}})^{2},\frac{c^{2}}{2(c-\frac{q_{1}}{2(q_{1}+q_{2})})}\right\}=(\frac{c}{c-\frac{q_{1}}{2(q_{1}+q_{2})}})^{2}\geq(\frac{c}{c-\frac{q_{1}}{2(q_{1}+q_{2})}})^{2}\bigg\arrowvert_{c= \frac{5q_{1}+4q_{2}}{2(q_{1}+q_{2})}}$$ Similarly, suppose that
$$(\frac{c}{c-\frac{q_{1}}{2(q_{1}+q_{2})}})^{2}<\frac{c^{2}}{2(c-\frac{q_{1}}{2(q_{1}+q_{2})})}$$
which is equivalent to $c>\frac{5q_{1}+4q_{2}}{2(q_{1}+q_{2})}$. Since the function $\frac{c^{2}}{2(c-\frac{q_{1}}{2(q_{1}+q_{2})})}$ is monotonically increasing in the region $c>\frac{5q_{1}+4q_{2}}{2(q_{1}+q_{2})}$, we have
$$\max\left\{(\frac{c}{c-\frac{q_{1}}{2(q_{1}+q_{2})}})^{2},\frac{c^{2}}{2(c-\frac{q_{1}}{2(q_{1}+q_{2})})}\right\}=\frac{c^{2}}{2(c-\frac{q_{1}}{2(q_{1}+q_{2})})}\geq\frac{c^{2}}{2(c-\frac{q_{1}}{2(q_{1}+q_{2})})}\bigg\arrowvert_{c= \frac{5q_{1}+4q_{2}}{2(q_{1}+q_{2})}}$$
Thus the minimization of \eqref{comdmaxexp} gives us $c=\frac{5q_{1}+4q_{2}}{2(q_{1}+q_{2})}$, which solves
$$(\frac{c}{c-\frac{q_{1}}{2(q_{1}+q_{2})}})^{2}=\frac{c^{2}}{2(c-\frac{q_{1}}{2(q_{1}+q_{2})})}$$
and we note that both sides of this equation is $\frac{q_{1}^{2}}{16(q_{1}+q_{2})^{2}}+\frac{q_{1}}{2(q_{1}+q_{2})}+1$. \hfill\Halmos
\endproof

\proof{Proof of Theorem~\ref{uncomd single}.}
We have
$$R^{rec}(\hat\theta(\cdot),d,\tilde d,c)=\frac{(\frac{c\tilde d^{q_{1}}}{c-\frac{q_{1}}{2(q_{1}+q_{2})}})^{2}B^{2}+\frac{c^{2}}{2\tilde d^{2q_{2}}(c-\frac{q_{1}}{2(q_{1}+q_{2})})}\sigma^2}{d^{2q_{1}}B^{2}+\frac{1}{d^{2q_{2}}}\sigma^2}$$
For any $d,\tilde d, B$ and $\sigma^2$, we have
\begin{align*}
R^{rec}(\hat\theta(\cdot),d,\tilde d,c)&\leq\max\left\{\frac{(\frac{c\tilde d^{q_{1}}}{c-\frac{q_{1}}{2(q_{1}+q_{2})}})^{2}}{ d^{2q_1}},\frac{\frac{c^{2}}{2\tilde d^{2q_{2}}(c-\frac{q_{1}}{2(q_{1}+q_{2})})}}{\frac{1}{d^{2q_2}}}\right\}\\
&=\max\left\{(\frac{c}{c-\frac{q_{1}}{2(q_{1}+q_{2})}})^{2}\left(\frac{\tilde d}{d}\right)^{2q_1},\frac{c^{2}}{2(c-\frac{q_{1}}{2(q_{1}+q_{2})})}\frac{1}{\left(\frac{\tilde d}{d}\right)^{2q_2}}\right\}
\end{align*}
Note that the right hand side above is approachable by choosing $B$ or $\sigma^2$ to be arbitrarily big. Therefore
\begin{equation}\label{uncomdmaxexp}
\max_{\hat\theta(\cdot)\in\Theta,d>0}R^{rec}(\hat\theta(\cdot),d,\tilde d,c)=\max_{d>0}\max\left\{(\frac{c}{c-\frac{q_{1}}{2(q_{1}+q_{2})}})^{2}\eta^{2q_1},\frac{c^{2}}{2(c-\frac{q_{1}}{2(q_{1}+q_{2})})}\frac{1}{\eta^{2q_2}}\right\}
\end{equation}
where we let $\eta=\frac{\tilde d}{d}$. We minimize the right hand side of \eqref{uncomdmaxexp} via minimizing
\begin{equation}\label{uncomdmaxexpfixd}
\max\left\{(\frac{c}{c-\frac{q_{1}}{2(q_{1}+q_{2})}})^{2}\eta^{2q_1},\frac{c^{2}}{2(c-\frac{q_{1}}{2(q_{1}+q_{2})})}\frac{1}{\eta^{2q_2}}\right\}
\end{equation}
for each $d$. First, for any $c$, since both of the expressions in \eqref{uncomdmaxexpfixd} are monotonic in $\eta$, we need
$$
(\frac{c}{c-\frac{q_{1}}{2(q_{1}+q_{2})}})^{2}\eta^{2q_1}=\frac{c^{2}}{2(c-\frac{q_{1}}{2(q_{1}+q_{2})})}\frac{1}{\eta^{2q_2}}$$
which upon solving leads to
$$\eta=\left(\frac{c-\frac{q_1}{2(q_1+q_2)}}{2}\right)^{1/(2(q_1+q_2))}$$
Thus \eqref{uncomdmaxexpfixd} becomes
$$\max\left\{(\frac{c}{c-\frac{q_{1}}{2(q_{1}+q_{2})}})^{2}\eta^{2q_1},\frac{c^{2}}{2(c-\frac{q_{1}}{2(q_{1}+q_{2})})}\frac{1}{\eta^{2q_2}}\right\}=(\frac{c}{c-\frac{q_{1}}{2(q_{1}+q_{2})}})^{2}\eta^{2q_1}=\frac{1}{2^{q_1/(q_1+q_2)}}\frac{c^2}{\left(c-\frac{q_1}{2(q_1+q_2)}\right)^{\frac{q_1+2q_2}{q_1+q_2}}}$$
We then optimize $c$ over the region $c>\frac{q_1}{2(q_1+q_2)}$, i.e,
$$c={\arg\min}_{c>\frac{q_1}{2(q_1+q_2)}}\frac{1}{2^{q_1/(q_1+q_2)}}\frac{c^2}{\left(c-\frac{q_1}{2(q_1+q_2)}\right)^{\frac{q_1+2q_2}{q_1+q_2}}}=1$$
This gives $\eta=(\frac{q_{1}+2q_{2}}{4(q_{1}+q_{2})})^{\frac{1}{2(q_{1}+q_{2})}}$ and \eqref{uncomdmaxexpfixd} is $2^{\frac{2q_{2}}{q_{1}+q_{2}}}(\frac{q_{1}+2q_{2}}{q_{1}+q_{2}})^{-\frac{q_{1}+2q_{2}}{q_{1}+q_{2}}}$. We note that the optimal $c,\eta$ are independent of $d$, and therefore the value of \eqref{uncomdmaxexp} is also $2^{\frac{2q_{2}}{q_{1}+q_{2}}}(\frac{q_{1}+2q_{2}}{q_{1}+q_{2}})^{-\frac{q_{1}+2q_{2}}{q_{1}+q_{2}}}$.
 \hfill\Halmos
\endproof

Next, we consider the uniform-averaging scheme:

\begin{theorem}\label{averagemse multi}
Under Assumption \ref{multi assumption},  the MSE of $\hat\theta_n^{avg}$ in estimating $\theta$ behaves as follows:
\begin{enumerate}
\item For $\beta<1$ and $\alpha\leq1/(2(q_1+q_2))$,
$$E\|\hat\theta_n^{avg}-\theta\|^2=\big(\frac{d^{q_{1}}}{1-q_{1}\alpha}\big)^{2}\|B\|^{2}n^{-2q_{1}\alpha}+\frac{1}{(1+2q_{2}\alpha)d^{2q_{2}}}tr(\Sigma) n^{2q_{2}\alpha-1}+o(n^{-2q_{1}\alpha}+n^{2q_{2}\alpha-1})\text{ as }n\to\infty$$

\item For $\beta<1$ and $\alpha>1/(2(q_1+q_2))$,
$$E\|\hat\theta_n^{avg}-\theta\|^2=\frac{1}{(1+2q_{2}\alpha)d^{2q_{2}}}tr(\Sigma) n^{2q_{2}\alpha-1}+o(n^{2q_{2}\alpha-1})\text{ as }n\to\infty$$
\end{enumerate}
\end{theorem}

\proof{Proof of Theorem \ref{averagemse multi}.}
We first analyze $E\hat\theta_n^{avg}-\theta$,
For $0<\alpha\leq\frac{1}{2(q_{1}+q_{2})}$, since $-1<-q_{1}\alpha<0$, we have that
$$\int_{1}^{n+1}s^{-q_{1}\alpha}ds\leq\sum_{i=1}^{n}i^{-q_{1}\alpha}\leq\int_{0}^{n}s^{-q_{1}\alpha}ds$$
Thus
$$\sum_{i=1}^{n}i^{-q_{1}\alpha}=\int_{0}^{n}s^{-q_{1}\alpha}ds+o(\int_{0}^{n}s^{-q_{1}\alpha}ds)=\frac{n^{1-q_{1}\alpha}}{1-q_{1}\alpha}+o(n^{1-q_{1}\alpha})$$
and
$$\frac{1}{n}\sum_{i=1}^{n}i^{-q_{1}\alpha}=\frac{1}{1-q_{1}\alpha}n^{-q_{1}\alpha}+o(n^{-q_{1}\alpha})$$
From \eqref{smallbetamean} we have $E(\hat\theta_n^{rec}-\theta)=Bd^{q_{1}}n^{-q_{1}\alpha}+o(n^{-q_{1}\alpha})$.
Thus
$$E\hat\theta_n^{avg}-\theta=\frac{1}{n}\sum_{i=1}^{n}E(\hat\theta_{i}^{rec}-\theta)=\frac{1}{n}\sum_{i=1}^{n}\left(Bd^{q_{1}}i^{-q_{1}\alpha}+o(i^{-q_{1}\alpha})\right)=\frac{d^{q_{1}}}{1-q_{1}\alpha}Bn^{-q_{1}\alpha}+o(n^{-q_{1}\alpha})$$
For $\alpha>\frac{1}{2(q_{1}+q_{2})}$, by a similar analysis we get
\begin{displaymath}
E\hat\theta_n^{avg}-\theta=\frac{1}{n}\sum_{i=1}^{n}E(\hat\theta_{i}^{rec}-\theta) = \left\{ \begin{array}{ll}
O(\frac{1}{n^{q_{1}\alpha}}) & \textrm{if $-q_{1}\alpha>-1$}\\
O(\frac{log(n)}{n}) & \textrm{if $-q_{1}\alpha=-1$}\\
O(\frac{1}{n}) & \textrm{if $-q_{1}\alpha<-1$}
\end{array} \right.
\end{displaymath}
Since $1/2-q_2\alpha<1$ and $1/2-q_2\alpha<q_{1}\alpha$, we have $$n^{1/2-q_2\alpha}E(\hat\theta_n^{avg}-\theta)=o(1)$$

We then analyze $tr(Cov(\hat\theta_n^{avg}))$. Rewrite the iteration \eqref{recursionequ} as
$$\hat\theta_n^{rec}-E\hat\theta_n^{rec}=(1-\gamma_n)(\hat\theta_{n-1}^{rec}-\hat\theta_{n-1}^{rec})+\gamma_n(\hat\theta(\delta_n)-E\hat\theta(\delta_n))$$
Let $U_{n}=\hat\theta_n^{rec}-E\hat\theta_n^{rec}$. Thus
$$U_{n}=(1-\gamma_{n})U_{n-1}+\gamma_{n}v(\delta_n)$$
Following \cite{polyak1992acceleration}, we can write
$$U_n=\prod_{i=1}^{n}(1-\gamma_{i})U_0+\sum_{i=1}^{n}(\prod_{j=i+1}^{n}(1-\gamma_{j}))\gamma_{i}v(\delta_i)$$
Thus $\hat\theta_n^{avg}-E\hat\theta_n^{avg}$ can be written as
\begin{align*}
\hat\theta_n^{avg}-E\hat\theta_n^{avg}&=\frac{1}{n}\sum_{k=1}^{n}U_k\\
&=\frac{1}{n}\sum_{k=1}^{n}\prod_{i=1}^{k}(1-\gamma_{i})U_0+\frac{1}{n}\sum_{k=1}^{n}\sum_{i=1}^{k}(\prod_{j=i+1}^{k}(1-\gamma_{j}))\gamma_{i}v(\delta_i)\\
&=\frac{1}{n}\sum_{k=1}^{n}\prod_{i=1}^{k}(1-\gamma_{i})U_0+\frac{1}{n}\sum_{i=1}^{n}(\sum_{k=i}^{n}\prod_{j=i+1}^{k}(1-\gamma_{j}))\gamma_{i}v(\delta_i)\\
\end{align*}
Let
$$p_{n}=\sum_{k=1}^{n}\prod_{i=1}^{k}(1-\gamma_{i})$$
$$q_{n}^{i}=\gamma_{i}\sum_{k=i}^{n}\prod_{j=i+1}^{k}(1-\gamma_{j})$$
and $w_{n}^{i}=q_{n}^{i}-1$. Then
\begin{equation}\label{avgrecursion}
\hat\theta_n^{avg}-E\hat\theta_n^{avg}=\frac{p_{n}}{n}U_0+\frac{1}{n}\sum_{i=1}^{n}v(\delta_i)+\frac{1}{n}\sum_{i=1}^{n}w_{n}^{i}v(\delta_i)
\end{equation}
From Lemma 1 and Lemma 2 in \cite{polyak1992acceleration}, we have that
$$\lim_{n\to\infty}\frac{1}{n}\sum_{i=1}^{n}|w_{n}^{i}|=0,\textrm{ and }|w_{n}^{i}|\leq C_1,\,|p_{n}|\leq C_1,\textrm{ for some }C_1>0$$
Multiplying \eqref{avgrecursion} by $n^{1/2-q_2\alpha}$, we have
$$n^{1/2-q_2\alpha}(\hat\theta_n^{avg}-E\hat\theta_n^{avg})=\frac{p_{n}}{n^{1/2+q_2\alpha}}U_0+\frac{1}{n^{1/2+q_2\alpha}}\sum_{i=1}^{n}v(\delta_i)+\frac{1}{n^{1/2+q_2\alpha}}\sum_{i=1}^{n}w_{n}^{i}v(\delta_i)$$
Since $p_{n}$ is bounded, $E\|\frac{p_{n}}{n^{1/2+q_2\alpha}}U_0\|^2=o(1)$. Besides, by independence,
$$E\|\frac{1}{n^{1/2+q_2\alpha}}\sum_{i=1}^{n}w_{n}^{i}v(\delta_i)\|^2=\frac{1}{n^{1+2q_2\alpha}}\sum_{i=1}^{n}(w_{n}^{i})^2E\|v(\delta_i)\|^{2}\leq\frac{C_2}{n^{1+2q_2\alpha}}\sum_{i=1}^{n}|w_{n}^{i}|i^{2q_2\alpha}\leq\frac{C_2}{n}\sum_{i=1}^{n}|w_{n}^{i}|$$
for some $C_2>0$. Therefore, $E\|\frac{1}{n^{1/2+q_2\alpha}}\sum_{i=1}^{n}w_{n}^{i}v(\delta_i)\|^2=o(1)$. Thus
\begin{align*}
n^{1-2q_2\alpha}tr(Cov(\hat\theta_n^{avg}))&=\frac{1}{n^{1+2q_2\alpha}}\sum_{i=1}^{n}tr(Cov(v(\delta_i)))+o(1)\\
&=\frac{1}{n^{1+2q_2\alpha}}\sum_{i=1}^{n}i^{2q_2\alpha}\frac{tr(\Sigma)+o(1)}{d^{2q_2}}+o(1)\\
&=\frac{tr(\Sigma)}{(1+2q_2\alpha)d^{2q_2}}+o(1)
\end{align*}
In conclusion, for $\alpha\leq1/(2(q_1+q_2))$, we have
\begin{align*}
E\|\hat\theta_n^{avg}-\theta\|^2&=\|E\hat\theta_n^{avg}-\theta\|^2+tr(Cov(\hat\theta_n^{avg}))\\
&=\left(\frac{d^{q_{1}}}{1-q_{1}\alpha}\right)^2\|B\|^2n^{-2q_{1}\alpha}+\frac{tr(\Sigma)}{(1+2q_2\alpha)d^{2q_2}}n^{2q_2\alpha-1}+o(n^{-2q_{1}\alpha}+n^{2q_2\alpha-1})
\end{align*}
and for $\alpha>1/(2(q_1+q_2))$, we have
$$E\|\hat\theta_n^{avg}-\theta\|^2=\frac{tr(\Sigma)}{(1+2q_2\alpha)d^{2q_2}}n^{2q_2\alpha-1}+o(n^{2q_2\alpha-1})$$
\hfill\Halmos\endproof

\proof{Proof of Theorem~\ref{averagemse}.}
This follows immediately from Theorem \ref{averagemse multi} by setting the dimension to 1.\hfill\Halmos
\endproof

\proof{Proof of Theorem~\ref{avg single}.}
The proof follows exactly that of Theorem \ref{uncomd single} and setting the dimension to 1, by noting the equivalence between the MSE expressions in Theorem \ref{averagemse multi} and Theorem \ref{onmse} with $c=1$, $\beta=1$ and $\alpha=1/(2(q_1+q_2))$. \hfill\Halmos
\endproof

\proof{Proof of Theorem~\ref{fixed parameter}.}
This follows immediately by noting that the proofs for Theorems \ref{comd single} and \ref{uncomd single} apply exactly the same when $d$ is fixed.\hfill\Halmos
\endproof

\section{Proofs in Section \ref{sec:opt}}\label{sec:proof main}
We prove Theorem \ref{main thm}. Note that part of the proof has been sketched in Section \ref{sec:opt}, and for clarity we will have slight amount of repetition to make this proof self-contained.

\proof{Proof of Theorem \ref{main thm}.}
Let $\alpha=1/(2(q_1+q_2))$. For convenience, we skip the second subscript of $w_{j,n}$ and write $w_j$, and denote $w=(w_j)_{j=1,\ldots,n}$, when no confusion arises. We also assume $n_0=0$ without loss of generality.

First, we argue that $\sum_{j=1}^{n} w_j\to1$. Suppose not, then there exists a subsequence $n_k$ such that $\left|\sum_{j=1}^{n_k}w_j-1\right|>\epsilon_{0}$ for some $\epsilon_{0}>0$. Assume without loss of generality that $\sum_{j=1}^{n_k}w_j-1>\epsilon_{0}$. Moreover, suppose the sequence
\begin{equation}\label{sumone}
\sum_{j=1}^{n_k}w_j\left(B\frac{g(d)^{q_1}}{j^{\alpha q_1}}+o\left(\frac{1}{j^{\alpha q_1}}\right)\right)
\end{equation}
is bounded. We can choose a sufficiently large $\theta$ such that
$$\liminf_{k\to\infty}\left(\left(\sum_{j=1}^{n_{k}}w_j-1\right)\theta+\sum_{j=1}^{n_{k}}w_j\left(B\frac{g(d)^{q_1}}{j^{\alpha q_1}}+o\left(\frac{1}{j^{\alpha q_1}}\right)\right)\right)^2>0$$
On the other hand, suppose \eqref{sumone} is unbounded. Then we can choose $\theta=0$ so that
$$\limsup_{k\to\infty}\left(\left(\sum_{j=1}^{n_{k}}w_j-1\right)\theta+\sum_{j=1}^{n_{k}}w_j\left(B\frac{g(d)^{q_1}}{j^{\alpha q_1}}+o\left(\frac{1}{j^{\alpha q_1}}\right)\right)\right)^2=\infty$$
Therefore, either way we would have $R_{n_k}\to\infty$.


Now, we consider a particular scheme $w,g(\cdot)$ such that $\sum_jw_j=1$ and $g(d)=\eta d$ for some $\eta>0$. Then
\begin{align}\label{onrawmse}
\text{MSE}_{1}&=\left(Bd^{q_1}\eta^{q_1}\sum_{j=1}^n\frac{w_j(1+o(1))}{j^{\alpha q_1}}\right)^2+\frac{\sigma^2}{d^{2q_2}\eta^{2q_2}}\sum_{j=1}^nj^{2\alpha q_2}w_j^2(1+o(1))\notag\\
&=\left(Bd^{q_1}\eta^{q_1}\sum_{j=1}^n\frac{w_j}{j^{\alpha q_1}}\right)^2+\frac{\sigma^2}{d^{2q_2}\eta^{2q_2}}\sum_{j=1}^nj^{2\alpha q_2}w_j^2+\varepsilon_n
\end{align}
where $\varepsilon_n$ is an error term.

We consider the following optimization problem to obtain $w,\eta$ that minimizes \eqref{onrawmse} asymptotically:
\begin{equation}
\begin{array}{ll}
\min_{w,\eta}&S\\
\text{subject to}&S=\left(\eta^{q_1}\sum_{j=1}^n\frac{w_j}{j^{\alpha q_1}}\right)^2=\frac{1}{\eta^{2q_2}}\sum_{j=1}^nj^{2\alpha q_2}w_j^2\\
&\eta\leq K\\
&\sum_{j=1}^nw_j=1
\end{array}\label{opt}
\end{equation}
We call $S_n^*$ the optimal value of \eqref{opt}. We will show that
$$\max_{\hat\theta(\cdot)\in\Theta,d>0}R^{gen}(\hat\theta(\cdot),d,g(d),W)=\lim_{n\to\infty}n^{q_1/(q_1+q_2)}S_n^*$$
is the asymptotic minimax risk ratio we seek for, and consequently the solution $w,\eta$ to \eqref{opt} is the optimal configuration. In the following, we first obtain a characterization of the solution to \eqref{opt}, and then verify that the solution also ensures the error term $\varepsilon_n$ is negligible. Then we argue that no other configurations, namely $w,g(\cdot)$ such that $\sum_jw_j\to1$ and $g(\cdot)\in\mathcal F_K$ that can give a better risk ratio. Although the solution $\eta$ to \eqref{opt} may depend on $n$, we will demonstrate that $\eta$ converges to a positive number as $n\to\infty$, and it will be clear that substituting $\eta$ with its limit will not affect the asymptotic risk ratio.

We first solve \eqref{opt}. From the first constraint in \eqref{opt}, we have
\begin{equation}\eta=\left(\frac{\sum_{j=1}^nj^{2\alpha q_2}w_j^2}{\left(\sum_{j=1}^n\frac{w_j}{j^{\alpha q_1}}\right)^2}\right)^{1/(2(q_1+q_2))}\label{interim1}
\end{equation}
so that the second constraint is equivalent to
$$\sum_{j=1}^nj^{2\alpha q_2}w_j^2\leq K^{2(q_1+q_2)}\left(\sum_{j=1}^n\frac{w_j}{j^{\alpha q_1}}\right)^2$$
Moreover, by plugging in \eqref{interim1} the objective function becomes
$$\left|\sum_{j=1}^n\frac{w_j}{j^{\alpha q_1}}\right|^{2q_2/(q_1+q_2)}\left(\sum_{j=1}^nj^{2\alpha q_2}w_j^2\right)^{q_1/(q_1+q_2)}$$
Therefore, \eqref{opt} can be rewritten as
\begin{equation}
\begin{array}{ll}
\min_{w}&\left|\sum_{j=1}^n\frac{w_j}{j^{\alpha q_1}}\right|^{2q_2/(q_1+q_2)}\left(\sum_{j=1}^nj^{2\alpha q_2}w_j^2\right)^{q_1/(q_1+q_2)}\\
\text{subject to}&\sum_{j=1}^nj^{2\alpha q_2}w_j^2\leq K^{2(q_1+q_2)}\left(\sum_{j=1}^n\frac{w_j}{j^{\alpha q_1}}\right)^2\\
&\sum_{j=1}^nw_j=1
\end{array}\label{opt1}
\end{equation}

We now set $a=\sum_{j=1}^n\frac{w_j}{j^{\alpha q_1}}$, and write \eqref{opt1} as
\begin{equation}
\begin{array}{ll}
\min_{w,a}&|a|^{2q_2/(q_1+q_2)}\left(\sum_{j=1}^nj^{2\alpha q_2}w_j^2\right)^{q_1/(q_1+q_2)}\\
\text{subject to}&\sum_{j=1}^nj^{2\alpha q_2}w_j^2\leq K^{2(q_1+q_2)}a^2\\
&\sum_{j=1}^n\frac{w_j}{j^{\alpha q_1}}=a\\
&\sum_{j=1}^nw_j=1
\end{array}\label{opt2}
\end{equation}
which can be further reformulated as
\begin{equation}\min_a|a|^{2q_2/(q_1+q_2)}Z_n^*(a)^{2q_1/(q_1+q_2)}\label{opt6}
\end{equation}
where
\begin{equation}
\begin{array}{lll}
Z_n^*(a)=&\min_{w}&\left(\sum_{j=1}^nj^{2\alpha q_2}w_j^2\right)^{1/2}\\
&\text{subject to}&\sum_{j=1}^nj^{2\alpha q_2}w_j^2\leq K^{2(q_1+q_2)}a^2\\
&&\sum_{j=1}^n\frac{w_j}{j^{\alpha q_1}}=a\\
&&\sum_{j=1}^nw_j=1
\end{array}\label{opt3}
\end{equation}

We rewrite \eqref{opt3} as
\begin{equation}
\begin{array}{ll}
\min_{w}&\|\Sigma^{1/2}w\|\\
\text{subject to}&\|\Sigma^{1/2}w\|^2\leq K^{2(q_1+q_2)}a^2\\
&\mu'w=a\\
&\mathbb 1'w=1
\end{array}\label{opt4}
\end{equation}
where $\Sigma=\text{diag}(j^{2\alpha q_2})_{j=1,\ldots,n}\in\mathbb R^{n\times n}$, $\mu=(1/j^{\alpha q_1})_{j=1,\ldots,n}\in\mathbb R^n$, $\mathbb 1=(1)_j\in\mathbb R^n$, and $\|\cdot\|$ is the $L_2$-norm.

We now consider further
\begin{equation}
\begin{array}{lll}
\tilde Z_n^*(a)=&\min_{w}&\|\Sigma^{1/2}w\|\\
&\text{subject to}&\mu'w=a\\
&&\mathbb 1'w=1
\end{array}\label{opt5}
\end{equation}
and note that
\begin{equation}
Z_n^*(a)=\left\{\begin{array}{ll}\tilde Z_n^*(a)&\text{\ if\ }{\tilde Z_n^*(a)}^{2}\leq K^{2(q_1+q_2)}a^2\\\infty&\text{\ otherwise}\end{array}\right.\label{interim2}
\end{equation}
Thus \eqref{opt6} can be written as
\begin{equation}
\min_{a:{\tilde Z_n^*(a)}^2\leq K^{2(q_1+q_2)}a^2}|a|^{2q_2/(q_1+q_2)}\tilde Z_n^*(a)^{2q_1/(q_1+q_2)}\label{opt7}
\end{equation}
Therefore, our strategy to solve \eqref{opt} is to first obtain an optimal solution $w^*(a)=(w_j^*(a^*))_{j=1,\ldots,n}$ to \eqref{opt5} and correspondingly $\tilde Z_n^*(a)$, and then solve for an optimal solution $a^*$ to \eqref{opt7}. The optimal value of \eqref{opt7} is equal to that of \eqref{opt}. Moreover, the optimal configuration is given by $w^*(a^*)$, and $\eta^*$, where
$$\eta^*=\left(\frac{\sum_{j=1}^nj^{2\alpha q_2}{w_j^*(a^*)}^2}{\left(\sum_{j=1}^n\frac{w_j^*(a^*)}{j^{\alpha q_1}}\right)^2}\right)^{1/(2(q_1+q_2))}
$$
by using \eqref{interim1}.


We now solve \eqref{opt5}. Consider the Lagrangian
$$\|\Sigma^{1/2}w\|-\lambda_1(\mu'w-a)-\lambda_2(\mathbb 1'w-1)$$
Differentiating with respect to $w$ and equating to 0, we get
$$\frac{\Sigma w}{\|\Sigma^{1/2}w\|}-\lambda_1\mu-\lambda_2\mathbb 1=0$$
which gives
$$w=\Sigma^{-1}(\lambda_1\mu+\lambda_2\mathbb 1)=\lambda_1\Sigma^{-1}\mu+\lambda_2\Sigma^{-1}\mathbb 1$$
for some $\lambda_1,\lambda_2$ (scaled by $\|\Sigma^{1/2}w\|$ compared to the ones displayed before). Note that this is equivalent to
\begin{equation}
w_j=\frac{\lambda_1}{j^{\alpha(q_1+2q_2)}}+\frac{\lambda_2}{j^{2\alpha q_2}}\label{interim3}
\end{equation}
for $j=1,\ldots,n$.

Now, for convenience, we write
\begin{equation}
w=\left[\Sigma^{-1}\mu\ \ \Sigma^{-1}\mathbb 1\right]\left[\begin{array}{c}\lambda_1\\\lambda_2\end{array}\right]\label{revised interm}
\end{equation}
so that
$$\mu'w=\left[\mu'\Sigma^{-1}\mu\ \ \mu'\Sigma^{-1}\mathbb 1\right]\left[\begin{array}{c}\lambda_1\\\lambda_2\end{array}\right]$$
and
$$\mathbb 1'w=\left[\mathbb 1'\Sigma^{-1}\mu\ \ \mathbb 1'\Sigma^{-1}\mathbb 1\right]\left[\begin{array}{c}\lambda_1\\\lambda_2\end{array}\right]$$
Setting $\mu'w=a$ and $\mathbb 1'w=1$, we get
$$\left[\begin{array}{c}\lambda_1\\\lambda_2\end{array}\right]=\left[\begin{array}{cc}\mu'\Sigma^{-1}\mu&\mu'\Sigma^{-1}\mathbb 1\\\mathbb 1'\Sigma^{-1}\mu&\mathbb 1'\Sigma^{-1}\mathbb 1\end{array}\right]^{-1}\left[\begin{array}{c}a\\1\end{array}\right]$$
Let $\phi(\kappa)=\sum_{j=1}^n1/j^\kappa$. We can write this as
\begin{align}
\left[\begin{array}{c}\lambda_1\\\lambda_2\end{array}\right]&=\left[\begin{array}{cc}\phi(\alpha(2q_1+2q_2))&\phi(\alpha(q_1+2q_2))\\\phi(\alpha(q_1+2q_2))&\phi(2\alpha q_2)\end{array}\right]^{-1}\left[\begin{array}{c}a\\1\end{array}\right]\notag\\
&=\left[\begin{array}{cc}\phi(1)&\phi\left(\frac{q_1+2q_2}{2(q_1+q_2)}\right)\\\phi\left(\frac{q_1+2q_2}{2(q_1+q_2)}\right)&\phi\left(\frac{q_2}{q_1+q_2}\right)\end{array}\right]^{-1}\left[\begin{array}{c}a\\1\end{array}\right]\label{interim4}
\end{align}

From \eqref{revised interm}, we can represent the optimal weights as
$$w^*(a)=\left[\Sigma^{-1}\mu\ \ \Sigma^{-1}\mathbb 1\right]\left[\begin{array}{cc}\mu'\Sigma^{-1}\mu&\mu'\Sigma^{-1}\mathbb 1\\\mathbb 1'\Sigma^{-1}\mu&\mathbb 1'\Sigma^{-1}\mathbb 1\end{array}\right]^{-1}\left[\begin{array}{c}a\\1\end{array}\right]$$
and write
\begin{align*}
\tilde Z_n^*(a)^2&=\|\Sigma^{1/2}w^*(a)\|^2\\
&=\left[a\ \ 1\right]\left[\begin{array}{cc}\mu'\Sigma^{-1}\mu&\mu'\Sigma^{-1}\mathbb 1\\\mathbb 1'\Sigma^{-1}\mu&\mathbb 1'\Sigma^{-1}\mathbb 1\end{array}\right]^{-1}\left[\begin{array}{c}\mu'\Sigma^{-1}\\\mathbb 1'\Sigma^{-1}\end{array}\right]\Sigma\left[\Sigma^{-1}\mu\ \ \Sigma^{-1}\mathbb 1\right]\left[\begin{array}{cc}\mu'\Sigma^{-1}\mu&\mu'\Sigma^{-1}\mathbb 1\\\mathbb 1'\Sigma^{-1}\mu&\mathbb 1'\Sigma^{-1}\mathbb 1\end{array}\right]^{-1}\left[\begin{array}{c}a\\1\end{array}\right]\\
&=\left[a\ \ 1\right]\left[\begin{array}{cc}\mu'\Sigma^{-1}\mu&\mu'\Sigma^{-1}\mathbb 1\\\mathbb 1'\Sigma^{-1}\mu&\mathbb 1'\Sigma^{-1}\mathbb 1\end{array}\right]^{-1}\left[\begin{array}{cc}\mu'\Sigma^{-1}\mu&\mu'\Sigma^{-1}\mathbb 1\\\mathbb 1'\Sigma^{-1}\mu&\mathbb 1'\Sigma^{-1}\mathbb 1\end{array}\right]\left[\begin{array}{cc}\mu'\Sigma^{-1}\mu&\mu'\Sigma^{-1}\mathbb 1\\\mathbb 1'\Sigma^{-1}\mu&\mathbb 1'\Sigma^{-1}\mathbb 1\end{array}\right]^{-1}\left[\begin{array}{c}a\\1\end{array}\right]\\
&=\left[a\ \ 1\right]\Xi\left[\begin{array}{c}a\\1\end{array}\right]
\end{align*}
where
\begin{equation}\label{interim6}
\Xi=\left[\begin{array}{cc}\xi_{11}&\xi_{12}\\\xi_{21}&\xi_{22}\end{array}\right]=\left[\begin{array}{cc}\mu'\Sigma^{-1}\mu&\mu'\Sigma^{-1}\mathbb 1\\\mathbb 1'\Sigma^{-1}\mu&\mathbb 1'\Sigma^{-1}\mathbb 1\end{array}\right]^{-1}
=\left[\begin{array}{cc}\phi(1)&\phi\left(\frac{q_1+2q_2}{2(q_1+q_2)}\right)\\\phi\left(\frac{q_1+2q_2}{2(q_1+q_2)}\right)&\phi\left(\frac{q_2}{q_1+q_2}\right)\end{array}\right]^{-1}
\end{equation}
Thus, \eqref{opt7} can be written as
\begin{equation}
\min_{a:(K^{2(q_1+q_2)}-\xi_{11})a^2-2\xi_{12}a-\xi_{22}\geq0}|a|^{2q_2/(q_1+q_2)}\left(\xi_{11}a^2+2\xi_{12}a+\xi_{22}\right)^{q_1/(q_1+q_2)}\label{opt8}
\end{equation}

We now find the asymptotic limit of \eqref{opt} scaled by $n^{q_1/(q_1+q_2)}$. First, we write $a$ as $\tilde a/n^{q_1/(2(q_1+q_2))}$. Then, reparametrizing by $\tilde a$ and denoting $\bar Z_n^*(\tilde a)=\tilde Z_n^*(\tilde a/n^{q_1/(2(q_1+q_2))})$, we have
$$\bar Z_n^*(\tilde a)^2=\left[\frac{\tilde a}{n^{q_1/(2(q_1+q_2))}}\ \ 1\right]\Xi\left[\begin{array}{c}\frac{\tilde a}{n^{q_1/(2(q_1+q_2))}}\\1\end{array}\right]$$
Note that $\phi(1)\sim\log n$ and $\phi(\kappa)\sim\frac{1}{1-\kappa}n^{1-\kappa}$ for $\kappa<1$ as $n\to\infty$. Thus,
\begin{eqnarray}
&&n^{q_1/(q_1+q_2)}\bar Z_n^*(\tilde a)^2\notag\\
&=&n^{q_1/(q_1+q_2)}\left[\frac{\tilde a}{n^{q_1/(2(q_1+q_2))}}\ \ 1\right]\left[\begin{array}{cc}(1+o(1))\log n&\frac{2(q_1+q_2)(1+o(1))}{q_1}n^{q_1/(2(q_1+q_2))}\\\frac{2(q_1+q_2)(1+o(1))}{q_1}n^{q_1/(2(q_1+q_2))}&\frac{(q_1+q_2)(1+o(1))}{q_1}n^{q_1/(q_1+q_2)}\end{array}\right]^{-1}\left[\begin{array}{c}\frac{\tilde a}{n^{q_1/(2(q_1+q_2))}}\\1\end{array}\right]\notag\\
&=&n^{q_1/(q_1+q_2)}\left[\frac{\tilde a}{n^{q_1/(2(q_1+q_2))}}\ \ 1\right]\frac{\left[\begin{array}{cc}\frac{(q_1+q_2)(1+o(1))}{q_1}n^{q_1/(q_1+q_2)}&-\frac{2(q_1+q_2)(1+o(1))}{q_1}n^{q_1/(2(q_1+q_2))}\\-\frac{2(q_1+q_2)(1+o(1))}{q_1}n^{q_1/(2(q_1+q_2))}&(1+o(1))\log n \end{array}\right]}{\frac{(q_1+q_2)}{q_1}n^{q_1/(q_1+q_2)}\log n(1+o(1))-\frac{4(q_1+q_2)^2}{q_1^2}n^{q_1/(q_1+q_2)}(1+o(1))}\left[\begin{array}{c}\frac{\tilde a}{n^{q_1/(2(q_1+q_2))}}\\1\end{array}\right]\notag\\
&=&\left[\tilde a\ \ 1\right]\frac{\left[\begin{array}{cc}\frac{(q_1+q_2)(1+o(1))}{q_1}&-\frac{2(q_1+q_2)(1+o(1))}{q_1}\\-\frac{2(q_1+q_2)(1+o(1))}{q_1}&(1+o(1))\log n\end{array}\right]}{\frac{q_1+q_2}{q_1}\log n(1+o(1))-\frac{4(q_1+q_2)^2}{q_1^2}(1+o(1))}\left[\begin{array}{c}\tilde a\\1\end{array}\right]\label{scaledzed}\\
&=&\left[\tilde a\ \ 1\right]\left(\tilde\Xi+o(1)\right)\left[\begin{array}{c}\tilde a\\1\end{array}\right]\notag
\end{eqnarray}
where
$$\tilde\Xi=
\left[\begin{array}{cc}0&0\\0&\frac{q_1}{q_1+q_2}\end{array}\right]$$
Rewriting \eqref{opt7} in terms of $\tilde a$, we have that \eqref{opt7}, when multiplying its objective value by $n^{q_1/(q_1+q_2)}$, becomes
\begin{equation}\label{fdoptimi}
\min_{\tilde a:n^{q_1/(q_1+q_2)}\bar Z_n^*(\tilde a)^2\leq K^{2(q_1+q_2)}\tilde a^2}|\tilde a|^{2q_2/(q_1+q_2)}\left(n^{q_1/(q_1+q_2)}\bar Z_n^*(\tilde a)^2\right)^{q_1/(q_1+q_2)}
\end{equation}
We consider an asymptotic version of \eqref{fdoptimi} given by
\begin{equation}\label{asympoptimi}
\min_{\tilde a:\frac{q_1}{q_1+q_2}\leq K^{2(q_1+q_2)}\tilde a^2}|\tilde a|^{2q_2/(q_1+q_2)}\left(\frac{q_1}{q_1+q_2}\right)^{q_1/(q_1+q_2)}=\frac{q_1}{q_1+q_2}\frac{1}{K^{2q_2}}
\end{equation}
We now argue that the absolute value of an optimal solution to \eqref{fdoptimi}, denoted $\tilde a_n^*$, converges to $\sqrt{q_1/(q_1+q_2)}(1/K^{q_1+q_2})$, from which it follows immediately that the value of \eqref{fdoptimi} converges to $(q_1/(q_1+q_2))(1/K^{2q_2})$, as $n\to\infty$. Suppose that $\left||\tilde a_{n_k}^*|-\sqrt{q_1/(q_1+q_2)}(1/K^{q_1+q_2})\right|>\epsilon$, for some $\epsilon>0$ and subsequence $n_k\to\infty$. If for infinitely many $k$ it holds that
$|\tilde a_{n_k}^*|<\sqrt{q_1/(q_1+q_2)}(1/K^{q_1+q_2})-\epsilon$, then $\tilde a_{n_k}^*$ is excluded from the feasible region of \eqref{fdoptimi}, namely
\begin{equation}\label{feasibleregion}
\tilde a_{n_k}^*\notin\left\{\tilde a:n_k^{q_1/(q_1+q_2)}\bar Z_{n_k}^*(\tilde a)^2\leq K^{2(q_1+q_2)}\tilde a^2\right\}
\end{equation}
infinitely often, which is a contradiction by the definition of $\tilde a_n^*$. Therefore we have $|\tilde a_{n_k}^*|>\sqrt{q_1/(q_1+q_2)}(1/K^{q_1+q_2})+\epsilon$ for all $k$ sufficiently large. Next, from \eqref{scaledzed}, we have that $n^{q_1/(q_1+q_2)}\bar Z_n^*(\tilde a)^2$ is bounded from below uniformly over $\tilde a$:
\begin{align*}
\min_{\tilde a}n^{q_1/(q_1+q_2)}\bar Z_n^*(\tilde a)^2&=\min_{\tilde a}\left[\tilde a\ \ 1\right]\frac{\left[\begin{array}{cc}\frac{(q_1+q_2)(1+o(1))}{q_1}&-\frac{2(q_1+q_2)(1+o(1))}{q_1}\\-\frac{2(q_1+q_2)(1+o(1))}{q_1}&(1+o(1))\log n\end{array}\right]}{\frac{q_1+q_2}{q_1}\log n(1+o(1))-\frac{4(q_1+q_2)^2}{q_1^2}(1+o(1))}\left[\begin{array}{c}\tilde a\\1\end{array}\right]\\
&=\frac{(1+o(1))\left(\log n-\frac{4(q_1+q_2)}{q_1}\right)}{\frac{q_1+q_2}{q_1}\log n(1+o(1))-\frac{4(q_1+q_2)^2}{q_1^2}(1+o(1))}\\
&=\frac{q_1}{q_1+q_2}(1+o(1))
\end{align*}
where in the second equality we have used the property for the minimum of a quadratic function. Suppose that $|\tilde a_{n_k}^*|$ is unbounded, then
$$\limsup_{k\to\infty}|\tilde a_{n_k}^*|^{2q_2/(q_1+q_2)}\left(n_k^{q_1/(q_1+q_2)}\bar Z_{n_k}^*(\tilde a_{n_k}^*)^2\right)^{q_1/(q_1+q_2)}=\infty$$
which is again a contradiction. Thus we are left with the case where $|\tilde a_{n_k}^*|>\sqrt{q_1/(q_1+q_2)}(1/K^{q_1+q_2})+\epsilon$ and $|\tilde a_{n_k}^*|$ is bounded. Note that since $|\tilde a_{n_k}^*|$ is bounded we have
$$\left|n_k^{q_1/(q_1+q_2)}\bar Z_{n_k}^*(\tilde a_{n_k}^*)^2-\frac{q_1}{q_1+q_2}\right|=o(1)$$
Thus
\begin{equation}\label{falseminimizor}
|\tilde a_{n_k}^*|^{2q_2/(q_1+q_2)}\left(n_k^{q_1/(q_1+q_2)}\bar Z_{n_k}^*(\tilde a^*_{n_k})^2\right)^{q_1/(q_1+q_2)}\geq\left(\sqrt{\frac{q_1}{q_1+q_2}}\frac{1}{K^{q_1+q_2}}+\epsilon\right)^{2q_2/(q_1+q_2)}\left(\frac{q_1}{q_1+q_2}+o(1)\right)^{q_1/(q_1+q_2)}
\end{equation}On the other hand, since the feasible region to \eqref{fdoptimi} admits $\tilde a$ such that $\tilde a=\sqrt{q_1/(q_1+q_2)}(1/K^{q_1+q_2})+o(1)$, we have for such $\tilde a$
$$|\tilde a|^{2q_2/(q_1+q_2)}\left(n_k^{q_1/(q_1+q_2)}\bar Z_{n_k}^*(\tilde a)^2\right)^{q_1/(q_1+q_2)}=\frac{q_1}{q_1+q_2}\frac{1}{K^{2q_2}}+o(1)$$
Comparing the above equation to \eqref{falseminimizor}, we again have a contradiction. Thus we have shown that the absolute value of a solution $\tilde a_n^*$ to \eqref{fdoptimi} converges to $\sqrt{q_1/(q_1+q_2)}(1/K^{q_1+q_2})$. Besides, we have
\begin{equation}\label{etalimit}
\eta^*=\left(\frac{\tilde Z_n^*(a^*)^2}{a^{*2}}\right)^{1/(2(q_1+q_2))}\to\left(\frac{q_1/(q_1+q_2)}{(q_1/(q_1+q_2))(1/K^{2(q_1+q_2)})}\right)^{1/(2(q_1+q_2))}=K
\end{equation}

We now show that the error term in \eqref{onrawmse} is asymptotically negligible, which is true if
\begin{equation}\label{negcondition1}
\sum_{j=1}^n\frac{w_j^*(1+o(1))}{j^{\alpha q_1}}=\sum_{j=1}^n\frac{w_j^*}{j^{\alpha q_1}}+o(\sum_{j=1}^n\frac{w_j^*}{j^{\alpha q_1}})
\end{equation}
and
\begin{equation}\label{negcondition2}
\sum_{j=1}^nj^{2\alpha q_2}{w_j^*}^2(1+o(1))=\sum_{j=1}^nj^{2\alpha q_2}{w_j^*}^2+o(\sum_{j=1}^nj^{2\alpha q_2}{w_j^*}^2)
\end{equation}
For \eqref{negcondition1}, let $\gamma=\left(o(\frac{1}{j^{\alpha q_1}})\right)_{j=1,\cdots,n}\in\mathbb R^{n}$. We first show that $\gamma^{'}\Sigma^{-1}\mu=o(\mu^{'}\Sigma^{-1}\mu)$. For any $\epsilon>0$, by the definition of $\gamma$ we have that $|\gamma_j|\leq\frac{\epsilon}{2}\mu_j$ for all $j> j_0$, for some $j_0=j_0(\epsilon)$. Thus for all $n>j_0$
$$\gamma^{'}\Sigma^{-1}\mu=\sum_{j=1}^n\gamma_j\Sigma^{-1}_{jj}\mu_j=\sum_{j=1}^{j_0}\gamma_j\Sigma^{-1}_{jj}\mu_j+\sum_{j=j_0+1}^n\gamma_j\Sigma^{-1}_{jj}\mu_j$$
where $\Sigma^{-1}_{jj}$ denote the $j$th diagonal element of $\Sigma^{-1}$. Since $\mu^{'}\Sigma^{-1}\mu\to\infty$ as $n\to\infty$, we have for all $n$ large enough
\begin{align*}
\left|\gamma^{'}\Sigma^{-1}\mu\right|&\leq\left|\sum_{j=1}^{j_0}\gamma_j\Sigma^{-1}_{jj}\mu_j\right|+\sum_{j=j_0+1}^n\left|\gamma_j\right|\Sigma^{-1}_{jj}\mu_j\\
&\leq \frac{\epsilon}{2}\mu^{'}\Sigma^{-1}\mu+\frac{\epsilon}{2}\sum_{j=j_0+1}^n\mu_j\Sigma^{-1}_{jj}\mu_j\\
&\leq \epsilon\mu^{'}\Sigma^{-1}\mu
\end{align*}
Thus $\gamma^{'}\Sigma^{-1}\mu=o(\mu^{'}\Sigma^{-1}\mu)$. Similarly we can show that $\gamma^{'}\Sigma^{-1}\mathbb1=o(\mu^{'}\Sigma^{-1}\mathbb1)$.
We note that
\begin{eqnarray*}
&&\sum_{j=1}^nw_j^*o(\frac{1}{j^{\alpha q_1}})\\
&=&\left[\gamma^{'}\Sigma^{-1}\mu\ \ \gamma^{'}\Sigma^{-1}\mathbb 1\right]\left[\begin{array}{cc}\phi(1)&\phi\left(\frac{q_1+2q_2}{2(q_1+q_2)}\right)\\\phi\left(\frac{q_1+2q_2}{2(q_1+q_2)}\right)&\phi\left(\frac{q_2}{q_1+q_2}\right)\end{array}\right]^{-1}\left[\begin{array}{c}a^*\\1\end{array}\right]\\
&=&\left[o(\log n)\ \ o(n^{q_1/(2(q_1+q_2))})\right]\frac{\left[\begin{array}{cc}\frac{(q_1+q_2)(1+o(1))}{q_1}n^{q_1/(q_1+q_2)}&-\frac{2(q_1+q_2)(1+o(1))}{q_1}n^{q_1/(2(q_1+q_2))}\\-\frac{2(q_1+q_2)(1+o(1))}{q_1}n^{q_1/(2(q_1+q_2))}&(1+o(1))\log n \end{array}\right]}{\frac{(q_1+q_2)}{q_1}n^{q_1/(q_1+q_2)}\log n(1+o(1))-\frac{4(q_1+q_2)^2}{q_1^2}n^{q_1/(q_1+q_2)}(1+o(1))}\left[\begin{array}{c}O(n^{-q_1/(2(q_1+q_2))})\\1\end{array}\right]\\
&=&\left[o(\log n)\ \ o(n^{q_1/(2(q_1+q_2))})\right]\frac{\left[\begin{array}{c}O(n^{q_1/(2(q_1+q_2))})\\O(\log n)\end{array}\right]}{\frac{(q_1+q_2)}{q_1}n^{q_1/(q_1+q_2)}\log n(1+o(1))-\frac{4(q_1+q_2)^2}{q_1^2}n^{q_1/(q_1+q_2)}(1+o(1))}\\
&=&\frac{o(n^{q_1/(2(q_1+q_2))}\log n)}{\frac{(q_1+q_2)}{q_1}n^{q_1/(q_1+q_2)}\log n(1+o(1))-\frac{4(q_1+q_2)^2}{q_1^2}n^{q_1/(q_1+q_2)}(1+o(1))}\\
&=&o(n^{-q_1/(2(q_1+q_2))})\\
&=&o(\sum_{j=1}^n\frac{w_j^*}{j^{\alpha q_1}})
\end{eqnarray*}
where we have used the expression for $w^*$.
For \eqref{negcondition2}, since
$$n^{q_1/(q_1+q_2)}\sum_{j=1}^n(w_j^*)^2o(j^{2\alpha q_2})\to0$$
we also have that
$$\sum_{j=1}^{n}(w_{j}^*)^{2}j^{2\alpha q_{2}}o(1)=o(\sum_{j=1}^{n}(w_{j}^*)^{2}j^{2\alpha q_{2}})$$

Next, to show that no other choices of $W,g(\cdot)$ can asymptotically dominate $w^*(a^*)$ and $g(\cdot)$ where $g(d)=Kd$ obtained above, we consider a configuration of $w,\eta$ obtained by solving $w$ in
\begin{equation}
\begin{array}{ll}
\min_{w}&Q=\frac{1}{K^{2q_2}}\sum_{j=1}^nj^{2\alpha q_2}w_j^2\\
\text{subject to}&\frac{1}{K^{2q_2}}\sum_{j=1}^nj^{2\alpha q_2}w_j^2>K^{2q_1}\left(\sum_{j=1}^n\frac{w_j}{j^{\alpha q_1}}\right)^2\\
&\sum_{j=1}^nw_j=1
\end{array}\label{optextra}
\end{equation}
and choosing $\eta=K$. Let $Q_n^*$ the optimal value of \eqref{optextra}.
We first solve \eqref{optextra} and show that it does not give a smaller optimal value than \eqref{opt} asymptotically. Consider
\begin{equation}
\begin{array}{lll}
\tilde L_n(a)=&\min_{w}&\|\Sigma^{1/2}w\|\\
&\text{subject to}&\|\Sigma^{1/2}w\|^2>K^{2(q_1+q_2)}a^2\\
&&\mu'w=a\\
&&\mathbb 1'w=1
\end{array}\label{degenerateopt}
\end{equation}
For any $a$, if the optimal solution to \eqref{opt5} satisfies
$$\tilde Z_n^*(a)^2>K^{2(q_1+q_2)}a^2$$
then the minimum in definition \eqref{degenerateopt} is attainable and $\tilde L_n(a)=\tilde Z_n^*(a)$. Otherwise, the minimum is possibly unattainable and $\tilde L_n(a)\geq K^{2(q_1+q_2)}a^2$.
Let $a=\tilde a/n^{q_1/(2(q_1+q_2))}$. Reparametrizing by $\tilde a$, we denote $\bar L_n(\tilde a)=\tilde L_n(\tilde a/n^{q_1/(2(q_1+q_2))})$. Multiplying the objective value of \eqref{optextra} by $n^{q_1/(q_1+q_2)}$, we have
$$n^{q_1/(q_1+q_2)}Q_n^*= n^{q_1/(q_1+q_2)}\inf_{\tilde a}\bar L_n(\tilde a)^2\frac{1}{K^{2q_2}}$$
regardless of whether the minimum in \eqref{optextra} is attainable.
Suppose that $n^{q_1/(q_1+q_2)}\bar Z_n^*(\tilde a)^2>K^{2(q_1+q_2)}\tilde a^2$. From \eqref{scaledzed} we have that $\tilde a$ is asymptotically bounded. Thus for some $o(1)$ uniform over such $\tilde a$, we have
$$n^{q_1/(q_1+q_2)}\bar L_n(\tilde a)^2\frac{1}{K^{2q_2}}=n^{q_1/(q_1+q_2)}\bar Z_n^*(\tilde a)^2\frac{1}{K^{2q_2}}\geq\frac{q_1}{q_1+q_2}(1+o(1))\frac{1}{K^{2q_2}}$$
On the other hand, suppose that $n^{q_1/(q_1+q_2)}\bar Z_n^*(\tilde a)^2\leq K^{2(q_1+q_2)}\tilde a^2$. Then
\begin{align*}
n^{q_1/(q_1+q_2)}\bar L_n(\tilde a)^2\frac{1}{K^{2q_2}}&\geq K^{2(q_1+q_2)}\tilde a^2\frac{1}{K^{2q_2}}\\
&\geq \left(K^{2(q_1+q_2)}\tilde a^2\right)^{q_2/(q_1+q_2)}\left(n^{q_1/(q_1+q_2)}\bar Z_n^*(\tilde a)^2\right)^{q_1/(q_1+q_2)}\frac{1}{K^{2q_2}}\\
&\geq \min_{\tilde a:n^{q_1/(q_1+q_2)}\bar Z_n^*(\tilde a)^2\leq K^{2(q_1+q_2)}\tilde a^2}|\tilde a|^{2q_2/(q_1+q_2)}\left(n^{q_1/(q_1+q_2)}\bar Z_n^*(\tilde a)^2\right)^{q_1/(q_1+q_2)}\\
&\geq\frac{q_1}{q_1+q_2}\frac{1}{K^{2q_2}}(1+o(1))
\end{align*}
for some $o(1)$ independent of $\tilde a$. Therefore, we have
$$\liminf_{n\to\infty}n^{q_1/(q_1+q_2)}Q_n^*\geq\lim_{n\to\infty}n^{q_1/(q_1+q_2)}S_n^*$$
Using \eqref{asympoptimi} we identify the AMRR in the first part of the theorem. Using \eqref{interim3}, \eqref{interim4}, \eqref{interim6}, \eqref{opt8} and \eqref{etalimit} we identify the solution in the second part of the theorem.

It remains to argue that no other configurations $w,g(\cdot)$ such that $\sum_j^n w_j\to1$ and $g(\cdot)\in\mathcal F_K$ that can give a better risk ratio. We first note that we can solve the variant of optimization \eqref{opt}
\begin{equation}
\begin{array}{ll}
\min_{w,\eta}&T\\
\text{subject to}&T=\left(\eta^{q_1}\sum_{j=1}^n\frac{w_j}{j^{\alpha q_1}}\right)^2=\frac{1}{\eta^{2q_2}}\sum_{j=1}^nj^{2\alpha q_2}w_j^2\\
&\eta\leq K\\
&\sum_{j=1}^nw_j=1+o(1)
\end{array}\label{optvariant}
\end{equation}
via solving \eqref{opt7} like before, but this time with the constraint $\mathbb 1'w=1$ in \eqref{opt5} replaced by $\mathbb 1'w=1+o(1)$. This additional $o(1)$ term can be seen, by following the arguments above, to eventually be absorbed with no effect on the analysis. This gives an optimal solution $T_n^*$ such that $\lim_{n\to\infty}T_n^*/S_n^*=1$. Similarly, the variant of optimization \eqref{optextra}
\begin{equation}
\begin{array}{ll}
\min_{w}&P=\frac{1}{K^{2q_2}}\sum_{j=1}^nj^{2\alpha q_2}w_j^2\\
\text{subject to}&\frac{1}{K^{2q_2}}\sum_{j=1}^nj^{2\alpha q_2}w_j^2>K^{2q_1}\left(\sum_{j=1}^n\frac{w_j}{j^{\alpha q_1}}\right)^2\\
&\sum_{j=1}^nw_j=1+o(1)
\end{array}\label{optextravar}
\end{equation}
gives an optimal value $P_n^*$ such that $\liminf_{n\to\infty}n^{q_1/(q_1+q_2)}P_n^*\geq\lim_{n\to\infty}n^{q_1/(q_1+q_2)}T_n^*=(q_1/(q_1+q_2))(1/K^{2q_2})$.

We aim to find $\hat\theta(\cdot)\in\Theta$ and $d>0$, such that
$$R^{gen}(\hat\theta(\cdot),d,g(d),W)\geq \frac{q_1}{q_1+q_2}\frac{1}{K^{2q_2}}$$
We will consider $\hat\theta(\cdot)\in\Theta$ with $\theta=0$ and without the higher order terms in the asymptotic expansion, i.e. $b(\delta)=B\delta^{q_1}$ for some $B\neq0$ and $v(\delta)=\frac{\epsilon(\delta)}{\delta^{q_2}}$ such that $Var(\epsilon(\delta))=\sigma^2>0$. In this case
$$\text{MSE}_{1}=\left(Bd^{q_1}\left(\frac{g(d)}{d}\right)^{q_1}\sum_{j=1}^n\frac{w_j}{j^{\alpha q_1}}\right)^2+\frac{\sigma^2}{d^{2q_2}}\left(\frac{d}{g(d)}\right)^{2q_2}\sum_{j=1}^nj^{2\alpha q_2}w_j^2$$
For any $W,g(\cdot)$, we note that two cases can arise:
\begin{enumerate}
\item For all large enough $n$, either
$$\left(\frac{g(d)}{d}\right)^{2q_1}\left(\sum_{j=1}^n\frac{w_j}{j^{\alpha q_1}}\right)^2=\left(\frac{d}{g(d)}\right)^{2q_2}\sum_{j=1}^nj^{2\alpha q_2}w_j^2$$
or
$$\left(\frac{g(d)}{d}\right)^{2q_1}\left(\sum_{j=1}^n\frac{w_j}{j^{\alpha q_1}}\right)^2\neq\left(\frac{d}{g(d)}\right)^{2q_2}\sum_{j=1}^nj^{2\alpha q_2}w_j^2$$
but there exists $\eta\leq K$, such that
$$\eta^{2q_1}\left(\sum_{j=1}^n\frac{w_j}{j^{\alpha q_1}}\right)^2=\frac{1}{\eta^{2q_2}}\sum_{j=1}^nj^{2\alpha q_2}w_j^2$$
\item There exists a subsequence $n_k$ such that
$$K^{2q_1}\left(\sum_{j=1}^{n_k}\frac{w_j}{j^{\alpha q_1}}\right)^2<\frac{1}{K^{2q_2}}\sum_{j=1}^{n_k}j^{2\alpha q_2}w_j^2$$
\end{enumerate}
For case 1, by the definition of $T_n^*$ we have
$$\max\bigg\{\left(\frac{g(d)}{d}\right)^{2q_1}\left(\sum_{j=1}^n\frac{w_j}{j^{\alpha q_1}}\right)^2,\left(\frac{d}{g(d)}\right)^{2q_2}\sum_{j=1}^nj^{2\alpha q_2}w_j^2\bigg\}\geq T_n^*$$
Thus
\begin{eqnarray*}
&&\max_{\hat\theta(\cdot)\in\Theta,d>0}R^{gen}(\hat\theta(\cdot),d,g(d),W)\\
&\geq&\max_{B\neq0,\sigma^2>0,d>0}\limsup_{n\to\infty} \frac{\left(Bd^{q_1}\left(\frac{g(d)}{d}\right)^{q_1}\sum_{j=1}^n\frac{w_j}{j^{\alpha q_1}}\right)^2+\frac{\sigma^2}{d^{2q_2}}\left(\frac{d}{g(d)}\right)^{2q_2}\sum_{j=1}^nj^{2\alpha q_2}w_j^2}{\frac{1}{n^{q_1/(q_1+q_2)}}\left(B^2d^{2q_1}+\frac{\sigma^2}{d^{2q_2}}\right)+o(\frac{1}{n^{q_1/(q_1+q_2)}})}\\
&\geq&\lim_{n\to\infty}n^{q_1/(q_1+q_2)}T_n^*\\
&\geq&\frac{q_1}{q_1+q_2}\frac{1}{K^{2q_2}}
\end{eqnarray*}
For case 2, we have
$$\left(\frac{g(d)}{d}\right)^{2q_1}\left(\sum_{j=1}^{n_k}\frac{w_j}{j^{\alpha q_1}}\right)^2\leq K^{2q_1}\left(\sum_{j=1}^{n_k}\frac{w_j}{j^{\alpha q_1}}\right)^2<\frac{1}{K^{2q_2}}\sum_{j=1}^{n_k}j^{2\alpha q_2}w_j^2\leq \left(\frac{d}{g(d)}\right)^{2q_2}\sum_{j=1}^{n_k}j^{2\alpha q_2}w_j^2$$
Thus by the definition of $P_n^*$
\begin{eqnarray*}
&&\max_{\hat\theta(\cdot)\in\Theta,d>0}R^{gen}(\hat\theta(\cdot),d,g(d),W)\\
&\geq&\max_{B\neq0,\sigma^2>0,d>0}\limsup_{k\to\infty} \frac{\left(Bd^{q_1}\left(\frac{g(d)}{d}\right)^{q_1}\sum_{j=1}^{n_k}\frac{w_j}{j^{\alpha q_1}}\right)^2+\frac{\sigma^2}{d^{2q_2}}\left(\frac{d}{g(d)}\right)^{2q_2}\sum_{j=1}^{n_k}j^{2\alpha q_2}w_j^2}{\frac{1}{{n_k}^{q_1/(q_1+q_2)}}\left(B^2d^{2q_1}+\frac{\sigma^2}{d^{2q_2}}\right)+o(\frac{1}{{n_k}^{q_1/(q_1+q_2)}})}\\
&\geq&\max_{B\neq0,\sigma^2>0,d>0}\limsup_{k\to\infty} {n_k}^{q_1/(q_1+q_2)}\frac{1}{K^{2q_2}}\sum_{j=1}^{n_k}j^{2\alpha q_2}w_j^2\frac{B^2d^{2q_1}\frac{\left(\left(\frac{g(d)}{d}\right)^{q_1}\sum_{j=1}^{n_k}\frac{w_j}{j^{\alpha q_1}}\right)^2}{\frac{1}{K^{2q_2}}\sum_{j=1}^{n_k}j^{2\alpha q_2}w_j^2}+\frac{\sigma^2}{d^{2q_2}}}{\left(B^2d^{2q_1}+\frac{\sigma^2}{d^{2q_2}}\right)+o(1)}\\
&\geq& \limsup_{k\to\infty} {n_k}^{q_1/(q_1+q_2)}\frac{1}{K^{2q_2}}\sum_{j=1}^{n_k}j^{2\alpha q_2}w_j^2\text{\quad (by considering $B$ arbitrarily close to $0$)}\\
&\geq&\limsup_{k\to\infty}n_k^{q_1/(q_1+q_2)}P_{n_k}^*\\
&\geq&\frac{q_1}{q_1+q_2}\frac{1}{K^{2q_2}}
\end{eqnarray*}
\hfill\Halmos\endproof

\proof{Proof of Corollary \ref{main cor}.}
This follows immediately by noting that the proof of Theorem \ref{main thm} applies exactly the same when $d$ is fixed.
\hfill\Halmos
\endproof

\end{APPENDICES}
\end{document}